\documentclass[acmlarge]{acmart}

\usepackage{fullpage}

\usepackage{bm}

\usepackage{booktabs}
\usepackage{lscape}
\usepackage{longtable}
\usepackage{amsmath}
\usepackage[ruled]{algorithm2e}
\usepackage{url}
\usepackage{enumitem}
\usepackage{natbib}

\usepackage{geometry}
\usepackage{setspace}

\setcitestyle{numbers, sort&compress}

\usepackage{soul} 
\usepackage{color, xcolor} 
\soulregister{\cite}7 
\soulregister{\citep}7 
\soulregister{\citet}7 
\soulregister{\ref}7 
\soulregister{\pageref}7 

\AtBeginDocument{%
  \providecommand\BibTeX{{%
    \normalfont B\kern-0.5em{\scshape i\kern-0.25em b}\kern-0.8em\TeX}}}

\setcopyright{acmcopyright} \acmJournal{CSUR} \acmYear{2022} \acmVolume{1} \acmNumber{1} \acmArticle{1} \acmMonth{1} \acmPrice{15.00}\acmDOI{10.1145/3524499}

\begin{document}

\captionsetup{font={tiny}}

\title{EEG based Emotion Recognition: A Tutorial and Review}

\author{Xiang Li}
\authornote{Xiang Li and Yazhou Zhang contribute equally and share the co-first authorship.}
\email{xiangli@sdas.org}
\orcid{0000-0001-5471-1236}
\affiliation{%
  \institution{Qilu University of Technology (Shandong Academy of Sciences), Shandong Computer Science Center (National Supercomputer Center in Jinan)}
  \city{Jinan}
  \country{China}
}

\author{Yazhou Zhang}
\email{yzzhang@zzuli.edu.cn}
\orcid{0000-0002-5699-0176}
\authornotemark[1]
\affiliation{%
  \institution{Software Engineering College, Zhengzhou University of Light Industry}
  \city{Zhengzhou}
  \country{China}}
  \affiliation{%
  \institution{State Key Lab. for Novel Software Technology, Nanjing University}
  \city{Nanjing}
  \country{China}}

\author{Prayag Tiwari}
\authornotemark[2]
 \email{prayag.tiwari@aalto.fi}
 \orcid{0000-0002-2851-4260}
\affiliation{%
  \institution{Department of Computer Science, Aalto University}
  \country{Finland}}

  \author{Dawei Song}
\authornote{Corresponding authors.}
  \email{dwsong@bit.edu.cn}
  \orcid{0000-0002-8660-3608}
\affiliation{%
  \institution{School of Computer Science and Technology, Beijing Institute of Technology}
  \city{Beijing}
  \country{China}}
  
  \author{Bin Hu}
\authornotemark[2]
  \email{bh@bit.edu.cn}
  \orcid{0000-0003-3514-5413}
\affiliation{%
  \institution{Institute of Engineering Medicine, Beijing Institute of Technology}
  \city{Beijing}
  \country{China}}
  
    \author{Meihong Yang}
  \email{yangmh@sdas.org}
  \orcid{0000-0002-6837-2690}
\affiliation{%
  \institution{Qilu University of Technology (Shandong Academy of Sciences), Shandong Computer Science Center (National Supercomputer Center in Jinan)}
  \city{Jinan}
  \country{China}}


    \author{Zhigang Zhao}
    \email{zhaozhg@sdas.org}
    \orcid{0000-0002-4144-8587}
\affiliation{%
  \institution{Qilu University of Technology (Shandong Academy of Sciences), Shandong Computer Science Center (National Supercomputer Center in Jinan)}
  \city{Jinan}
  \country{China}}

    \author{Neeraj Kumar}
\authornotemark[2]
  \email{neeraj.kumar@thapar.edu}
   \orcid{0000-0002-3020-3947}
\affiliation{%
  \institution{Department of Computer Science and Engineering, Thapar Institute of Engineering and Technology (Deemed University)}
  \city{Patiala (Punjab)}
  \country{India}}

\author{Pekka Marttinen}
\authornotemark[2]
 \email{pekka.marttinen@aalto.fi}
 \orcid{0000-0001-7078-7927}
\affiliation{%
  \institution{Department of Computer Science, Aalto University}
  \country{Finland}}

\renewcommand{\shortauthors}{Li and Zhang, et al.}

\begin{abstract}
Emotion recognition technology through analyzing the EEG signal is currently an essential concept in Artificial Intelligence and holds great potential in emotional health care, human-computer interaction, multimedia content recommendation, etc. Though there have been several works devoted to reviewing EEG-based emotion recognition, the content of these reviews needs to be updated. In addition, those works are either fragmented in content or only focus on specific techniques adopted in this area but neglect the holistic perspective of the entire technical routes. Hence, in this paper, we review from the perspective of researchers who try to take the first step on this topic. We review the recent representative works in the EEG-based emotion recognition research and provide a tutorial to guide the researchers to start from the beginning. The scientific basis of EEG-based emotion recognition in the psychological and physiological levels is introduced. Further, we categorize these reviewed works into different technical routes and illustrate the theoretical basis and the research motivation, which will help the readers better understand why those techniques are studied and employed. At last, existing challenges and future investigations are also discussed in this paper, which guides the researchers to decide potential future research directions.
\end{abstract}

\ccsdesc[500]{Human-centered computing~Human computer interaction (HCI)}
\ccsdesc[500]{Computing methodologies~Artificial intelligence}
\ccsdesc[500]{Human-centered computing~Ubiquitous and mobile computing}

\keywords{EEG, emotion recognition, affective computing, psychophysiological computing}

\maketitle

\small
\section{Introduction}
Emotion (or affect) recognition (or detection) has increasingly drawn attention from researchers with a multidisciplinary background. It is the leading scientific problem in Affective Computing, which is a comparatively new research field proposed by Picart  \cite{picard2000affective}, namely how to empower computer systems to precisely process, recognize and comprehend emotional information expressed by a human for natural human-computer interactions (HCI) \cite{picard2001building}. It is an important concept both in Artificial Intelligence and Ambient Intelligence \cite{dunne2021survey}. Interdisciplinary knowledge is needed in Affective Computing, and the findings can further promote the development of the various disciplines, including Computer Science, Electronic Engineering, Human Factors Engineering, Psychology, Neuroscience, Medical Science, etc.


As a complex psychological state, emotion is reflected in physical behaviors and physiological activities \cite{adolphs2018neuroscience}. In the past decade, much effort has been made to recognize emotions based on affective information gathered from various physical behaviors and physiological activities, such as voices from the microphone, signals from neurophysiological activity measuring devices, videos from cameras, and texts from the website, etc. The essence of emotion detection research is utilizing statistical machine learning techniques (e.g., classification, regression, or clustering) to identify users' different emotional states in real-time or offline. The problem is challenging as we have to dig and utilize the latent components embedded in the weak and noisy emotion-related data sources, including natural language, facial expressions, speech, body gestures, bio-signals, text, eye gaze, etc., collected from multiple monitor platforms mentioned above. Currently, judging emotional states based on physiological activities (physiological clue) is a hot topic in Affective Computing. Some psycho or physiological researches have manifested there exists dependencies between the physiological process and the emotion cognition process even though there still exists debates on the order of appearance of these two processes \cite{cannon1927james}. Hence, Computational Psychophysiology-based approaches are supposed to be effective complements for facial or speech information (non-physiological clue) based recognition methods, whose performance could be greatly influenced when users intentionally dissemble their true feelings by wearing `social masks'. Considering the central nervous system (brain) regulates and controls the autonomic nervous system to participate in emotional processes, directly utilizing the brain activities information (e.g., EEG) to study the emotional cognition mechanism and recognize emotional state is especially worth studying.

The EEG based emotion recognition has wide application prospects. For example, developing emotion-aware driver assistance systems for cars is currently recognized as a potential way to enhance driving safety \cite{halim2020identification}. In the field of neurology, identified emotions in response to specific stimuli and the corresponding neural activities can be analyzed for diagnosing some affective disorders, such as PTSD (post straumatic stress disorder) \cite{rozgic2014multi} and depression \cite{valstar2013avec,cai2020feature}. The psychological studies have found an attentional bias phenomenon in depressed individuals, in which increased attention to negative or dysphoric stimuli rather than positive contents \cite{beck1979cognitive}. Besides, the detected emotion can be utilized to guide various emotion disorder therapy, e.g. robot-assisted therapy \cite{gumuslu2020emotion} and music-assisted therapy \cite{chang2017personalized,ramirez2018eeg}. In the field of Information Retrieval (IR), emotion recognition (or called sentiment analysis) has always been an active research field. The detected emotion states can be used for emotion-associated IR needs, e.g., for implicit tagging of the multimedia contents \cite{soleymani2011multimodal,moshfeghi2013effective,koelstra2013fusion}, or for enriching user profiles to improve the topical relevance of the recommended multimedia contents \cite{arapakis2009enriching}. Emotion recognition contributes to building the human-centered information retrieval (IR) system \cite{moshfeghi2012role,yadollahi2017current}. In the field of leisure and entertainment, e.g., the computer gaming, researchers sought to detect gamers' emotional states in order to adjust to game's level of difficulty, punishment, and encouragement \cite{alakus2020database}. In Virtual Reality (VR) applications, e.g., VR in education, the influence of emotional states on memory has been verified, the positive mood has beneficial effects on spatial learning. Hence, the emotion should be recognized during learning in VR environment \cite{stavroulia2019assessing,vesisenaho2019virtual}.

Essentially, EEG-based emotion recognition belongs to one kind of pattern recognition research. As we know, resolving a pattern recognition problem usually contains several main steps, namely as follows:
\begin{itemize}[leftmargin=*]
\item Firstly, the definition and quantification of the recognition target should be determined, by which the problem can be resolved as one computable problem.
\item Secondly, acquiring sufficient and valid research data is vitally important in preparing comprehensive space for searching model decision boundaries.
\item Thirdly, preprocessing the data and acquiring the representation (e.g., feature extraction or feature learning) are typically needed in building pattern recognition models. Target-related representative characteristics extracted from raw data can eliminate redundant information that may influence model construction.
\item Finally, the recognition models are designed, trained, and evaluated based on the processed data iteratively until a model with acceptable recognition accuracy can be determined.
\end{itemize}

Domestic and overseas research on EEG-based emotion recognition studies also covers these topics. Such being the case, in this paper, we choose to outline the review covering these topics as mentioned above. Before writing this survey, there have been several works devoted to reviewing EEG-based emotion recognition. It's necessary to introduce those related survey papers that have been published in recent three years and expound on the motivation and the necessity to make a new survey paper for this research field. Firstly, the contents of several survey papers need to be updated. For example, the surveys of \citet{lotte2018review} and \citet{alarcao2017emotions} were published before 2019, and the surveyed works were developed between 2006 and 2017. \citet{dadebayev2021eeg} mainly introduces the application of consumer-grade EEG acquisition devices (e.g., the Emotive, OpenBCI, and NeuroSky) in emotion recognition. It mainly reviews works between 2014 and 2019. Although the surveys conducted by \citet{suhaimi2020eeg}, \citet{arya2021affect}, and \citet{rahman2021recognition} were published after 2021, most of the reviewed methodologies were developed before 2019. For example, \citet{arya2021affect} and \citet{suhaimi2020eeg} choose to survey works between 2009 and 2018. 
Secondly, these surveys emphasize traditional feature engineering and Machine Learning-based approaches. The Deep Learning-based approaches have not been systematically introduced. 
We think these surveys should be improved by adding more content about Deep Learning. Thirdly, several surveys are not written specifically for the affective brain-computer interface (BCI) tasks. For example, although \citet{craik2019deep} focuses on introducing the Deep Learning-based methodologies applied to EEG modeling. 
The reviewed EEG classification tasks are not restricted to emotion recognition. Other tasks related to mental workload, motor imagery, event-related potential, seizure detection, and sleep stage scoring are also included. In addition, its content needs to be updated considering its publication year is 2019. Similarly, the survey of \citet{lotte2018review} mainly discusses the classification algorithms for motor imagery-based BCI.

In our review, we focus on the EEG-based emotion recogntion tasks, and try to add more introduction about the up-to-date methodologies, especially various Deep Learning-based approaches that have not been systematically reviewed in these two papers. Further, we will discuss several up-to-date research problems, such as the `domain shift' problem, the few-shot learning problem, etc. We also discuss potential routes in this research field, such as the large-scale pre-trained EEG model applied in emotion recognition. In addition, other surveys only focus on specific techniques adopted in this area but neglect the holistic technical perspective. Hence, in this paper, we review from the perspective of researchers who try to take the first step on this topic. In addition to reviewing the recent advances in the EEG-based emotion recognition research, we also provide a tutorial to guide the researchers to start from the very beginning. For example, we introduce the scientific basis of EEG-based emotion recognition in the psychological and physiological levels. Further, we categorize these reviewed works into different technical routes and illustrate the theoretical basis and the research motivation, which will help the readers better understand why those techniques are studied and employed. At last, existing challenges and future investigations are also discussed in this paper, which guides the researchers to decide potential future research directions. We also carefully discuss various EEG segmentation strategies when discussing the evaluation methods, which need to be treated carefully but have been overlooked in related surveys. We believe this review provides valuable research references and research problems for researchers who are new to this field. Overall, this review paper draws on the advantages of many survey articles, and is a high-quality complementation to these classical surveys.


\section{Preliminaries and Basic Knowledge}

\subsection{Definition and quantification of emotional state}
Acquiring users' behavioral or physiological data with quantified emotional state labels is vital for statistical analysis-based psychological research and intelligent computing-based emotion recognition systems. Hence, the emotional state characterized by target samples should be effectively recorded and quantitatively evaluated. All the quantitative methods of emotion can be divided into two main categories, as follows.
\subsubsection{Discrete type of emotion quantification model}
According to Darwin's theory of evolution, human emotions are discrete, and they are preserved by natural selection \cite{ekman2009darwin}. \citet{ekman1999basic} and  \citet{plutchik2003emotions} approved the point of view of Darwin and then proposed that the emotion was composed of six or eight basic states, e.g., anger, anticipation, fear, sadness, disgust, trust, surprise, joy, etc., which could be further expanded to fifteen or more types. On this basis, an improved model was put forward by scholars to improve the quantification of emotions further. For example, the Palette Theory indicated that the basic emotional states could be further taken as Primary Colors, and then other emotional states would be generated by mixing the Primary Colors. For instance, the two basic emotions of surprise and sadness can be compounded into disappointment. Subsequently, Plutchik developed an emotion wheel representation method \cite{plutchik2013theories}. In addition, a tree type or hierarchical type of emotion quantification method was proposed by Parrott and Gerrod \cite{Parrott2001Emotions}. In essence, these quantification models can all be incorporated into the discrete-type emotion quantization models.


When building a recognition system on the discrete type of emotion quantification model, we typically regard this task as a classification modeling task, in which classification models are applied and studied, including support vector machine (SVM) algorithm, K-nearest neighbor (KNN) algorithm, decision tree algorithm, etc.

\subsubsection{Continuous dimensional type of emotion quantification model}
\begin{figure*}[!htbp]
\centering
\includegraphics[width=0.6\textwidth]{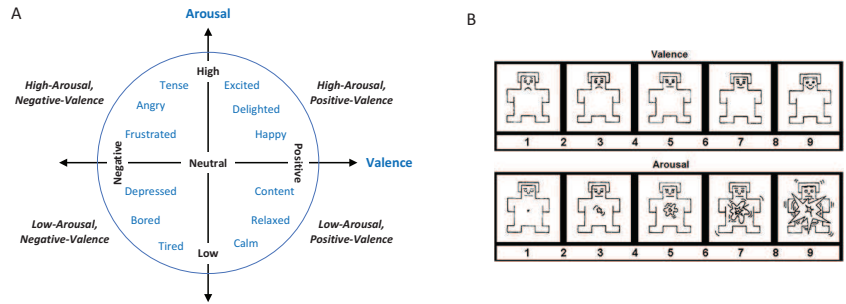}\\
\caption{Valence-Arousal Bipolar Coordinate System Porposed by Russell (A) and the corresponding Self-Assessment Manikins scale (B) \cite{langeslag2018effects}.}
\label{fig:valence_arousal}
\end{figure*}

The boundaries for distinguishing the emotional states are vague, and the changes and evolutions of states are continuous without breakpoint. Partitioning the emotional states into a dozen discrete types can only show the main aspects of emotion and fail in the accurate quantification of emotional state. In addition, discrete emotional labels are not consistent among various cultures and nationalities. For example, we can not find a corresponding translation in Polish for the emotion of ‘disgust’. Hence, the continuous dimensional type of emotion quantification method is proposed. This quantification method uses several mutually orthogonal basic axes to display different dimensions of emotion, which solves the contradiction between the discrete quantification method and the rich emotional connotation. The Valence-Arousal bipolar emotional quadrant system was put forward by Russell \cite{Russell1979Affective}, which has been widely accepted in Affective Computing. As shown in Figure \ref{fig:valence_arousal}(A), the classical two dimensions of Valence and Arousal are used to depict the Valence level and Arousal level of emotion.

The values (or ratings) of the Valence axis from positive to negative refer to the measurement for individuals' happy and sad degrees. Likewise, a positive value in Arousal indicates an activated state (excitement), while the negative value indicates an unactivated state (calmness). In addition to the two standard base axes, adding more dimensions for a comprehensive measurement of emotions is feasible. The Dominance dimension represents the dominance control degree of the individual in the emotional process. When an external environment controls a user, the emotional state is at a lower dominance level (e.g., surprise, fear, etc.). Conversely, when a user can master the external environment, the emotional state is at a higher dominance level. It should be pointed out that the various discrete emotional states can be located to specific locations in the continuous dimensional state space with a one-to-one correspondence. For example, the sadness emotion is located in the low Arousal-low Dominance-low Valence coordinate space in the continuous emotional coordinate system, while the happiness emotion is located in a high Arousal-high Dominance-high Valence coordinate space. A broadly adopted method for evaluating continuous emotional states is the Self-Assessment Manikins (SAM) scale-based approach. SAM is designed by introducing the manikins into the questionnaire to visually evaluate the degree of Valence and arousal. The SAM questionnaire generally sets a discrete scale from 1 to 9, as shown in Figure \ref{fig:valence_arousal}(B).

When building a recognition system on the continuous type of emotion quantification model, researchers can either tackle this task through classification modeling or regression modeling. Briefly speaking, constructing a model directly based on samples with continuous emotional ratings is a regression modeling task, in which the built model (e.g., ridge regression, recurrent neural networks) should be able to precisely predict the unknown samples' emotional ratings. Whereas, when regarding this task as a classification problem, researchers typically need to divide the emotional ratings into several levels, the samples are assigned with specific class labels according to at which level the emotional ratings are located. After that, classification algorithms are trained on the labeled samples and applied to inferring the unknown samples' emotional classes.

\subsection{EEG activity's specificity and neural correlate in the emotional process}
\begin{figure*}[!htbp]
\centering
\includegraphics[width=0.35\textwidth]{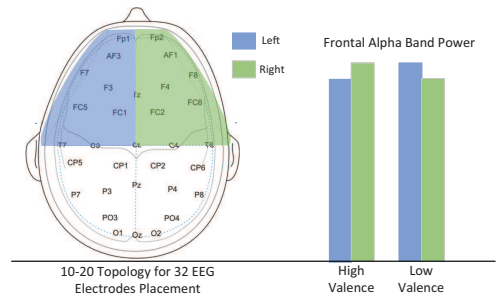}\\
\caption{Differences between high and low Valence on left and right frontal EEG alpha power in music listening (Note that EEG power is inversely related to activity, thus lower power reflects more activity \cite{schmidt2001frontal}) .}
\label{fig:asymmetry}
\end{figure*}

The critical function of the central nervous system (brain) is to regulate the whole-body physical or psycho activities to participate in the emotional process. Intracranial EEG or non-implanted EEG can record the physiological activities of the central nervous system. For the non-implanted EEG, the EEG signals are acquired by deploying multiple electrodes on different brain regions according to the standard 10-20 topology, as shown in Figure \ref{fig:asymmetry}.    

The correlation between neural sources and emotions is one of the key scientific questions of cognitive neuroscience research. There are two different views on neural sources and emotions. The locationist view postulates that some discrete emotions reflect the discrete anatomical structure in the brain (e.g., amygdala) \cite{adolphs1994impaired,lench2011discrete}. The distributionist view argues that no single anatomical structure uniquely specializes for individual emotion categories \cite{lindquist2012brain,ledoux2012rethinking}. Human emotion is a product of the cooperation of multiple cortex regions \cite{britton2006neural}. Localizing neural sources from scalp EEG is problematic. We should not assume that the electrical activity comes from the adjacent cortex. Hence, cognitive neuroscience adopts intracranial electrophysiology to study these questions. Intracranial electrophysiology-based studies provide evidence for the distributionist view. For example, negative emotions are processed in multiple brain regions. Stimulation of subcortical nuclei, the temporal lobe and gyrus, temporal-parietal junction, inferior frontal gyrus, etc, can affect the perception of sadness, fear, and anger \cite{gordon1996mapping,peron2010subthalamic}. The positive emotions (e.g., joy and mirth) are also processed in multiple brain regions \cite{satow2003mirth,vaca2011mirth}. It is found that the forebrain takes part in the regulation process of emotion \cite{Davidson1992Anterior,Davidson2004What,koelsch2014brain}, and asynchronous activities would occur in different locations of the brain during the emotional process. A higher left prefrontal activity is the reflection of an `approach model' of emotion process (e.g., positive emotions), whereas a higher right prefrontal activity is the reflection of the `withdrawal model' of emotion process (e.g., negative emotions) \cite{palmiero2017frontal}. Researches on people with depression also found greater activation in the right forebrain than other brain regions.

The correlation between EEG components and various emotions is also one of the key scientific questions in cognitive neuroscience and is critical for building effective recognition models. Joyful music has been reported to be positively correlated with the power energy in the theta band (4$\sim$7 Hz) near the midline of the prefrontal cortex \cite{sammler2007music}. As shown in Figure \ref{fig:asymmetry}, the valence and arousal of musical stimuli have been reported to correlate with frontal alpha (8$\sim$13 Hz) asymmetry \cite{schmidt2001frontal}. When playing happy music for subjects, the EEG activity in the left front area of the brain is more active than that in the right front area of the brain. However, the opposite result appears if playing sad music. Nevertheless, some works found no effects in the alpha band and instead found a relationship in the beta-2 band (18$\sim$22 Hz) \cite{daly2014neural}. The effect of high-frequency EEG, including the Beta-2 band (18$\sim$22 Hz), Beta-3 band (22$\sim$30 Hz), and Gamma band (30$\sim$45 Hz), in emotion also have been verified. The experience of happiness results in the decreases in beta-2 power in the front central regions and beta-3 and gamma power over the entire cortex. Whereas, when feeling anger, the Beta-2, Beta-3, and Gamma bands get increased power in the front of both hemispheres \cite{aftanas2006neurophysiological}. Subjects in the DEAP experiments produced higher gamma and frontal midline theta power while watching emotion provoking music videos \cite{koelstra2012deap}. \citet{balconi2008consciousness} reported the Gamma band activity is a marker of the subject's evaluation of the Arousal dimension in emotional faces stimuli. \citet{yang2020high} revealed that the network connections in the high Gamma band have significant differences among the positive, neutral, and negative emotional states. In addition, the specific phenomena related to emotion in other brain areas and other frequency components have also been reported \cite{hagemann1999eeg,coan2004frontal}.

Critical EEG bands and brain regions in recognizing emotions are also extensively studied in pattern recognition. Earlier traditional machine learning-based researches have reported that the features extracted from high-frequency bands (the Gamma and Beta bands) are more effective for an algorithm to recognize both Valence and Arousal dimensions \cite{li2009emotion}. Currently, extensive Deep Learning-based recognition approaches also verify the Beta and Gamma band information are the most suitable bands for emotion recognition. For example, the differential entropy (DE) features extracted from these bands lead to a higher recognition performance compared with the other bands when adopting the deep belief networks \cite{zheng2015investigating} and the convolutional neural networks \cite{li2018hierarchical}. Even so, these studies also approve that combining the information from all the bands can achieve the best performance. Feature Selection based approaches are suitable for analyzing the key EEG variables in emotion recognition. \citet{Li2018Exploring} studies the key frequency bands and channels by analyzing features selected out by the L1-norm logistic regression model. They find the Hjorth parameter of mobility in the Beta band achieves the best performance. They also find that the electrodes on the bilateral temporal and left anterior regions help get a higher performance for cross-subject emotion recognition, especially when the information in the beta band was utilized. \citet{naser2021influence} adopt a Minimum Redundancy Maximum Relevance (MRMR) feature selection method on the asymmetrical features. Besides the Beta and Gamma bands, they also find features selected from Alpha bands promote the Arousal and Valence classification. Also adopting the mRMR method, \citet{zhuang2018investigating} find that the DE of the Gamma band have good classification performance, and important electrodes are distributed in the bilateral temporal, the prefrontal, and the occipital regions. In addition to directly compare the performance between different bands and regions in emotion recognition, analyzing the model weights associated with the variables is another common way. For example, \citet{zheng2015investigating} explore the critical EEG electrodes by analyzing the weights of the trained DBNs. The results show the lateral temporal and prefrontal brain areas activate more than other brain areas in beta and gamma frequency bands.


To sum up, it can be seen from these studies that the central nervous system is directly related to the emotional process. As an external manifestation of the brain's cognitive psychological activity, EEG can be taken as an effective measurement to study emotional psychology and provides the feasibility for studying emotion recognition technology. Nevertheless,  the research findings mentioned above indicate that EEG activity's specificity and neural correlate in the emotional process are not yet fully explored and understood. It's better to build recognition models on EEG from all critical frequency bands and cortex regions.

\subsection{Classical research methodologies of EEG based emotion recognition studies}

\begin{figure*}[!htbp]
\centering
\includegraphics[width=0.5\textwidth]{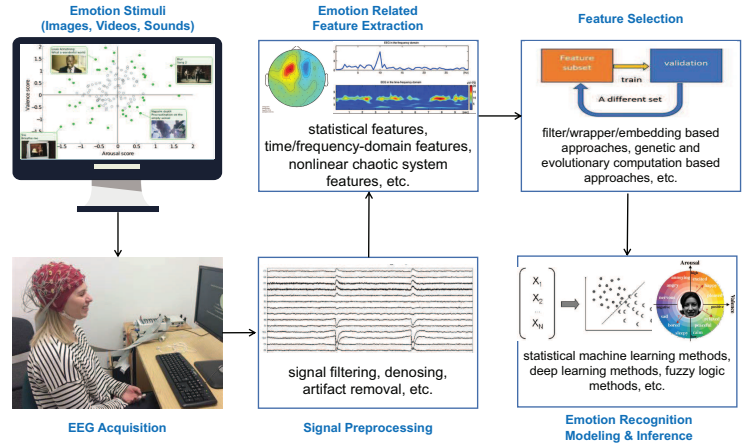}\\
\caption{The classical methods in the EEG based Emotion recognition study\protect\footnotemark[1]}
\label{fig:workflow}
\end{figure*}

\footnotetext[1]{The bottom-left figure is obtained from the following web link: https://levita-lab.group.shef.ac.uk/eeg/}

Classical research methodologies in this field follow the procedure shown in Figure \ref{fig:workflow}, namely applying pattern recognition models to the handcrafted emotion-related EEG features to distinguish different emotional states. Firstly, if you want to build and validate the models on self-collected EEG data, you should design a user experiment to collect emotion-induced EEG data. There are two main categories of methods for emotion induction. The first one aims to measure EEG during viewing emotionally provoking stimuli. Recommended stimuli include presentation of faces with emotional facial expressions, display of emotional pictures (e.g., IAPS pictures \cite{constantinescu2017cluster}), and emotionally provoking video or audio (e.g., film, music, music video, IADS sounds \cite{greco2016arousal}). The classical benchmark datasets introduced in Section 5.1 were collected under these types of stimuli. Building Virtual Reality scenes to induce emotions is reported effective and has been extensively adopted in recent years. For example, \citet{zhang2017affective} proposed the Affective Virtual Reality System, which combines IAPS, IADS, and China Affective Video System (CAVS) to produce a virtual environment that would accommodate VR headset for emotion induction. Another method asks the test subject to imagine the target emotion (e.g., recalling the past experience). After each trial, self-reports of evoked emotional states are required. Then, classical methods need conducting data preprocessing for the acquired raw EEG data, based on which domain knowledge (signal processing, sequential pattern mining, emotional psychology, etc.) guided feature extraction is further conducted to distill as many emotion-related characteristics as possible. The refined characteristics are called data features in machine learning that is further organized to construct the training samples. Furthermore, the features closely related to emotional state and helpful for improving the emotion recognition performance can be screened out from all the candidate features set by the feature selection method. Finally, we attempt to select various statistical machine learning models or Deep Learning models to build on the chosen features and iteratively evaluate their performance by computing the evaluation metrics until achieving the goal that the model outputs approximate the ground truth. Many classical studies that follow this research methodologies have verified the feasibility of building an emotion recognition system on EEGs. In addition, based on the classical methodologies, great advances have been made in this field in recent years. This review will give a more detailed introduction of the classical and latest methodologies in the following sections.

\section{EEG preprocessing and feature engineering}

Its high time resolution characterizes EEG. The high-sampled EEG data contains much emotion-related information, which has excellent potential for building precise and real-time recognition systems. Generally, EEG with a higher sampling rate may cover more details, however meanwhile may introduce lots of noise and increase the computation cost for signal processing, feature engineering, and modeling training. Regarding the device features and the problems mentioned above, the EEGs are usually sampled with a frequency of 256Hz, 512Hz, or 1024Hz. Theoretically speaking, a sampling frequency of 128 Hz would give a Nyquist frequency of 64 Hz, which is adequate for extracting sufficient emotion-related features. As discussed in Section 2.2, the Beta-2 band (18$\sim$22 Hz), Beta-3 band (22$\sim$30 Hz), and Gamma band (30$\sim$45 Hz) are reported closely correlated with emotion recognition.

It is challenging to represent the emotional EEG signals effectively due to the many noises embedded in EEG signals. Besides, it is difficult to capture the implicit correlations between EEG and a specific cognitive process. Hence, preprocessing and feature engineering usually need to be included in the workflow.

\subsection{EEG Preprocessing}
EEG signals are mixed with various noises of the human body and environment, thus bringing challenges to the anti-interference and robustness of the recognition algorithm. Therefore, the collected EEG signals are not directly used to build the recognition models and systems. The research paradigm should not overlook preprocessing the signals and extracting representative features. The preprocessing for the acquired EEG data is mainly to remove EOG artifacts with a frequency less than 4Hz that caused by eye blink, ECG artifacts with a frequency about 1.2Hz, EMG artifacts with a frequency more than 30Hz, power frequency artifacts in the environment with a frequency between 50 to 60Hz, and so on. These above artifacts can be removed with the independent component analysis (ICA), discrete wavelet, or band-pass filters (such as the Butterworth filter) to retain the rhythmic components associated with the emotional activity. The filtering process cannot remove all the artifacts from the EEG signals, and thus additional processing is needed. We can utilize the Artifact Subspace Reconstruction (ASR) method for enhanced artifact removal. The ASR method consists of a sliding window Principal Component Analysis (PCA), which statistically interpolates any high-variance signal components exceeding a threshold. Furthermore, the Common Average Reference (CAR) method is recommended to compute the average value over all electrodes and subtract it from each sample of each electrode \cite{katsigiannis2018dreamer}.

Even though the emotion-provoking experiments strictly follow the experimental protocol, some problems may arise with the recordings from some subjects or trials due to technical issues or personal issues, resulting in incomplete or high-noise data. This kind of data is recommended to be entirely abandoned without further processing and analysis. By the way, we could choose only to select out a subset of the whole EEG channels for studying if we can determine the specific emotion-related brain areas, e.g., the forehead. Channel selection will decrease the computation cost and may increase the recognition performance.

\subsection{Time Domain Features}
The most direct method for extracting EEG features is to calculate statistics such as mean value, variance, skewness, kurtosis and peak-to-peak interval, etc., which can characterize the time-domain properties of EEG signal. Some studies also adopt the higher-order crossing (HOC) method, which calculates the number of zero-crossing points of EEG as a feature after being processed by different filters \cite{petrantonakis2010emotion,lan2016real}. Event-related potentials (ERPs) reflect the underlying cognitive process, including emotion. Early ERP components have been verified to correlate with Valence \cite{olofsson2008affective}, whereas the late ERP components have been verified to correlate with Arousal \cite{bernat2001event}. Hence, characteristics of ERP components, e.g, P100, N100, N200, P200, P300, were also taken into studies \cite{frantzidis2010toward}. It should be noted, ERPs are obtained by averaging multiple EEG trials, which may not be feasible for online applications.

\subsection{Frequency Domain Features}
The frequency-domain attribute is a description of the signal from another perspective. Moreover, it has a better anti-interference to noise and can reflect the details of each component of the signal. Therefore, the frequency-domain features, e.g., the power spectral density (PSD), are widely extracted in physiological emotion calculation \cite{wang2014emotional, zheng2015investigating}. PSD reflects the signal power of a specific frequency band. It is usually obtained through fast Fourier transform (FFT), by which the raw EEG signal can be decomposed into several distinct frequency bands, e.g., the Delta (1-4 Hz), Theta (4-8 Hz), Alpha (8-13 Hz), Beta (13-30 Hz), and Gamma (>30 Hz). Then the average power of a specific frequency band is computed and adopted as the feature. The power spectra density (PSD) can also be estimated using Welch's method. For simplicity, we can directly take advantage of the MATLAB Signal Processing Toolbox\protect\footnotemark[2] to get this feature.

\footnotetext[2]{https://www.mathworks.com/help/signal/ref/dspdata.psd.html}

\subsection{Time-Frequency Domain Composition Features}
However, the single domain of time or frequency can not fully depict the characteristics of the signal. For the non-steady signal, the frequency components often change with the cranial neural, cognitive process. It needs a transition from static frequency component analysis to dynamic time-frequency joint analysis, for example, adopting the short-time Fourier transform (STFT) \cite{Lin2009EEG,liu2018real} and wavelet analysis (e.g., discrete wavelet transform) \cite{sorkhabi2014emotion,mohammadi2017wavelet}. Wavelet analysis is usually considered as a type of `Sparse Representation'. It is good at manifesting subtle local characteristics in both signal domains. Hence, it is broadly applied to process non-steady neural signals, especially for the EEG \cite{subasi2005automatic}. STFT is only suitable for analyzing steady signals, and it is hard to get a high resolution simultaneously in frequency and time domain based on Heisenberg's Uncertainty Principle. Hence, we recommend adopting wavelet analysis to extract fine-grained EEG features.

\begin{figure*}[!htbp]
\centering
\includegraphics[width=0.55\textwidth]{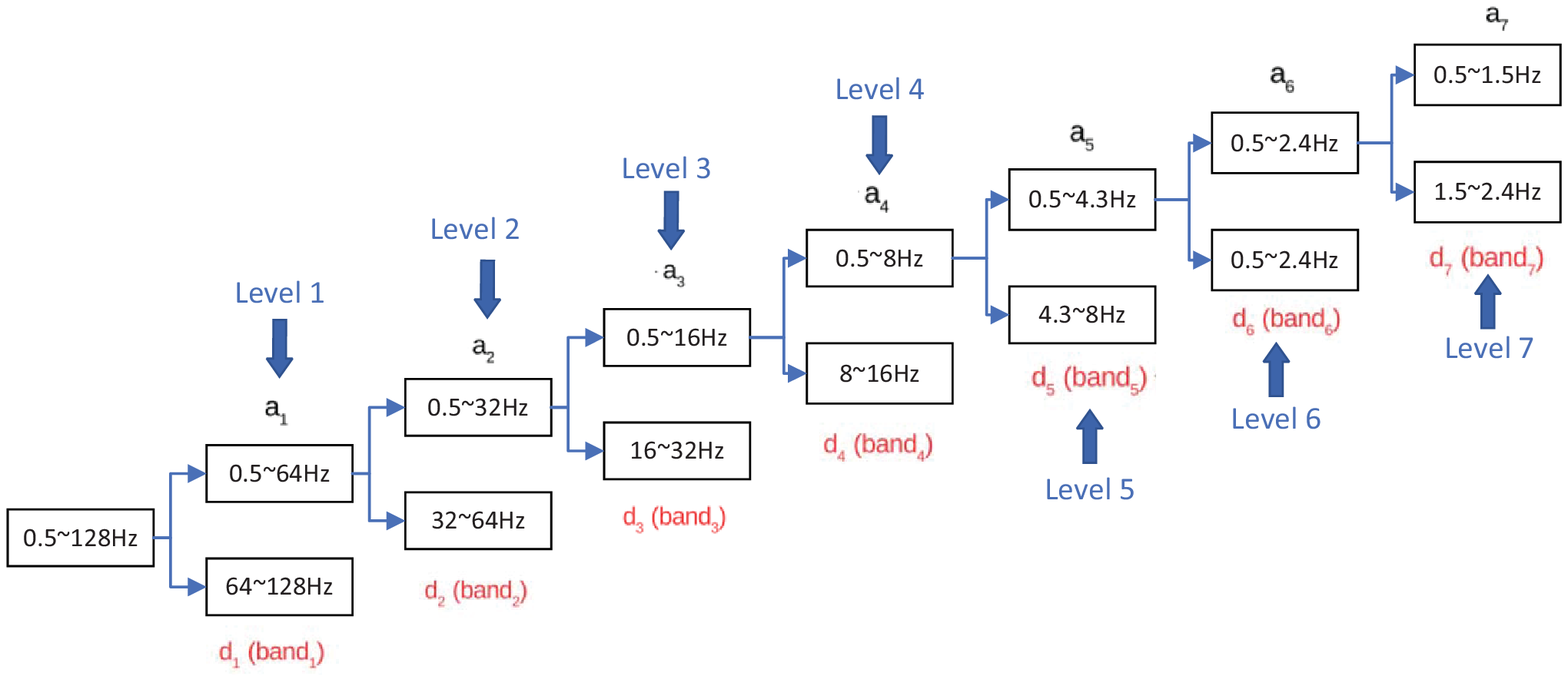}\\
\caption{Examples of 7-level decomposition and corresponding frequency bands of EEG with Sample Frequency of 128Hz.}
\label{fig:dwt_level}
\end{figure*}

The raw EEG signal is decomposed by the basis functions obtained through scaling and translating the mother wavelet. Specifically, the discrete wavelet transform (DWT) decomposes the signal into approximation and detail coefficients. Approximation coefficients describe the high-scale, low-frequency parts of the signal, while the detail coefficients describe the low-scale, high-frequency parts of the signal. The decomposition process is an iteration process, i.e., the obtained approximation coefficients of one scale can be further decomposed into more granular coefficients corresponding to the next scale. This process is iterative, yielding a series of approximation coefficients and detail coefficients belonging to different scales, as shown in Figure \ref{fig:dwt_level}. The effects of decomposition will be affected by what kind of mother wavelet is predetermined, e.g., the Haar wavelet, Daubechies wavelet, Bior Wavelet, etc. \cite{gandhi2011comparative}. For example, the Daubechies4-based DWT helps to extract the A5, D5, D4, D3, D2, and D1 signal components, the corresponding frequency range and decomposition level is listed in Table \ref{tabel_dwt}. The frequency ranges of these components and their correlation with various human states are provided.

\tiny
\begin{longtable}{p{2cm}p{2.5cm}p{2.5cm}p{5cm}}
\caption{The Decomposition Levels and Corresponding Frequency Bands for EEG with 128Hz Sampling Rate.}
\label{tabel_dwt} \\
\toprule
    \textbf{Brain Wave} & \textbf{Frequency Range (Hz)}  & \textbf{Decomposition Level} & \textbf{Association} \\
\midrule
    Delta Rhythm & 0-4 & A5 (0.5-4.3Hz) & Deep sleep \\
    \midrule
    Theta Rhythm & 4-8 & D5 (4.3-8Hz) & Drowsiness; Creative inspiration; Deep meditation \\
    \midrule
    Alpha Rhythm & 8-13 & D4 (8-16 Hz) & Relaxed awareness; Eye closing \\
    \midrule
    Beta Rhythm & 13-30 & D3 (16-32 Hz) & Active thinking; Attention; Behavior; Settling problems\\
    \midrule
    Gamma Rhythm & 30-64 & D2 (32-64 Hz) & Sensory processing; High-level Information Processing, Certain cognitive; Motor function \\
    \midrule
    Noise & 64-128 & D1 (64-128 Hz) & \\
\bottomrule
\end{longtable}
\normalsize

\small
Then, the wavelet engery $ENG_j$ feature and the wavelet entropy $ENT_j$ featuer can be computed in each frequency range according to the following formula \ref{eq_dwt_energy} and \ref{eq_dwt_entropy}, respectively, where $j$ is the decomposition scale, $k$ is the number of wavelet coefficients. DWT can be easily obtained through utilizing the MATLAB Wavelet Toolbox\protect\footnotemark[3].

\footnotetext[3]{http://www.mathworks.com/help/wavelet/ref/dwt.html}

\begin{tiny}
\begin{equation}
\label{eq_dwt_energy}
\begin{aligned}
& ENG_j = \frac{1}{N}\sum_{k=1}^{N}(d_j(k)^2) \\
\end{aligned}
\end{equation}
\end{tiny}

\begin{tiny}
\begin{equation}
\label{eq_dwt_entropy}
\begin{aligned}
& ENT_j = -\sum_{k=1}^{N}(d_j(k)^2)log(d_j(k)^2) \\
\end{aligned}
\end{equation}
\end{tiny}

In addition, several mode decomposition approaches, e.g., the  empirical mode decomposition (EMD), the multiple empirical mode decomposition (MEMD) and the variational mode decomposition (VMD), are also quite suitable for the nonlinear unsteady EEG feature extraction, in which the multi-channel EEG is decomposed into multiple intrinsic mode functions (IMFs), based on which the more representational features can be extracted \cite{mert2018emotion}.

\subsection{Nonlinear Dynamical System Features}

\begin{figure*}[!htbp]
\centering
\includegraphics[width=0.6\textwidth]{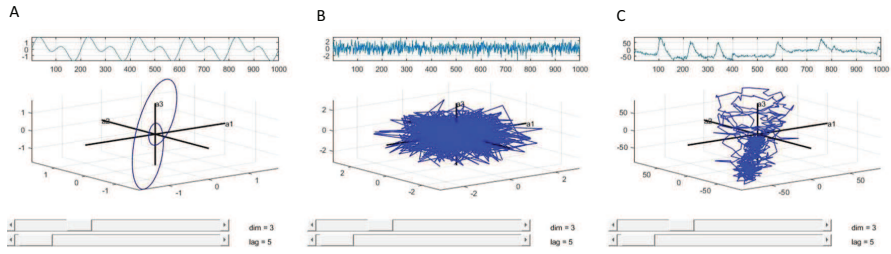}\\
\caption{Phase diagram example for different types of signals: A. Periodic Signal, B. White Noise Signal, C. EEG Signal}
\label{fig:phase_diagram}
\end{figure*}

In addition to time-frequency-domain analysis, the researchers pay more and more attention to the chaotic characteristics in the neural system activities \cite{Stam2005Nonlinear}. For instance, the brain and heart are considered non-linear dynamical systems. As shown in Figure \ref{fig:phase_diagram}, the neurophysiological signals present the properties of a non-linear dynamic system so that they can be studied and analyzed by the non-linear dynamic method. The fractal dimension is an index to describe the complexity and self-similarity of a chaotic non-linear system. EEG has multi-fractal dimensions \cite{Kulish2006Human}.The brain states under different cognitive tasks correspond to specific fractal dimension \cite{wang2011fractal}. For example, Liu has found that the fractal dimension value corresponding to the high Arousal level was higher than that corresponding to the low Arousal level, so it could be used for emotion recognition \cite{Liu2010Real}. In addition, the features describing the non-linear dynamical system,  such as correlation dimension, approximate entropy, Lyapunov index, K-C complexity, etc., have all appeared in some studies on emotion recognition \cite{hosseinifard2013classifying,Liu2014EEG}. One less studied property called Recurrence also reflects the characteristics of dynamical systems and can help to predict its evolution. Recurrence Plots accurately depict the distance correlation among trajectories in the non-linear system \cite{eckmann1995recurrence}. Hence, several studies also focus on Recurrence Plots-based EEG representation and feature extraction \cite{yu2016encoding,yang2018recurrence}. However, the non-linear feature calculation cost is higher, which is disadvantageous to build a real-time recognition system. Moreover, its calculation is sensitive to parameter setting (such as embedded dimension, etc.). Simultaneously, the parameter setting is still needed to be studied and explored. In addition, differential entropy (DE) also is a kind of representative non-linear feature. It is the extension of Shannon entropy on continuous EEG variable $x$, which are assumed to obey the Gaussian distribution $N(\mu, \sigma^2)$. Hence, it can be easily computed according to the following Formula \ref{eqDE}.

\begin{tiny}
\begin{equation}
\begin{aligned}
\label{eqDE}
   DE&=-\int_{-\infty}^{+\infty}p(x)log(p(x))dx \\
   &=-\int_{-\infty}^{+\infty} \frac{1}{\sqrt{2\pi\sigma^2}}e^{-\frac{(x-\mu)^2}{2\sigma^2}} log(\frac{1}{\sqrt{2\pi\sigma^2}}e^{-\frac{(x-\mu)^2}{2\sigma^2}}) dx \\
   &=\frac{1}{2}(2{\pi}e{\sigma}^2)
\end{aligned}
\end{equation}
\end{tiny}

The effectiveness and robustness of DE has been extensively verified in related works, and have advantage over PSD \cite{zheng2015investigating}. Some researches indicate that the computation of DE is equivalent to the logarithm power spectrum in a certain frequency range for a fixed length signal \cite{shi2013differential}. More details about the non-linear dynamical features in EEG based emotion recognition can be found in the review paper of \citet{garcia2019review}.

\subsection{Asymmetry Features}
In addition, specific phenomena found in brain cognition studies can also be taken as the feature to infer emotional states. Some cognitive studies based on brain imaging technology have seen the lateralization phenomenon of the brain in different locations in emotion processing \cite{Jones1992Electroencephalogram}. Clinical studies have also found that there is the lateralization phenomenon of brain activity in the brains of patients with mood disorders such as anxiety and depression \cite{Mathersul2008Investigating}. Asymmetry exists throughout the brain, and the hemispheres are not strictly symmetric in structure and function \cite{greve2013surface}. Therefore, some studies have adopted the differences in the EEG characteristics among electrodes in symmetrical positions of the brain to indicate the emotional states. For example, the Theta, Alpha, Beta, and Gamma spectral power asymmetry (SPA) derived from fourteen electrode pairs in the left and right lobes can be extracted as features \cite{koelstra2012deap}. The index asymmetry feature is generally obtained by calculating the difference \cite{Liu2010Real,Thammasan2016Continuous} or the ratio \cite{Huang2012Asymmetric} of the indexes of two signal sources such as power spectrum, fractal dimension and so on. In addition, \citet{Petrantonakis2012Adaptive} proposed the $AsI$ measure by estimating the mutual information shared between the brain hemispheres and further expanded $AsI$ to be applicable in the time-frequency domain.

\subsection{Brain network Features}
High-level cognition function depends on subtle cooperation between local and global brain activities and is inseparable from a network of brain neurons and brain regions \cite{bullmore2012economy}. There exist intrinsic correlations between EEG signals from different brain regions. Researchers believe that the functional connectivity graph and the derived structural characteristics could significantly enhance the distinctiveness of various emotions \cite{wang2019convolutional}. Several studies have found the effectiveness of brain network indexes such as correlation, coherence, and synchronization in emotion recognition \cite{lee2014classifying}. Therefore, the study of the brain from the perspective of the brain network has received widespread attention \cite{betzel2017multi}. It provides a kind of `Graph Theory' based research basis for studying the cognitive process of emotion. For example, \citet{rotem2017infants} analyzed several indicators, including connection density, clustering coefficient through the EEG-derived brain network. As far as we know, it takes the first step on utilizing graph theory to analyze infants' brain processing of emotional facial expressions.

\begin{figure*}[!htbp]
\centering
\includegraphics[width=0.7\textwidth]{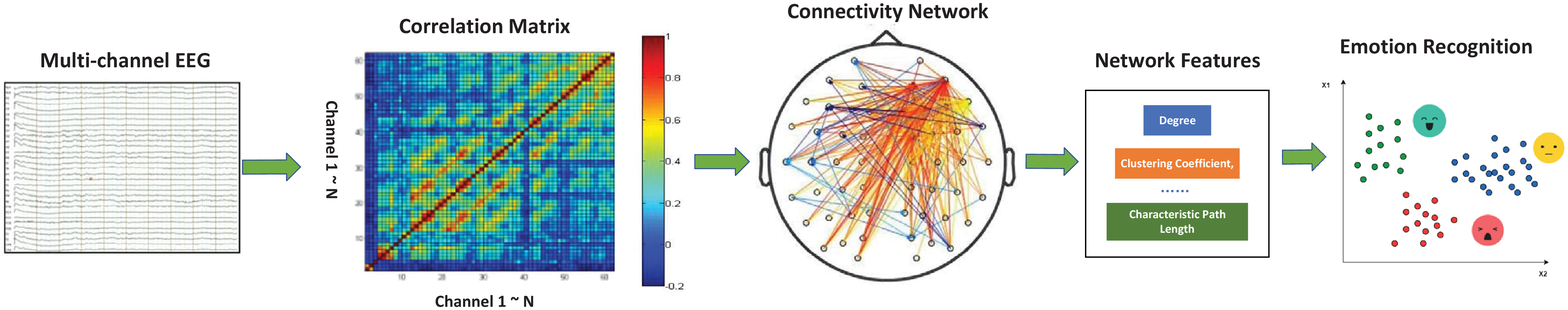}\\
\caption{Illustration of the brain network construction and the derived network features for classification \cite{chen2018assessment}.}
\label{fig:brain_network}
\end{figure*}

The estimation and construction of brain networks are usually achieved by studying the time correlation or spectral coherence between multichannel brain signals, including cross-spectrum \cite{kinney2019analysis}, Pearson correlation coefficients, mutual information \cite{morabito2015longitudinal}, synchronization likelihood \cite{de2009functional}, phase-lag index (PLI), phase-locking value (PLV) \cite{wang2019convolutional}, Granger causality \cite{franciotti2019cortical}, etc. For example, the PLV reflects the mean difference between the instantaneous phases of two channels of EEGs over time. The value of PLV ranges from 0 to 1, where 0 indicates the inexistence of phase coupling, while 1 reflects a strict phase coupling. PLV has been verified effective in evaluating the cooperation over different brain areas \cite{varela2001brainweb}. Then, based on the constructed brain network, specific indicators related to network topology can be derived to build the recognition models \cite{chen2018assessment}, including the modularity, clustering coefficient, degree (in-degree, out-degree, average-degree), characteristic path length, closeness centrality, local or global-efficiency, and small-world property, etc. \cite{chen2019deep} (see Figure \ref{fig:brain_network}). The constructed brain network reflects the coupling correlation between two EEG channels. Hence it is not sensitive to amplitude variations. This property reduces the influence of inter-person difference and helps to build robust and accurate EEG-based recognition models.

\begin{table}[!htbp]
\tiny
\caption{List of representative EEG features extracted in related works.}
\label{table_features}
\begin{tabular}{p{3cm}p{12cm}}
\toprule
\textbf{Feature Type} & \textbf{Extracted Features} \\
\midrule
Time+Frequency Domain Features & 1. Peak-to-Peak Interval. 2. Mean Square Value. 3. Variance. 4. Mean Value. 5. Skewness. 6. Kurtosis. 7. 1st/2nd Difference. 8. Hjorth Parameter: Mobility, Complexity, Activity. 9. Higher-order Crossing. 10. Maximum Power Spectral Frequency. 11. Power Sum. 12. Maximum Power Spectral Density. 13. Wavelet Energy. 14. Wavelet Entropy. 15. Amplitude and latency of ERPs. 16. Shannon Entropy.\\
\midrule
Nonlinear Dynamical System Features & 1. Approximate Entropy. 2. C0 Complexity. 3. Correlation Dimension.  4. Kolmogorov Entropy. 5. Lyapunov Exponent. 6. Permutation Entropy. 7. Singular Entropy.  8. Spectral Entropy. 9. Sample Entropy. 10. Differential Entropy. 11. Fractal Dimension. 12. Hurst Exponent. 13. Lyapunov Complexity. 14. Recurrence Plot: recurrence rate, determinism, entropy, averaged diagonal length, length of the longest diagonal line, laminarity, trappping time, length of longest vertical line, recurrence time of 1st type, recurrence time of 2nd type\\
\midrule
Brain Asymmetry Features & 1. Difference Between Channels. 2. Ratio Between Channels. 3. Asymmetry Index (AsI) \\
\midrule
Brain Network Features & 1. Correlation. 2. Coherence. 3. Clustering Coefficient. 4. Degree. 5. Characteristic Path Length. 6. Local/Global Efficiency. 7. Connectivity Density. 8. Modularity. 9. Closeness Centrality \\
\bottomrule
\end{tabular}
\end{table}

The representative EEG features utilized in related emotion recognition researches are listed in Table \ref{table_features}. Even though so many candidates handcrafted EEG features can be extracted, it should be pointed out that those traditional handcrafted features are obtained based on a quantity of domain knowledge, thus improving the learning cost of researchers, especially for those only majoring in computer science. In addition, most features of the current neural signals are based on the traditional time-series signal analysis theory and method. The correlation between those signal features and the emotional states is unknown and still needs to be explored, and the effects are also limited. Furthermore, EEG variations triggered by physiologic or psychological factors can easily disturb those features, e.g., cardiac activity, eye movement, etc.

\subsection{Feature Processing}
\subsubsection{Automatic Feature Selection Method}
Automatic feature selection techniques can be categorized into `filter method' based selection and the `wrapper method' based selection \cite{guyon2003introduction}. Either way, the obtained EEG features need to be ranked according to specific criteria, e.g., by evaluating the relationship between the features and the target emotions or assessing the feature importance derived from the model parameters. The top vital features can be reserved for further model design, while the others will be abandoned \cite{huang2006filter,Maldonado2008A}.

The `filter method' based selection does not depend on the built recognition models, and its computation cost is usually less than the `wrapper method.' Hence we recommend utilizing it in real-time and big-data scenarios. The most widely used `filter methods' are the chi-squared ($\chi^2$) test-based approach, mutual information-based approach, ANOVA F-test-based approach, etc. The $\chi^2$ test-based approach tests the independence of two variables by measuring the distribution difference between the feature variable and the emotion classes. Features with a higher $\chi^2$ value have a close relationship with the target emotions that will be reserved. The mutual information is calculated to evaluate the interdependence between the specific feature and the emotions. The most representative mutual information-based approach is referred to as minimal-redundancy-maximal-relevance (MRMR) \cite{ding2005minimum}. For example, by combining the MRMR selection method with the kernel function classifier, the recognition performance can be improved \cite{atkinson2016improving}. ANOVA F-test measures the difference over multiple distributions by calculating the ratio of the between-class variance to within-class variance, which reflects the degree of discrepancy. Features with higher F-ratio values can better differentiate the different emotions.

The `wrapper method' needs to work with a specific machine learning model, among which the recursive feature elimination (RFE) method is one representative algorithm. It is originally designed by \citet{Guyon2002Gene} for gene selection. It works based on a sequential backward abandon scheme. The algorithm initially starts on the entire feature set. Then, some features with smaller feature weights are abandoned from the feature set. The process iterates several times until the desired objective is achieved. In addition, the highly efficient L1-norm penalty-based method is also recommended. It adds an L1-norm regularization term at the end of the original loss function to encourage weight sparsity. Regularization is one strategy adopted in machine learning in case that the feature dimensions are larger than the size of the samples. It guarantees to produce small-value model parameters to prevent over-fitting. We can abandon those features with 0 weight from the current feature set. According to some research findings, the L1-norm-based method has a big advantage over the L2-norm-based method when faced with lots of redundant features \cite{Ng2004Feature}.

\subsubsection{Manually Operated Feature Selection}
The definitions of feature importance differ with the change of the research objectives. For recognition tasks, the automatically selected features contribute to enhancing the recognition performance. However, the automatically selected features are not recommended for cognition studies to analyze cognition phenomenons. For example, \citet{Haufe2014On} pointed out that the research on the parameters of backward methods (multivariate classifiers) is not recommended in brain imaging data analysis. The derived findings may be inaccurate. Specifically, in machine learning, if several features are highly related to each other, only one feature may be reserved in model construction. Other features could be neglected by assigning relatively low weights without influencing the model performance. Simply considering the high-weight features may lose valuable information related to emotional cognition. Hence, in some studies, a manually operated feature selection method is recommended instead of the automation methods. For example, the `searchlight' strategy can be adopted as an alternative method, by which you manually distill features from different angles, e.g., electrode groups, brain regions, rhythms, feature types, etc. \cite{Li2018Exploring}. The features in critical brain regions or EEG rhythms have more impacts on the recognition performance.

\subsubsection{Feature Smoothing}
We know that EEG is a mixture of various neuronal activities in the brain and various noises from the body or environment. The features extracted from EEG will vary within short periods, but the human emotions may be relatively stable, which means the extracted features are still a not precise reflection of the emotional patterns. In addition, though increasing the kinds of features extracted may improve the recognition accuracy, it will introduce more noise and computational cost in feature extraction, model training, and the inference task. Regarding this, feature smoothing methods are also recommended in the feature processing process to decrease the influence of the emotion-irrelevant patterns and improve the emotion recognition accuracy without increasing the feature dimension. For example, we can first divide EEG data into non-overlapping windows and extract features from each window, and further adopt the Savitzky-Golay smoothing method or the moving average method to smooth the features in time sequence \cite{pham2015enhancing,tang2017eeg}.

\section{Pattern recognition technical routes applied in the field}
\begin{figure*}[!htbp]
\centering
\includegraphics[width=0.8\textwidth]{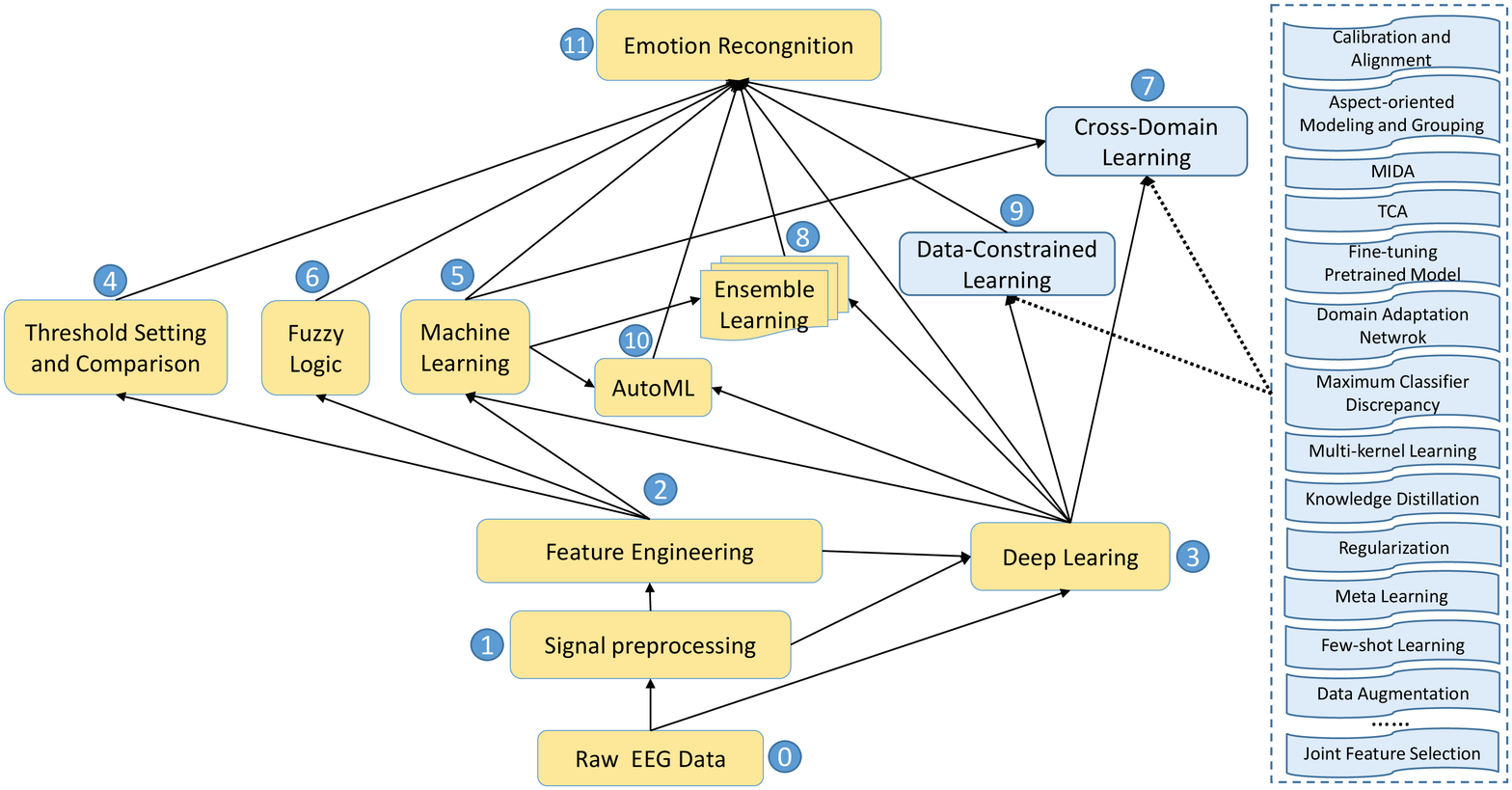}\\
\caption{Flow Chart of the Pattern Recognition Technical Routes Applied in The Field.}
\label{fig:tech_route}
\end{figure*}

Emotion recognition follows the nature of pattern recognition research: judging target samples' emotion categories based on existing data and some measurement criteria. We summarize existing pattern recognition approaches adopted in most related works using a flow chart, shown in Figure \ref{fig:tech_route}, in which different technical routes are clearly divided. With the rapid development of Deep Learning (DL) in graph and image processing and natural language processing, the DL-based technical route has begun to attract the attention of researchers in this field, and existing works indeed have been proved effective \cite{zhang2020investigation}. Hence, In this review, we pay more attention to Deep Learning-based studies. The following parts provide a summary of these technical routes and representative works.

\subsection{Route: \textbf{0\texorpdfstring{$\rightarrow$}.1\texorpdfstring{$\rightarrow$}.2\texorpdfstring{$\rightarrow$}.4\texorpdfstring{$\rightarrow$}.11}}

The simplest method follows the route \textbf{$0\rightarrow1\rightarrow2\rightarrow4\rightarrow11$}, namely setting a threshold for a specific feature, and if the feature value exceeds the default threshold, the sample is determined to belong to a particular emotional state \cite{Liu2010Real}. However, the threshold is fixed for the subject and depends on experience, leading to a lack of adaptability. Hence, this technical route is not mainstream, and we will not describe much on this route.

\subsection{Route: \textbf{0\texorpdfstring{$\rightarrow$}.1\texorpdfstring{$\rightarrow$}.2\texorpdfstring{$\rightarrow$}.5\texorpdfstring{$\rightarrow$}.11}}
At present, machine learning have been extensively studied in this field, including the traditional supervised learning models, e.g., discriminant analysis \cite{lee2014classifying}, support vector machine \cite{lan2016real}, K-nearest neighbor \cite{mohammadi2017wavelet}, Bayesian method \cite{Chung2012Affective}, Random forest \cite{ackermann2016eeg}, perceptron \cite{bhatti2016human}, etc, as well as unsupervised learning methods such as manifold learning \cite{wang2014emotional}, clustering \cite{Murugappan2008Lifting}, and so on. Traditional statistical machine learning based approaches follow the route \textbf{$0\rightarrow1\rightarrow2\rightarrow5\rightarrow11$} are detailedly illustrated in Figure \ref{fig:workflow}.

Here are some representative approaches. \citet{wang2014emotional} extracted two kinds of power spectrum features, two types of wavelet characteristics, and three types of nonlinear features from EEG. Afterward, they reduced the feature noise by the feature smoothing method and adopted linear discriminant analysis (LDA) to conduct feature dimension reduction. They finally utilized the linear SVM classifier to classify the two types of emotions. Besides, they predicted the change trajectory of emotional states based on the manifold learning method. \citet{jenke2014feature} extracted time-frequency-domain features as well as channel combination features from multichannel EEGs. Furthermore, they utilized feature selection methods based on ReliefF, minimum redundancy maximum relevance (MRMR), and statistical test method with the quadratic discriminant analysis (QDA) modeling method for enhanced emotion classification. In addition, \citet{atkinson2016improving} combined the MRMR feature selection method and kernel function classifier to promote the emotion recognition performance. \citet{lan2016real} and \citet{ackermann2016eeg} used SVM and random forest models, respectively, to construct recognition models based on statistical features, nonlinear features, spectral features, etc. Simultaneously, they compared the recognition effects of the method for the same period data and different periods data of users. \citet{li2019eeg} proposed graph regular linear regression model (GRSLR), where the sparse regularization is introduced for channel selection. Besides, the graph regularization can preserve inherent manifold topology after data embedding, thus preventing model over-fitting. Recently, \citet{cheng2020emotion} proposed to use a deep forest model named Multi-Grained Cascade Forest, termed as gcForest, for EEG-based emotion recognition task. The gcForest algorithm has fewer hyperparameters and is robust to hyperparameter settings. In addition, its model complexity adapts with different sizes of data, thus is worthy of our attention.

\subsection{Route: \textbf{0\texorpdfstring{$\rightarrow$}.1\texorpdfstring{$\rightarrow$}.2\texorpdfstring{$\rightarrow$}.3\texorpdfstring{$\rightarrow$}.11}}

\begin{figure*}[!htbp]
\centering
\includegraphics[width=0.4\textwidth]{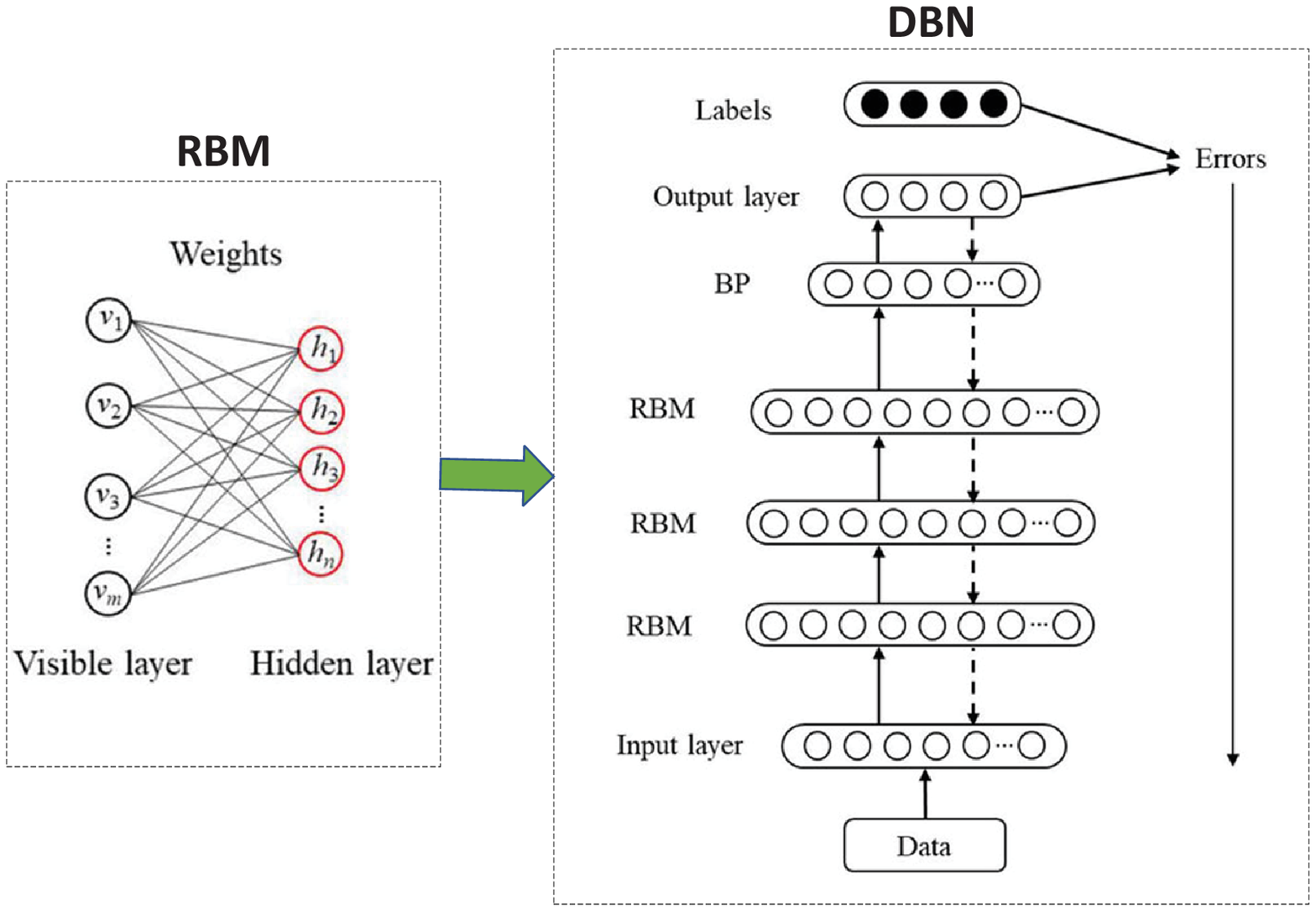}\\
\caption{Illustration of a deep belief network (DBN) consists of restricted Boltzmann machines (RBMs) and the corresponding training process.}
\label{fig:DBN}
\end{figure*}

In parts of DL-based works, the DL models are purely regarded as classifiers that play the same roles as the traditional machine learning models. The advantage of this Route compared with Route: \textbf{$0\rightarrow1\rightarrow2\rightarrow5\rightarrow11$} is the representation learning and the universal approximation property of the DL model that can nonlinearly transform the original features into any vector space \cite{scarselli1998universal}. For example, \citet{zheng2015investigating} and \citet{thammasan2016application} utilized the deep belief network (DBN) to classify the EEG emotions based on the handcrafted features extracted from EEG, e.g., the PSD and discrete wavelet. The DBN is proposed by \citet{hinton2006fast}, it builds on multiple restricted Boltzmann machines (RBMs) to solve the training problem of deep neural networks and promotes the rapid development of Deep Learning. RBMs are a two-layered artificial neural network with generative capabilities. Compared to Boltzmann Machines, RBMs are restricted in terms of the connections between the visible layer and the hidden layer. They are able to learn a probability distribution over the input data. Typically, the DBN training includes three main steps: 1) pre-train the DBN through Gibbs sampling method; 2) the DBN is transformed into an encoder-decoder network, thus fine-tune the DBN through unsupervised back-propagation; and 3) train the DBN through supervised back-propagation,  as shown in Figure \ref{fig:DBN}.

\begin{figure*}[!htbp]
\centering
\includegraphics[width=0.6\textwidth]{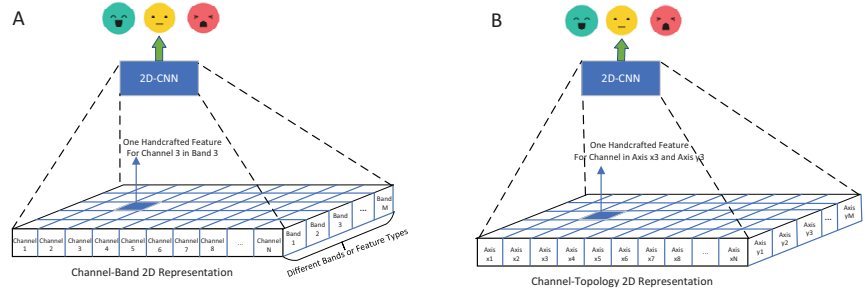}\\
\caption{Two representative 2D EEG feature representation methods in emotion recognition when adopting the 2D-CNN-based approaches.}
\label{fig:CNN}
\end{figure*}

The implicit correlations over different channels are a significant indicator to recognize emotions. Convolutional neural networks (CNN) are ideally suitable for processing two-dimensional data and extracting inter-channel joint information. Applying CNN to detecting emotions based on multi-channel EEG is worthy of study. Two central problems need to solve: 1) transforming the EEG data into proper representation to fit the input format of the CNN model; 2) building effective representation learning models based on various CNN modules for feature transformation. As shown in Figure \ref{fig:CNN}, we illustrate two possible representation approaches when applying 2D CNNs. For example, \citet{yang2018recurrence} proposed one channel-frequency convolutional neural network (CFCNN), which works with the recurrence quantification analysis (RQA). The entropy characteristics in different EEG frequency ranges derived through RQA are taken as the input of the CFCNN model. The input frame does not reserve the channel topology information. The rows of the feature map correspond to the channels, and the columns are the extracted feature in different frequency bands. Similarly, \citet{tripathi2017using} extracted nine types of statistical EEG characteristics of signals as the input of the CNN model, and finally, the effect of this method reached and exceeded those of mainstream methods. For better data representation, the input feature map can also reserve the channel spatial topology information, for example, \citet{li2018hierarchical} organized the differential entropy features from different channels as 2D sparse graphs, which maintains information of the electrode spatial topology, and finally used for CNN training and inferring. The constructed input feature maps could be comprehended as the input images. Hence the emotion recognition task can be resolved by the approaches adopted in computer vision tasks.


\begin{figure*}[!htbp]
\centering
\includegraphics[width=0.6\textwidth]{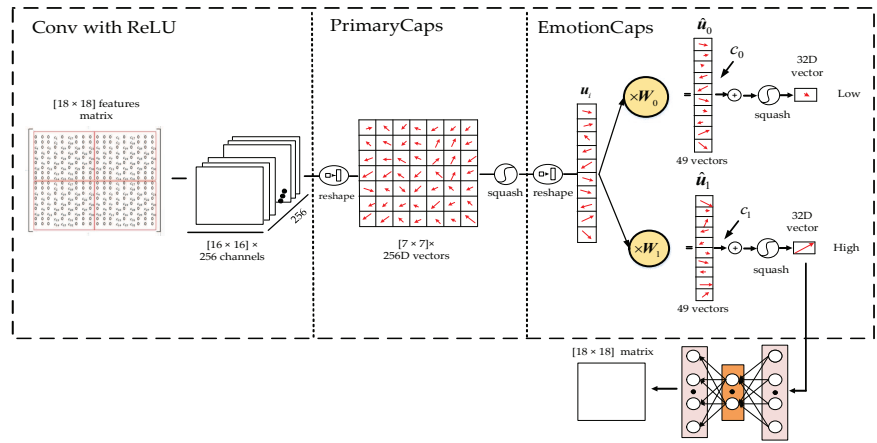}\\
\caption{The architecture of the CapsNet based emotion recognition model proposed by \citet{chao2019emotion}, in which the channel signal's PSD features are mapped into the model input, namely the multiband feature matrix.}
\label{fig:Capsule}
\end{figure*}

The CNNs are not good at recognizing features of input data when they are in different orientations. Specifically, through downsampling, pooling decreases the computation cost and can fit the variations in images. Nevertheless, the advantage of pooling is at the expense of neglecting precise spatial correlations between high-level parts, which is critical for recognizing objects with abundant spatial information \cite{sabour2017dynamic, liu2020multi}. To tackle this problem, recently, a new type of neural network called Capsule Network (CapsNet) inspired by neuroscience has been proposed. The brain is organized into modules, which can be considered capsules. An artificial neuron processes scalars, a capsule deals with vectors. The CapsNet can model the implicit correlations between local parts and whole objects. Besides, the CapsNet can be trained with a faster speed and requires a fewer amount of training samples compared with the CNN model. Hence, researchers have started to introduce CapsNet into this technical route. For example, \citet{chao2019emotion} point out the salient correspondence between the various emotions and cortex regions can be distinguished by the CapsNet. They also proposed one input representation structure, called the multiband feature matrix (MFM), which contains the topology correlation between EEG channels and the distinction of various EEG frequency ranges. Thus, it contributes to mining emotion-related information in spatial and frequency domains. The MFM-CapsNet based approach is illustrated in Figure \ref{fig:Capsule}, where the length and direction of each primary capsule indicate the existence and characteristic of the low-level representations correlated to emotions, respectively.

\begin{figure*}[!htbp]
\centering
\includegraphics[width=0.7\textwidth]{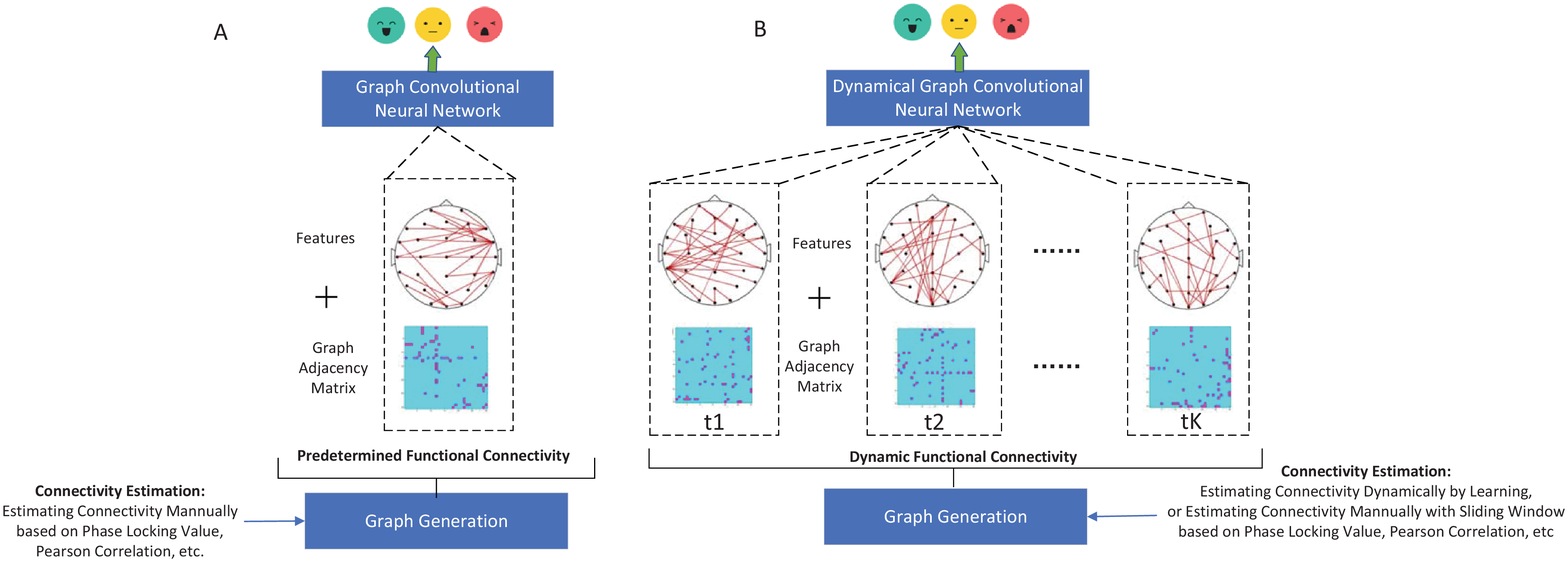}\\
\caption{Two representative approaches of applying graph convolutional neural network (GCNN). A: The traditional GCNN model, the input functional connectivity graph is pre-determined and is static in the process of model training. B: The dynamical GCNN model, the functional connectivity graph continues evolving in the process of model training.}
\label{fig:GCNN}
\end{figure*}

Some researchers also believe that the traditional CNN model may be not optimal for feature learning from EEG, which is discrete in the spatial domain. Besides, a closer spatial relationship may not guarantee a closer functional relationship. Hence, adopting the 2D representation and the CNN model may neglect he complex relationship among different channels, the relationship between the functional brain network patterns and the emotion process. The graph-based description method is advantageous in extracting signals' discriminative features in the discrete spatial domain \cite{such2017robust}, the structural representations learned from the functional connectivity graph could capture those correlation information mentioned above. For example, the graph convolutional neural network (GCNN) allows exploring the implicit correlations among the multiple graph nodes that represent the EEG channels. Similar to the approach shown in Figure \ref{fig:GCNN}(A), \citet{wang2019convolutional} built one typical GCNN model on the EEG derived graph. The graph is a fusion of the within-frequency functional connectivity graph (FCG) and the cross-frequency FCG. The within-frequency FCG is obtained through computing PLV for every pair of channel signals in the same frequency band, while the cross-frequency FCG does not require the signals to come from the same frequency bands. Those two graph representations are concatenated into a big graph with $N\times M$ nodes, where $N$ and $M$ denote the counts of the channel and the frequency band, respectively. The experiments verify that GCNN performs better than CNN on the FCG representation. In view of the dynamic process of functional network, similar to the approach shown in Figure \ref{fig:GCNN}(B), \citet{song2018eeg} proposed a dynamical graph convolutional neural networks (DGCNN) model, by which the discriminative characteristics and the functional connectivity information can be simultaneously extracted. Unlike the traditional GCNN, the adjacency matrix is not static but is adaptively updated dynamically during the DGCNN training. It is supposed that the learned adjacency matrix can capture the intrinsic correlation of the EEG channels. The graph's vertex representation is the handcrafted features, including the DE, PSD, DASM, and RASM, extracted from five rhythms. For DE and PSD features. The performance was evaluated based on each kind of graph. The experimental results indicated the graph of DE feature guide to the best performance and outperformed the GCNN based approach.

\begin{figure*}[!htbp]
\centering
\includegraphics[width=0.6\textwidth]{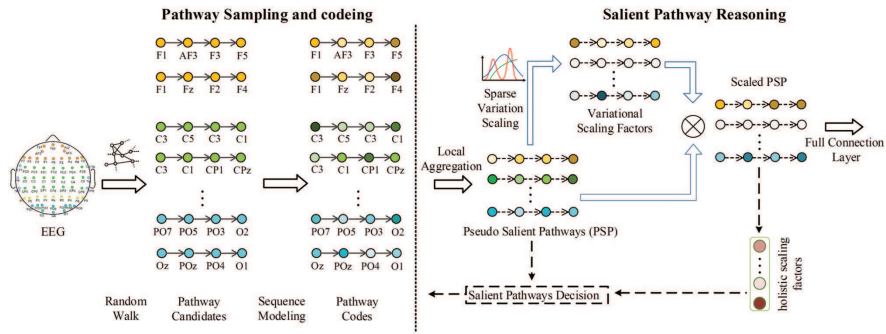}\\
\caption{The framework of the heuristic variational pathway reasoning (VPR) method that can adaptively determine salient pathways to facilitate EEG emotion recognition \cite{zhang2020variational}.}
\label{fig:VPR}
\end{figure*}

Also encouraged by the research findings that connections and pathways exist between spatially-adjacent and functional-related areas during emotion expression \cite{adolphs2002neural, bullmore2009complex}. As shown in Figure \ref{fig:VPR}, \citet{zhang2020variational} proposed a heuristic variational pathway reasoning (VPR) method that introduced random walk to generate candidate pathways along electrodes. LSTM was used to encode their ordered connectivity into high-level features of pathways that indicate between-electrode dependency to represent each pathway. They also proposed a salient pathway reasoning method, which includes two basic modules named pathway aggregation and sparse variational scaling. It can adaptively determine salient pathways to facilitate EEG emotion recognition and provide some explanation for emotion analysis. Based on the interpretable model, they explored salient interaction pathways w.r.t. different emotions. This research sets a new SOTA for the SEED dataset.

\begin{figure*}[!htbp]
\centering
\includegraphics[width=0.5\textwidth]{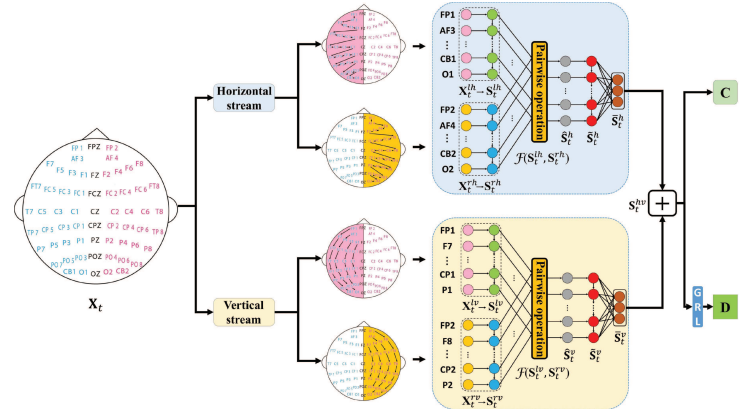}\\
\caption{Handcrafted feature based bi-hemispheric discrepancy information integrated model (BiHDM)\protect\footnotemark[4] \cite{li2020novel}. BiHDM utilizes four RNNs to capture the information of EEG channels on each hemisphere from horizontal and
vertical streams.}
\label{fig:BiHDM}
\end{figure*}

Another similar work is illustrated in Figure \ref{fig:BiHDM}, considering the importance of the asymmetrical information between the hemispheres in emotion cognition, one bi-hemispheric discrepancy model (BiHDM) is developed by \citet{li2020novel}, in which two individual horizontal and vertical traversing RNNs were employed to scan all left separately- and right-hemisphere channels' EEG features to learn the deep features of two hemispheres. Different from the prior work in Figure \ref{fig:VPR}, the electrode pathways here were predefined. After the deep representations of each channel above have been obtained, they performed pairwise subtraction, division, and inner product on the paired channels on symmetric locations of the brain region as the asymmetry information between two hemispheres is supposed to be more discriminative recognizing emotions from EEG. Another RNN models the obtained asymmetric representations, and the two-directional streams information are fused for final classification. In addition, they added a domain discriminator into the model to extract domain-invariant data representation. Before this work, \citet{li2018novel} has developed a bi-hemispheres domain adversarial neural network (BiDANN) based on the same neuroscience hypothesis. They fitted the cerebral hemisphere asymmetry information into the framework and took the domain adaptation. The framework has two feature extractors. Two local discriminators reduce the distribution discrepancy between the left and right hemisphere domains, respectively. Then, the global discriminator lessens the overall distribution discrepancy between two domains. The left and right hemispheric features are extracted through LSTM modules. To the best of our knowledge, for the first time, researchers introduced the hemisphere' asymmetry theory into the DL model design and verified that prior neuroscience knowledge is beneficial for guiding the modeling training but usually neglected by prior works. 

\footnotetext[4]{Reprinted from [Li Y, Wang L, Zheng W, et al. A novel bi-hemispheric discrepancy model for eeg emotion recognition. IEEE Transactions on Cognitive and Developmental Systems, 2020, 13(2): 354-367] with the permission of IEEE Publishing.}

\subsection{Route: \textbf{0\texorpdfstring{$\rightarrow$}.1\texorpdfstring{$\rightarrow$}.3\texorpdfstring{$\rightarrow$}.11} and \textbf{0\texorpdfstring{$\rightarrow$}.3\texorpdfstring{$\rightarrow$}.11}}
The approaches mentioned above take handcraft feature maps as the input, somewhat underestimate DL’s \textbf{‘end-to-end’} representation learning ability, actually disagrees with the data-driven model building philosophy in Deep Learning, the handcrafed representation may lose lots of precious information implied in raw EEGs.

\begin{figure*}[!htbp]
\centering
\includegraphics[width=0.8\textwidth]{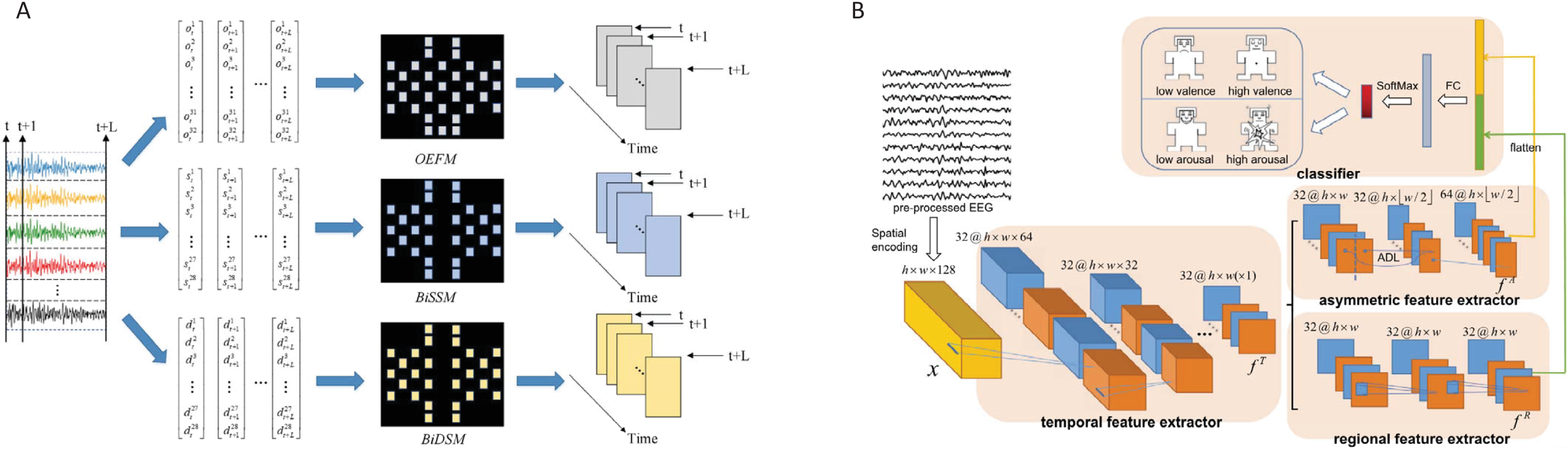}\\
\caption{End-to-end bi-hemispheric discrepancy information integrated model. A: BiDCNN\protect\footnotemark[5] \cite{huang2021differences}, B: RACNN\protect\footnotemark[6] \cite{cui2020eeg}.}
\label{fig:BiDCNN}
\end{figure*}
\footnotetext[5]{Reprinted from [Huang D, Chen S, Liu C, et al. Differences first in asymmetric brain: A bi-hemisphere discrepancy convolutional neural network for EEG emotion recognition[J]. Neurocomputing, 2021, 448: 140-151] with the permission of Elsevier Publishing.}
\footnotetext[6]{Reprinted from [Cui H, Liu A, Zhang X, et al. EEG-based emotion recognition using an end-to-end regional-asymmetric convolutional neural network. Knowledge-Based Systems, 2020, 205: 106243] with the permission of Elsevier Publishing.}

To tackle this issue, similar to the hypothesis of brain asymmetry in emotional processing adopted in the BiHDM and BiDANN models mentioned above, as shown in Figure \ref{fig:BiDCNN}(A), \citet{huang2021differences} proposed an end-to-end bi-hemisphere discrepancy convolutional neural network model (BiDCNN) that recognize the different emotions based on the asymmetry information between the two hemispheres. They transformed the multi-channel EEGs into 2D frames with a shape of $9\times9$, which reserves the knowledge of channel topology. Three different kinds of feature frames are proposed, namely are the original EEG value matrix (OEFM), the bi-hemisphere symmetric matrix (BiSSM) derived by subtracting the symmetrical electrode pairs' values, and the bi-hemisphere division symmetric matrix (BiDSM) derived by dividing the symmetrical electrode pairs' values. In BiDCNN, a 2D convolutional layer is utilized to learn from each of the three preprocessed data. Another end-to-end regional-asymmetric convolutional neural network model (RACNN) was proposed by \citet{cui2020eeg}. As shown in \ref{fig:BiDCNN}(B), it consists of three parts of feature learners to extract temporal, regional, and asymmetric features, respectively. Three-dimensional convolution functions are utilized in temporal feature extractors to mine time-frequency characteristics. The regional feature extractor uses two-dimensional convolution functions to mine regional characteristics from neighboring electrodes. At last, the asymmetric differential layer (ADL) is designed to capture long-distance information between symmetric positions. The original multi-channel EEGs are transformed into the 3D tensor $X$, which reserves the topology information of the electrodes.

\begin{figure*}[!htbp]
\centering
\includegraphics[width=0.6\textwidth]{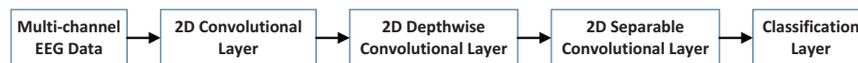}\\
\caption{The end-to-end 2D convolutional neural networks based spatial-temporal Deep Learning model (EEGNet) \cite{lawhern2018eegnet}, in which the input EEGs are randomly arranged into a 2D frame.}
\label{fig:EEGNet}
\end{figure*}

Furthermore, as shown in Figure \ref{fig:EEGNet}, \citet{lawhern2018eegnet} designed an EEG-specific ConvNet model (EEGNet) by integrating depthwise and separable convolutions. Even though the original EEGNet was only validated on the motor imagery classification tasks, its idea was further verified and compared with the EmotioNet model proposed by \citet{wang2018emotionet}, which adopted the 3D EEG representation method shown in \ref{fig:3DCNN}(B). \citet{islam2021eeg} proposed one efficient recognition method with lower computational complexity, lower memory requirement, and lower time consumption. Only applying the traditional CNN model to the channel correlation matrix of Pearson's correlation coefficient can achieve ideal performance. Inspired by the works shown in Figure \ref{fig:Capsule}, \citet{liu2020multi} developed one end-to-end CapsNet based approach that builds directly on the raw EEG signals and judge the emotions. It has three modules, namely, ConvReLU, PrimaryCaps, and EmotionCaps. Compared with the prior MFM-CapsNet approach, it works without the need for any feature design and extraction stages. It incorporates multi-level learned representations into the primary capsules.

\begin{figure*}[!htbp]
\centering
\includegraphics[width=0.3\textwidth]{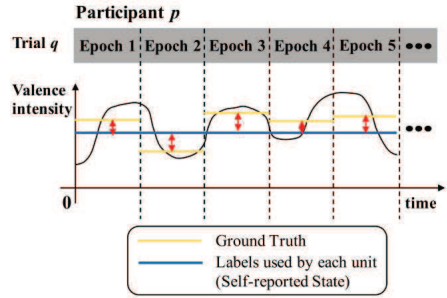}\\
\caption{The fluctuation of the ground truth emotion during one stimulus trial that leads to the problem of unreliability of the ground truth emotional label\protect\footnotemark[7] \cite{wang2018emotionet}.}
\label{fig:unreliability}
\end{figure*}

\footnotetext[7]{Reprinted from [Wang Y, Huang Z, McCane B, et al. EmotioNet: A 3-D convolutional neural network for EEG-based emotion recognition. 2018 International Joint Conference on Neural Networks (IJCNN). IEEE, 2018: 1-7] with the permission of IEEE Publishing.}

One drawback of the approach mentioned above is it's only suitable for processing short input signal segments with only a few second lengths. The dependencies of a long trial signal can not be fully mined. Hence, In addition to the `end-to-end modeling' problem, researchers also pay attention to the \textbf{'context modeling'} problem to mine long signal dependencies. Specifically, the works mentioned above are only suitable for modeling global static information. Nevertheless, the human's emotional cognitive process is not static but continuously evolving. As shown in Figure \ref{fig:unreliability}, the subjects' specific emotions generally evolve over the experiment with the fluctuations of the EEGs. Hence, the reported so-called ground truth emotional label of one trial only reflects their overall evaluation of their emotional experience. The temporal and fluctuant property of the EEG has been neglected in several prior related works. Contextual learning ability should be considered in the DL model study.

\begin{figure*}[!htbp]
\centering
\includegraphics[width=0.6\textwidth]{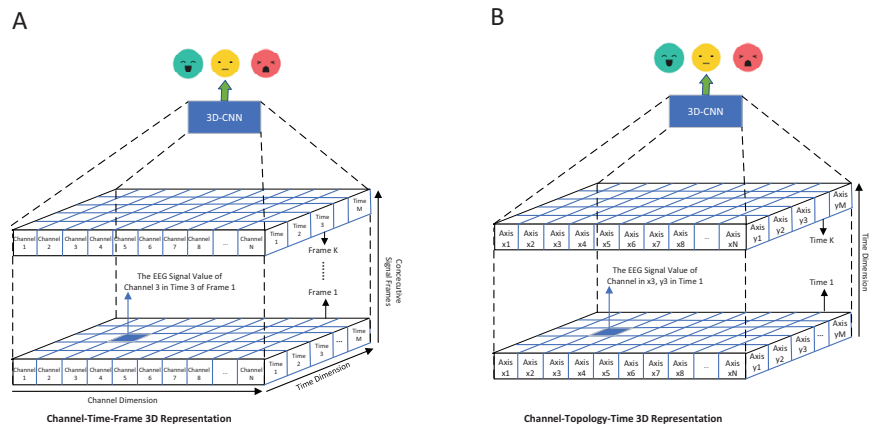}\\
\caption{Two representative 3D EEG representation methods in emotion recognition when adopting the 3D-CNN-based approaches. A: The raw multi-channel EEGs are randomly arranged into a 2D frame and 3D cube \cite{salama2018eeg}. B: The raw multi-channel EEGs are arranged into a 2D frame and 3D cube according to the 10-20 topology of electrodes \cite{wang2018emotionet, cho2020spatio}.}
\label{fig:3DCNN}
\end{figure*}

\begin{figure*}[!htbp]
\centering
\includegraphics[width=0.55\textwidth]{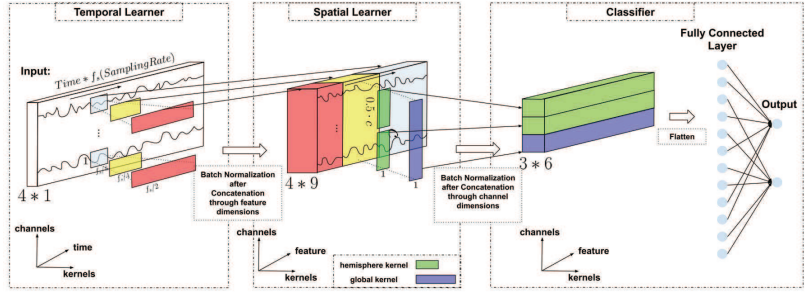}\\
\caption{End-to-end 1D-CNN model TSception\protect\footnotemark[8] \cite{ding2020tsception}, which consists of two types of 1D convolution based learners: temporal learner and
spatial learner. Correspondingly, the channels in the input are deliberately arranged according to the left and right hemispheres.}
\label{fig:1DCNN}
\end{figure*}

Various types of recurrent neural network (RNN) have been successfully applied in EEG-based emotion recognition, including the GRU, the LSTM, and the simple recurrent units (SRU). For example, \citet{wei2020eeg} proposed to use ensemble SRU networks to learn from the features sequences of different EEG rhythms obtained through wavelet transform. Although the recurrent neural networks (RNN) is good as sequential modeling tasks, you also can conduct end-to-end contextual learning only based on CNN without the help of the RNN. Utilizing 3D CNN for sequence modeling has been extensively explored in video analysis, e.g., action recognition \cite{ji20123d}. Hence, inspired by those studies, 3D CNN also has been introduced into this area. For example, \citet{salama2018eeg} proposed one 3D-CNN model that multi-channel EEGs are randomly arranged into frames. As shown in Figure \ref{fig:3DCNN}(A), consecutive frames are further concatenated together into one 3D cube. Besides, since those current open-source EEG datasets do not collect enough trials for each subject, the data augmentation strategy is adopted by adding white Gaussian noise to the original signals. The effectiveness of the 3D-CNN on long sequence modeling is verified in this work. \citet{wang2018emotionet} also proposed one 3D CNN model, called EmotioNet, which integrates batch normalization and dense prediction mechanism for resolving issues of covariance shift and the unreliability of ground truth labels. Specifically, as shown in Figure \ref{fig:3DCNN}(B), they transformed the 2-D frames (channels $\times$ time) into 3-D cubes (channel topology $\times$ time ) as the input fed to the model. The first two layers employ 3-D convolution to learn spatial and temporal characteristics, then followed by a fusion operation, which fuses these learned high-level representations. Consequently, the output of this layer only has temporal characteristics, which are fed into the following two layers to conduct high-level temporal representation learning. At the end of the model, a dense prediction is utilized to make a time-varying emotion state prediction. Experiments show that the EmotioNet performs better than the aforementioned 2D EEGNet (see Figure \ref{fig:EEGNet}) proposed by \citet{lawhern2018eegnet}. \citet{jia2020sst} proposed one spatial-spectral-temporal-based attention 3D dense network, called SST-EmotionNet, which consists of the spatial-spectral stream and spatial-temporal stream. Each stream consists of several attention-based 3D dense blocks. In the end, the two parallel streams are fused for classification. Although the EmotioNet and SST-EmotionNet seem alike in name, they are different models. In addition, the input of SST-EmotionNet is differential entropy feature-based 3D representation instead of raw EEG signals that the EmotioNet can process. Hence, it may be the weakness of the SST-EmotionNet. \citet{cho2020spatio} also introduced two types of 3D-CNN models, namely C3D and R(2 + 1)D. They adopted the 3D EEG representation method shown in Figure \ref{fig:3DCNN}(B). The input raw EEGs are arranged into 2D frames according to electrode topology, and the interpolated 2D EEG frames are further concatenated into 3D cubes. Unfortunately, the aforementioned 3D CNN-based approaches only verified on the few second long sequences. Conducting consecutive 1D-CNN operations also could effectively extract spatial-temporal information from raw multichannel EEGs. Inspired by the Inception block of GoogleNet, \citet{ding2020tsception} proposed the TSception model. As shown in Figure \ref{fig:1DCNN}, it consists of two types of 1D convolution-based learners for end-to-end temporal-spatial information modeling. Correspondingly, the channels in the input frame are deliberately arranged according to which hemisphere they locate. Then the spatial learner adopts a multi-scale 1D convolution operation to learn the asymmetry features from both hemispheres. The temporal learner adopts multi-scale 1D convolution operations that help to extract multiple temporal and frequency patterns.

\begin{figure*}[!htbp]
\centering
\includegraphics[width=0.8\textwidth]{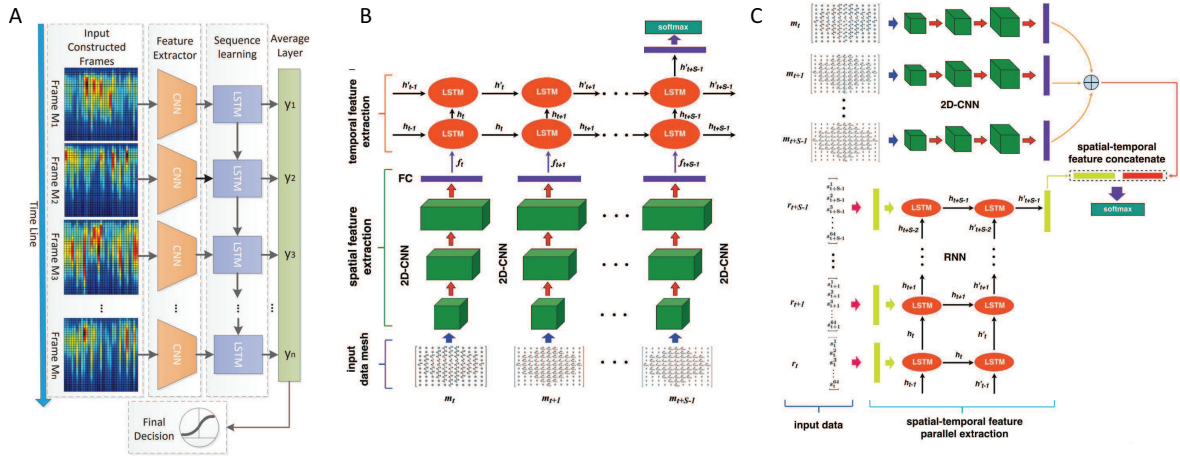}\\
\caption{Three representative CNN-RNN hybrid models. A: The CRNN hybrid DL model proposed by \citet{li2016emotion} that the input data representation are 2D wavelet scalogram representation\protect\footnotemark[9]. B: The cascade hybrid model introduced by \citet{zhang2018cascade} that similar with the CRNN model, but the input are original EEG signal and reserves the channel spatial topology information. C: The parallel hybrid model also introduced in work of \citet{zhang2018cascade}. The spatial-temporal features are extracted in parallel by CNN and LSTM, respectively, and fused for final recognition.}
\label{fig:CNN-LSTM}
\end{figure*}
\footnotetext[8]{Reprinted from [Ding Y, Robinson N, Zeng Q, et al. Tsception: a deep learning framework for emotion detection using EEG. 2020 International Joint Conference on Neural Networks (IJCNN). IEEE, 2020: 1-7] with the permission of IEEE Publishing.}
\footnotetext[9]{Reprinted from [Li X, Song D, Zhang P, et al. Emotion recognition from multi-channel EEG data through convolutional recurrent neural network. 2016 IEEE international conference on bioinformatics and biomedicine (BIBM). IEEE, 2016: 352-359] with the permission of IEEE Publishing.}

Integrating both the ability of CNN and LSTM to build hybrid Deep Learning models is a natural choice. For example, as shown in Figure \ref{fig:CNN-LSTM}(A), \citet{li2016emotion} propose a wavelet transformation-based preprocessing that transforms the multi-channel EEG into scalogram based 2D frame representation. Each frame reflects the energy distribution of the multi-channel EEG in a time slice. Further, they designed one hybrid DL model, called C-RNN, which fuses CNN and RNN. Specifically, the CNN module can decode inter-channel relationships, and the RNN (LSTM) module helps capture contextual information from sequential data. Even though this work does not achieve very high performance, this work contributes to the further development of end-to-end and hybrid EEG emotion recognition models. \citet{zhang2018cascade} introduce cascade (see Figure \ref{fig:CNN-LSTM}(B)) and parallel (see Figure \ref{fig:CNN-LSTM}(C)) hybrid DL models integrate CNN and RNN, in which the input is the raw EEG signal arranged according to electrode topology, each input map corresponds to a signal timestamp. The model can effectively learn the joint spatio-temporal representations from raw EEGs, the complex dependencies between adjacent signals, and the contextual information can be fully mined. For cascade model, it follows the same mechanism as the works shown in Figure \ref{fig:CNN-LSTM}(A). It first learns the spatial representation from each data frame, and the sequence of the learned spatial representations is further carried to the RNN to learn temporal representations. Unlike the cascade structure-based model, the parallel structure-based model learns the spatial and temporal representations from EEG parallelly. At last, the concatenated representations are utilized for final recognition. Both the cascade and parallel models consistently outperform the SOTA methods. Almost at the same time, a similar parallel hybrid model is proposed by \citet{yang2018emotion}. They introduce a preprocessing method that removes the non-stimulus pre-trial baseline signal from the stimulus trial signal. Based on the preprocessed data, the parallel hybrid model's accuracy is greatly improved by 32\%. Attention is a special mechanism in the information processing of the human brain, hence, inspired by those neuroscience findings, \citet{tao2020eeg} introduced attention mechanisms into the aforementioned CNN-LSTM hybrid DL models. They integrated channel-wise Attention and self-attention into the CNN and RNN, respectively, to learn attention characteristics among the channels and the attention characteristics within a sequence. There are several potential weaknesses in the works mentioned above: Firstly, both the cascade and parallel models mentioned above require a 2D representation of EEG channels. If we represent them according to their topology, it may cause information loss because channels are actually arranged in the 3D space, and the 2D frame has multiple positions of the null electrodes that need to be padded with zeros. Secondly, these approaches utilize RNNs to capture inter-channel and inter-time relations. However, each 2D frame corresponds to a time step rather than a time window. Hence, if the signal is extremely long with a high sampling frequency, the model computation burden will be largely increased, especially for RNN models. Thirdly, such a two-stage approach is somewhat inconvenient to implement. The whole process is time-consuming and highly dependent on domain knowledge. Hence, developing approaches modeling directly on the original multi-channel EEGs regardless of considering the topology is worth studying.

\begin{figure*}[!htbp]
\centering
\includegraphics[width=0.5\textwidth]{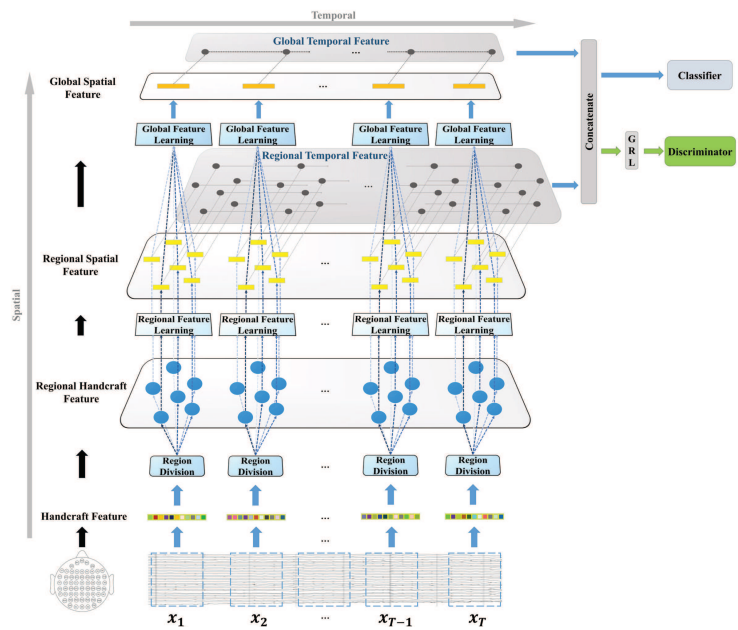}\\
\caption{The framework of LSTM based model R2G-STNN\protect\footnotemark[10]. The region to global (R2G) process includes two streams. The spatial stream constructs the relation in and among all the brain regions hierarchically, while the temporal stream captures the EEG signal's dynamic information as well as learning from the brain regions' time sequences \cite{li2019regional}.}
\label{fig:R2G-STNN}
\end{figure*}

\footnotetext[10]{Reprinted from [Li Y, Zheng W, Wang L, et al. From regional to global brain: A novel hierarchical spatial-temporal neural network model for EEG emotion recognition. IEEE Transactions on Affective Computing, 2019] with the permission of IEEE Publishing.}

When building a spatial-temporal model, the spatial information can also be effectively processed by the RNN model without the help of CNN. For example, \citet{zhang2019spatial} designed a spatial-temporal hybrid DL model called STRNN that only integrated RNN modules. It utilized RNN to learn the temporal dependencies and to capture the spatial dependencies in the multi-channel context. Firstly, a quad-directional spatial RNN (SRNN) layer scans each slice from different angles. Following the SRNN, a bi-directional temporal RNN layer (TRNN) learns the long-term temporal dependencies by the forward and backward processing of the sequences. \citet{li2019regional} proposed one LSTM based regional-to-global brain spatial-temporal neural network model (R2G-STNN) that realizes the regional to global hierarchical feature learning. A bidirectional long short-term memory (BiLSTM) network was adopted to learn spatial characteristics to model the regional correlations among EEG channels. Further, the regional attention layer is also introduced in the R2G-STNN model to differentiate the importance of different brain regions in the emotion process. The attention layer learns and assigns weights to increase or reduce the influence of different regions. At last, BiLSTM is adopted to learn the temporal dependencies of regional and global spatial representations. This work also adds one discriminator to solve the domain shift problem. \citet{lew2020eeg} proposed one regionally-operated domain adversarial network (RODAN) based on the GRU-RNN model and attention mechanism that considers the spatial-temporal relationships among brain regions and across time as well as resolves the distribution shift between training and testing data by integrating the domain adversarial network.

\begin{figure*}[!htbp]
\centering
\includegraphics[width=0.4\textwidth]{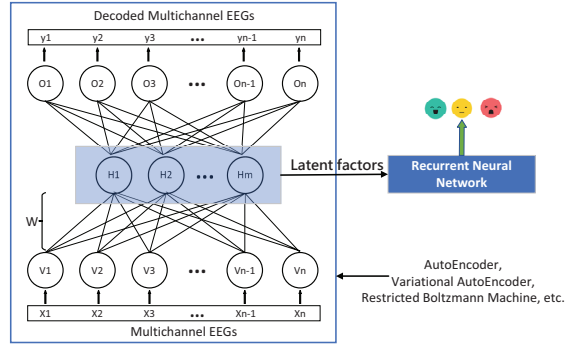}\\
\caption{Illustration of a hybrid model that fuses unsupervised decoding of latent source factors and the recurrent neural network (RNN).}
\label{fig:AE-LSTM}
\end{figure*}

\citet{li2020latent} introduced a raw EEG decoding-based recognition approaches. Specifically, they hypothesize that the EEG signals are a mixture of the multiple latent signals produced by the internal brain processes. Hence, the automatically learned representation of the source signals must contribute to building robust recognition models. They first utilized different AutoEncoder-like networks, including stacked AutoEncoder (non-generative model), restricted Boltzmann machines (generative model), and variational AutoEncoder (generative model) to decode the source signals from the raw EEGs, then further utilized the LSTM for sequence learning and emotion recognition. One weakness of this work is the input sequences for LSTM processing are a sampled sequence of the entire decoded latent source signals that reduce the computation cost meanwhile will produce information loss. A similar AE+LSTM model was also proposed, whereas handcrafted feature extraction is needed to construct sequence before fed into the LSTM model \cite{xing2019sae} that also may lead to information loss. They are not strict end-to-end models, still need extracting intermediate latent EEG source signals. Such a two-stage approach is somewhat inconvenient for practical application.

\begin{figure*}[!htbp]
\centering
\includegraphics[width=0.4\textwidth]{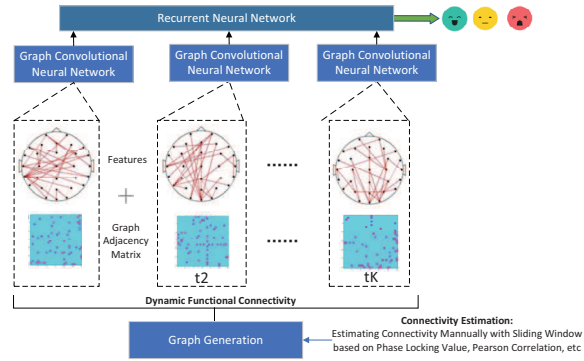}\\
\caption{Illustration of a hybrid model that fuses graph convolutional neural network (GCNN) and RNN.}
\label{fig:ECLGCNN}
\end{figure*}

As mentioned in Route: \textbf{$0\rightarrow1\rightarrow2\rightarrow3\rightarrow11$}, the graph neural networks (GNN) is capable of decoding the intrinsic correlations among the multi-source signal. Nevertheless, the functional connectivity between two channels is not static but continuously changing with the evolution of the emotional process. Hence, developing GNN based models that can capture the functional connectivity change between EEG channels may greatly enhance the emotion recognition effect. Combining the GNN model with some sequence modeling methods is one direct way, which is similar to the ideas shown in Figure \ref{fig:CNN-LSTM}. For example, \citet{yin2021eeg} proposed one hybrid DL model (named ECGGCNN) that integrates GCNN and LSTM. The GCNN module helps to learn the channel connectivity within a time slice. A parallel GCNN computing mode is designed to receive data frames in sequential order and transports the learned representations to the LSTM layer, which is used to model the evolution of the channel connectivity. At last, the dense layer predicts final emotions according to the LSTM's learned contextual information.

\begin{figure*}[!htbp]
\centering
\includegraphics[width=0.5\textwidth]{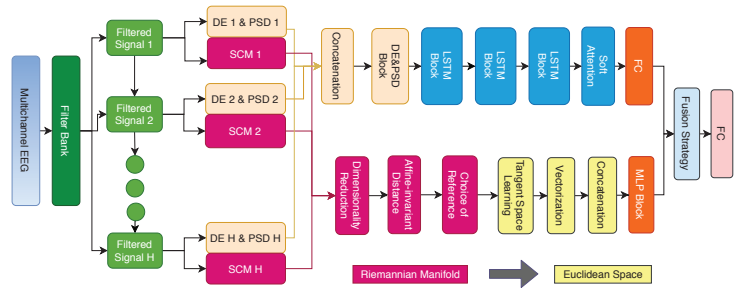}\\
\caption{A temporal-spatial EEG information Riemannian fusion network (RFNet) for affective BCI \cite{zhang2020rfnet}.}
\label{fig:RFNet}
\end{figure*}

Traditional BCI solutions rely on Riemannian geometry, in which the spatial covariance matrices (SCM) derived from raw EEGs contribute to developing BCI algorithms. As the SCM is symmetric positive definite, it lies in Riemannian space rather than the Euclidean space, the models designed in Euclidean space can not be directly employed \cite{kalunga2015euclidean}. To tackle this problem, as shown in Figure \ref{fig:RFNet}, \citet{zhang2020rfnet} proposed one end-to-end Riemannian fusion network (RFNesst), which separately extract spatial representation in the Riemannian space and temporal representation in the Euclidean space. Finally, the attention strategy guides the fusion of different learned embeddings. One key operation is mapping the spatial representations extracted from SCM in Riemannian space to the embeddings in Euclidean space through tangent space learning. Finally, an MLP is connected to process the spatial information. For the temporal representation learning in Euclidean space, they utilize the attention-LSTM network to learn from the EEG frequency sub-bands. The learned temporal representation is transported forward to a fully connected layer to get the high-level representations. At last, the latent spatial and temporal representations in Euclidean space are concatenated for final emotion recognition tasks.

Though lots of works mentioned above adopted the RNN based approaches, they do not talk about the computation cost. Actually, the RNNs tend to be more computationally intensive than CNNs, especially when the target sequence is quite long. The reason for that is because RNNs are very memory intensive with backpropagation through time.


\subsection{Route: \textbf{3\texorpdfstring{$\rightarrow$}.7\texorpdfstring{$\rightarrow$}.11} and \textbf{5\texorpdfstring{$\rightarrow$}.7\texorpdfstring{$\rightarrow$}.11}}
We hope the developed AI system can have consistent and robust performance on a wide range of data domains. Nevertheless, the difference in data distribution among users would result in degraded recognition performance. For example, \citet{kim2008emotion} firstly built a subject-dependent model with the specific user data, the 4-class emotion recognition rates can reach 95\% on intra-subject data. However, when one subject-independent recognition model was established with the mixed data of three users, the data distribution deviation among three users caused the recognition accuracy to be reduced to 70\%, suggesting the simple and crudely built user-independent models will inevitably have low robustness. Likewise, \citet{Petrantonakis2012Adaptive} verified the proposed method in the specific individual data and non-specific individual data, respectively. The experimental recognition accuracy of the subject-dependent model was 70\%$\sim$100\%, while for the subject-independent model, the performance decreased to 10\%$\sim$20\%. \citet{alzoubi2012detecting} analyzed the physiological signals of 27 students under eight kinds of emotions. They found that the consistency of emotional physiological response patterns of different individuals was poor and even proposed that user-independent modeling methods were not feasible in EEG-based emotion recognition research.

This problem is typically referred to as `\textbf{domain shift}'. The EEG data exhibit `domain shift' problems due to various factors. The differences among source users (such as gender, culture, gene, etc.) would lead to different neurophysiological activity patterns. Some studies have shown that the asynchronous activity of the brain presents different patterns in different individuals \cite{Hamann2004Individual}. Regarding gender factors, there has been a tremendous amount of research on the difference between men and women in processing emotional stimuli. For example, researchers found that men and women showed different scalp activity patterns and distributions in processing emotional information of music \cite{flores2009differential}. \citet{bilalpur2017discovering} adopted the EEG to examine the gender difference in facial emotion recognition. They found that women were more sensitive and faster to recognize negative emotions than men, irrespective of age, even when only partial information was provided. \citet{goshvarpour2019eeg} assessed EEG powers in depressing, fun, and sad music videos for women's and men's groups, respectively. They found the mean power of all frequency bands in the women's group was significantly higher than that of the men's group. There were significant gender-related differences in parietal lobe activation for depressing and sad music videos and limbic lobe activation for fun stimuli. It is believed that biological and sociocultural factors cause the differences \cite{bradley2001emotion}. In terms of biological factors, \citet{lee2005neural} revealed that the right insula and left thalamus were consistently activated for men but not for women during the emotional experience. They also suggested that men evaluate current emotional experiences by recalling past emotional experiences, whereas women tended to evaluate current emotional experiences rapidly according to the immediate stimuli. In terms of the genetic factor, \citet{raab2016understanding} revealed that serotonin transporter gene (5-HTTLPR) polymorphisms are closed correlated with brain activation during facial emotion processing. In terms of sociocultural factors, women socialize differently than men, which is not decided by genetic factors but by social norms defined by politics, culture, and religion \cite{fischer2004gender}. In Western culture, at least, women are more emotional than men and more reactive to unpleasant events \cite{bradley2001emotion}. \citet{zhu2015cross} studied the cross-gender EEG modeling, the recognition performance of female models is better than male models. It indicates that women share more stable EEG patterns during emotional experience than men. \citet{pava2018gender} conducted a special study on the gender differences present in the EEG-based emotion recognition system. They found the gender differences in the classification performance, the gender differences in differential entropy features extracted from EEG, and the gender differences in evaluating the emotion experienced in the Valence dimension. Regarding cultural factors, \citet{huang2004native} studied the evaluation of emotional images by Chinese and foreign groups. The finding showed that the viewpoint difference between Western and Chinese users would lead to large differences in emotion evaluation. \citet{kurbalija2018emotion} conducted experiments with several Serbian and Chinese subjects. They observed that cultural differences between the subjects did not significantly impact the recognition tasks and models. Nevertheless, \citet{gan2019cross} found that French has higher mean accuracy on beta frequency band while Chinese tends to perform better on gamma frequency band on tasks of recognizing emotions. They also found similarities and differences in connectivity patterns between Chinese and French subjects. Hence, we can not say the demographic factors will not affect the model building. We should pay more attention to the these factors in developing and assessing EEG-based emotion recognition systems.

\begin{figure*}[!htbp]
\centering
\includegraphics[width=0.6\textwidth]{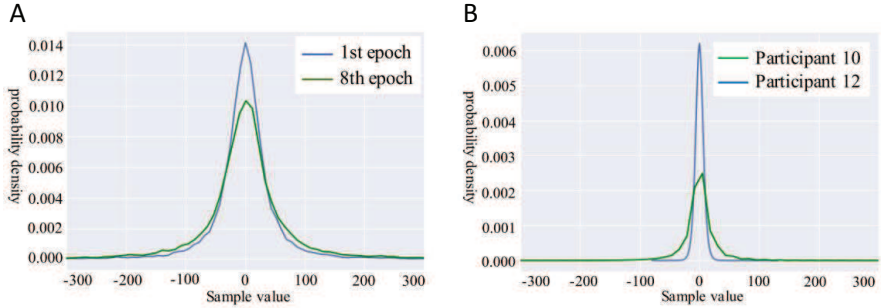}\\
\caption{Domain shift problems in DEAP data set\protect\footnotemark[11]. A: non-stationary EEG distribution between two epochs. B: inter-subject EEG variability under the same trial \cite{wang2018emotionet}.}
\label{fig:shift}
\end{figure*}

The domain shift problem will appear not only in different sources of EEG data, but they could also appear in the same source of EEG. Take the DEAP dataset for example, as shown in Figure \ref{fig:shift}(A), the instantaneous distribution of EEG continuously evolves that causing the data non-stationarity issues. The reason lies in the mental change of a participant or the technical factors, e.g., drying electrode gel changes. Therefore, the distributions of different epochs might be different. As shown in Figure \ref{fig:shift}(B), inter-subject variability also is a representative domain shift problem, which indicates there also exist discrepancies among the statistical characteristics of different subjects.

Though much research has studied subject-dependent modeling, the construction of a user-independent recognition system can meet practical application requirements. This drives the relevant researchers to focus on and improve the effect of user-independent approaches. At present, there are mainly four ways to solve the above problems, as follows:

(1) One way is calibrating or aligning the physiological signals among the participants. The calibration-based methods reduce the difference in physiological measurement among the participants by using the baseline characteristics of participants. For example, \citet{mohammad2010using} calibrated the level of physiological signals of each participant and took the physiological characteristics at calm state as baseline characteristics. The physiological characteristics under various emotional states subtract the baseline characteristics to obtain the calibrated characteristics. Then, the relative physiological characteristics were used to establish the prediction model of emotion and obtained a good experimental effect. The alignment-based methods are widely employed in brain-computer interface studies. Researchers proposed to align the multichannel EEGs of the source domain and target domain in the common Riemannian manifold space and judge the states according to the Riemannian distance between each state center and the EEG covariance matrix \cite{zanini2017transfer}. \citet{fernandez2021cross} studied different feature normalization methods combined with the deep neural network. The results show that the proposed stratified normalization-based neural network significantly outperforms batch normalization-based approaches in cross-subject emotion recognition settings.

\begin{figure*}[!htbp]
\centering
\includegraphics[width=0.5\textwidth]{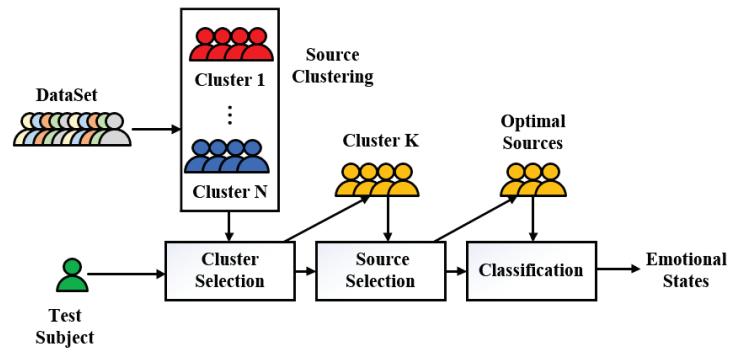}\\
\caption{The framework of the subject clustering based cross-subject recognition method\protect\footnotemark[12]. Cluster selection is required for selecting the optimal source cluster. Source selection is utilized for further selecting the optimal sources from the optimal cluster. The emotional state of the target is decided by those optimal sources \cite{liu2021domain}.}
\label{fig:subject-clustering}
\end{figure*}

(2) Another way is to establish aspect-oriented models according to specific factors that cause domain shift. \citet{zhou2011affect} established culture-specific model and gender-specific model for 46 participants, as well as compared their performance with the general models. The experiments showed that the specificities in gender and culture would affect recognition performance. The emotion model built on the user of the same culture or gender specificity improves the recognition accuracy. Similarly, \citet{bailenson2008real} introduced the individual-specific model, gender-specific model, and general model, respectively. The experiments suggested that the individual-specific model and gender-specific model had a higher recognition accuracy than the universal model. \citet{chen2017subject} proposed a user grouping-based approach, specifically the modeling workflow contains three stages, including user grouping, emotional state pool partition, and final state discrimination. \citet{liu2021domain} also introduced subject clustering into cross-subject emotion recognition, as shown in Figure \ref{fig:subject-clustering}. Based on the clustering results, cluster selection was conducted to match the target subject with one optimal source cluster, whose source subjects have similar emotional EEG activity patterns. Subspace alignment method is further utilized for selecting the optimal sources with possibly `positive transfer'. Finally, the emotional state of the target data is decided by majority voting of the optimal sources.

\footnotetext[11]{Reprinted from [Wang Y, Huang Z, McCane B, et al. EmotioNet: A 3-D convolutional neural network for EEG-based emotion recognition. 2018 International Joint Conference on Neural Networks (IJCNN). IEEE, 2018: 1-7] with the permission of IEEE Publishing.}
\footnotetext[12]{Reprinted from [Liu J, Shen X, Song S, et al. Domain Adaptation for Cross-Subject Emotion Recognition by Subject Clustering. 2021 10th International IEEE/EMBS Conference on Neural Engineering (NER). IEEE, 2021: 904-908] with the permission of IEEE Publishing.}

(3) In recent years, the Transfer Learning method has been paid more and more attention by scholars. For a domain $D=\{X,P(X)\}$ with a feature space $X$ and the corresponding marginal probability distribution $P(X)$. When source domain $D_S\{X_S,P(X_S)\}$ and target domains $D_T=\{X_T,P(X_T)\}$ lie in the same feature space, namely $X_S=X_T$, and modeled for the same type of task, the domain shift issue can be resolved through Transfer Learning or called domain adaptation approaches. Transfer Learning maps the EEG features from source and target domains into the common feature representation space, where the inter-subject or inter-session shifts of the EEG data are adjusted, and distinctive features across subjects or sessions are obtained. For example, \citet{lan2018domain} verified various domain adaptation methods, including maximum independence domain adaption (MIDA) and transfer component analysis (TCA), which is believed able to decrease inter-subject variance as well as the inter-dataset discrepancies. By utilizing those domain adaptation methods, only a simple logistic regression model can have a significant performance improvement in both cross-subject and cross-dataset experimental settings compared to the baselines that have no domain adaptation capabilities.

\begin{figure*}[!htbp]
\centering
\includegraphics[width=0.8\textwidth]{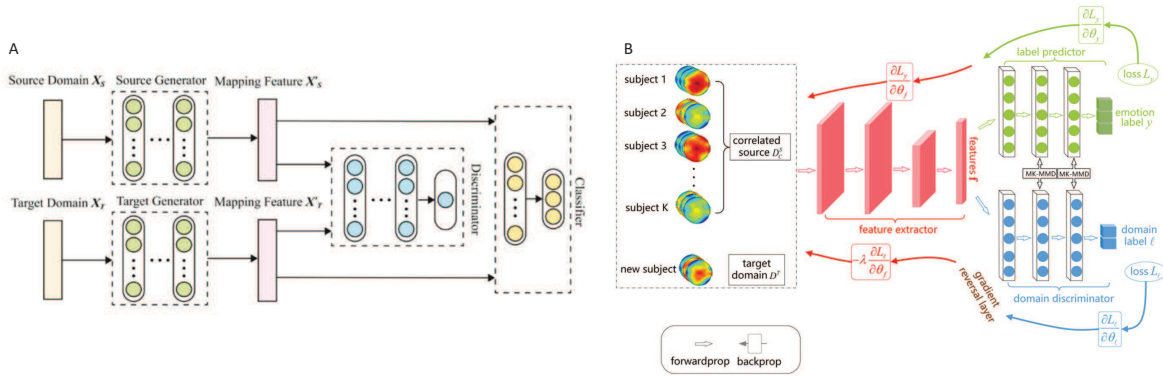}\\
\caption{Two representative domain adaptation neural networks (DANN). A: The Wasserstein generative adversarial network based domain adaptation approach\protect\footnotemark[13] \cite{luo2018wgan}. B: The multi-source adaptation transfer network (DMATN) based domain adaptation approach that integrates the deep adversarial network and the multi-kernel maximum mean discrepancies (MK-MMDs) measurement\protect\footnotemark[14] \cite{wang2021deep}.}
\label{fig:DAN}
\end{figure*}

\footnotetext[13]{Reprinted from [Luo Y, Zhang S Y, Zheng W L, et al. WGAN domain adaptation for EEG-based emotion recognition. International Conference on Neural Information Processing. Springer, Cham, 2018: 275-286] with the permission of Springer Publishing.}

\footnotetext[14]{Reprinted from [Wang F, Zhang W, Xu Z, et al. A deep multi-source adaptation transfer network for cross-subject electroencephalogram emotion recognition. Neural Computing and Applications, 2021: 1-13] with the permission of Springer Publishing.}

Transfer Learning techniques have also been extensively studied in Deep Learning. Among them, a naive approach draws lessons from experience in computer vision. They realize cross-domain application by fine-tuning the source domain model on the target domain data, including fine-tuning the whole network model and fine-tuning only part of the network structure in the target domain. For example, \citet{cimtay2020investigating} proposed to use the Inception Resnet model that pre-trained from the multi-subject raw EEG data, and they obtained promising cross-subject recognition performance on three benchmark datasets. \citet{wang2020emotion} proposed a residual block-based CNN, which is trained on the electrode-frequency distribution maps (EFDMs) with short-time Fourier transform (STFT), the pre-trained model on SEED dataset can be successfully transferred to apply on DEAP dataset.

Integrating domain adaptation mechanisms into the Deep Learning model is increasingly gained attention. For example, as shown in Figure \ref{fig:DAN}(A), \citet{luo2018wgan} proposed the Wasserstein generative adversarial network domain adaptation (WGANDA), transferred the differential entropy characteristics of different domains into the common space, which is helpful to improve the emotion recognition effect across participants. Inspired by the same idea, \citet{li2018cross} also developed one domain adaptation neural network (DANN) based on a deep adversarial network. This model contains components of one feature extractor, one label predictor, and one domain discriminator. The feature extractor is trained in the direction for deceiving the domain discriminator by maximizing the domain discrimination losses. In this way, the feature extractor eliminates the domain-specific characteristics of the input for the purpose of increasing the domain identification loss. The multi-kernel maximum mean discrepancies (MK-MMDs) were utilized for measuring the domain discrepancies. By simultaneously optimizing the loss functions of the MK-MMDs and the task, the DAN can reduce domain shift across domains, meanwhile preserving domain-invariant and task-related features. A multi-source adaptation transfer network (DMATN) for cross-subject emotion recognition is proposed by \citet{wang2021deep}, as shown in Figure \ref{fig:DAN}(B). The mechanism of this model is exactly similar to the model proposed by \citet{li2018cross}. The difference between these two approaches is the DMATN needs to select target domain-related source domains before modeling, and the features are automatically learned by the networks instead of handcrafted features. \citet{cai2021cross} also follows the DAN-based approach and proposed one model called maximum classifier discrepancy (MCD) for domain adversarial neural networks (MCD\_DA). MCD\_DA not only adopts the GAN module to adapt the feature distribution between the source and target domains, but also it maximizes the classifier difference between the source and target domains. \citet{zhao2021plug} proposes a plug-and-play domain adaptation method based on LSTM-Encoder-Decoder, in which the subject-invariant representations are modeled by the shared encoder and the subject-private representations are modeled by the private encoders. The target prediction is the integration of the shared classifier with those of individual classifiers ensemble.

In the aforementioned methods, the DANN regards each domain as a whole, ignoring the class boundary in each domain. The MCD considers the specific class-boundary and trains adversarially to relocate the target feature to be inside the source features. However, since the original feature space of source and target are related but distinguishable, MCD will eliminate useful features, especially when the two domains are far more than similar. To eliminate the problems of DANN and MCD simultaneously, \citet{ding2021eeg} proposed the task-specific domain adversarial neural network (T-DANN). Another problem that needs to be mentioned here is that although many works adopt discriminator-based domain adaptation approaches, it's a challenge to apply on the target domain with few-labeled data. Hence, \citet{wang2021cross} proposed one few-label adversarial domain adaptation (FLADA) approach for cross-subject emotion recognition task. The FLADA originates from Meta-learning, which is to find a feature representation that is broadly suitable for the target subject and source subject with limited labels. This approach can be applied to all Deep Learning models.
 
\begin{figure*}[!htbp]
\centering
\includegraphics[width=0.45\textwidth]{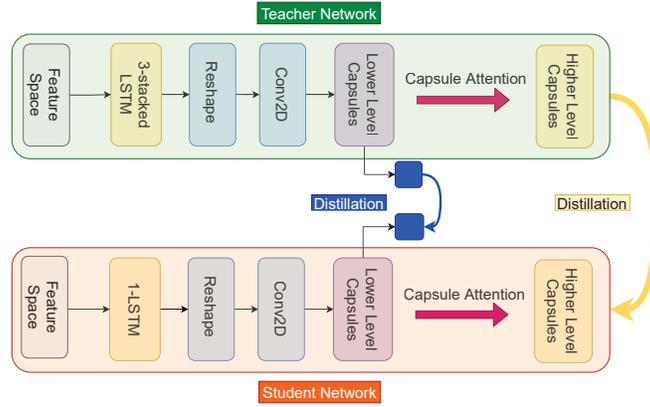}\\
\caption{One novel knowledge distillation DL framework for Affective BCI based on LSTM-Capsule structure that compresses the large model without performance degradation \cite{zhang2021distilling}.}
\label{fig:distillation}
\end{figure*}

Recently, \citet{zhang2021distilling} proposed a novel knowledge distillation-based knowledge transfer pipeline to distill EEG representations via capsule-based architectures, as shown in Figure \ref{fig:distillation}, the pipeline contains a teacher network and a student network. They first pre-train a large model (teacher network) on the large amounts of available cross-subject data. Then, using the pre-trained teacher to learn information embedded in capsules with intra-subject data. At last, the training of the lightweight student network on intra-subject data can be guided by the privileged information learned by the teacher via capsules. This knowledge distillation-based approach improves the robustness when faced with limited training samples and maximally compresses the model with minimal loss in performance. This approach follows the modeling idea of `generalization to concretization'. \citet{zhong2020eeg} proposed a regularized graph neural network (RGNN) to simultaneously resolve the domain shift problems of inter-subject variability and inconsistent/noisy emotion labels. Specifically, two regularizers are integrated into the model. One regularizer is the node-wise domain adversarial training (NodeDAT) mechanism, which regularizes RGNN to generalize well in inter-subject recognition scenarios. NodeDAT is a fine-grained regularization method to correct domain shift for each channel. Another regularizer is the emotion-aware distribution learning (EmotionDL) mechanism, which solves the problem of inconsistent emotion labels by learning the label distribution instead of the hard labels. To improve the recognition performance when facing large amounts of noisy labels.

\begin{figure*}[!htbp]
\centering
\includegraphics[width=0.9\textwidth]{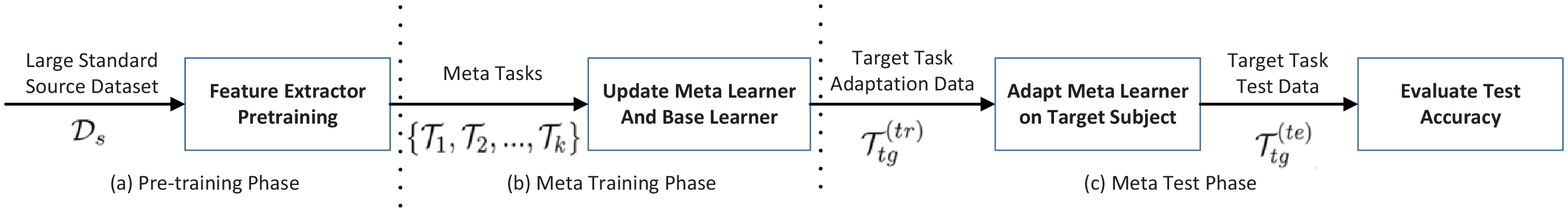}\\
\caption{The Workflow of the Meta Learning on constrained Transfer Learning (MLCL) \cite{duan2020meta}.}
\label{fig:meta}
\end{figure*}

The transfer process also could be accelerated by applying Meta Learning. For example, \citet{duan2020ultra} introduced the meta update mechanism (MUPS-EEG) for cross-subject classification. MUPS-EEG involves interaction between a base learner and a meta learner during meta training, each formed with a representation learning network and a prediction learning network. \citet {duan2020meta} proposed to utilize the model-agnostic meta-learning (MAML) algorithm to perform under limited target data, as shown in Figure \ref{fig:meta}. Experiments show it keeps enough flexibility to adapt to the new subject while significantly reducing the number of parameters to transfer. Considering existing domain adaptation approaches may become sensitive where a low discriminative feature space among classes is given. \citet{jimenez2021standardization} proposed a Standardization-Refinement Domain Adaptation (SRDA) method, which trains a target neural network model using Adaptive Batch Normalization (AdaBN) and introducing a novel loss based on the Variation of Information (VOI). Using AdaBN, SRDA makes the marginal distributions similar in source and target domains.

(4) In addition to calibration, alignment, aspect-oriented modeling, and transfer learning, some researchers directly explore the subject-independent robustness features. For example, \citet{soleymani2012multimodal} studied the importance of different EEG features in emotion recognition. The findings showed that for Arousal, the PSD extracted from the low-frequency Alpha rhythm in the occipital EEGs is the most distinguishable feature. But for Valence, the key features are mainly from the Beta and Gamma rhythm in the temporal lobe EEGs. \citet{zheng2015investigating} investigated the stable EEG activity patterns to promote the effect of emotion recognition across participants and time periods. The results showed that the Beta and Gamma rhythms of both sides of temporal lobes under positive emotion generated stronger activation than negative ones. Moreover, the subject-independent EEG features were mainly from these scalp channels and frequency bands. \citet{Li2018Exploring} extracted 18 widely used EEG features and studied the contribution of each feature in cross-subject emotion recognition according to the results derived from multiple feature selection methods. After analysis, the features of Hjorth parameters in Beta rhythm yield the best cross-subject recognition results. \citet{yin2020locally} proposed a locally-robust feature selection (LRFS) method for individual-independent emotion recognition. Kernel density estimation (KDE) first modeled the extracted EEG features. The inter-individual consistency of the EEG features is described by evaluating the similarity of all density functions between every two subjects, and the locally-robust EEG features could be further determined.

In addition, there has been one review paper proposed by \citet{wan2021review} that focuses on the Transfer Learning techniques in solving the `domain shift' problem in EEG analysis. As a complement to our review, we recommend the readers reference this review paper for detailed guidance of building EEG-based cross-subject emotion recognition models.

\subsection{Route: \textbf{3\texorpdfstring{$\rightarrow$}.8\texorpdfstring{$\rightarrow$}.11} and \textbf{5\texorpdfstring{$\rightarrow$}.8\texorpdfstring{$\rightarrow$}.11}}
\begin{figure*}[!htbp]
\centering
\includegraphics[width=0.6\textwidth]{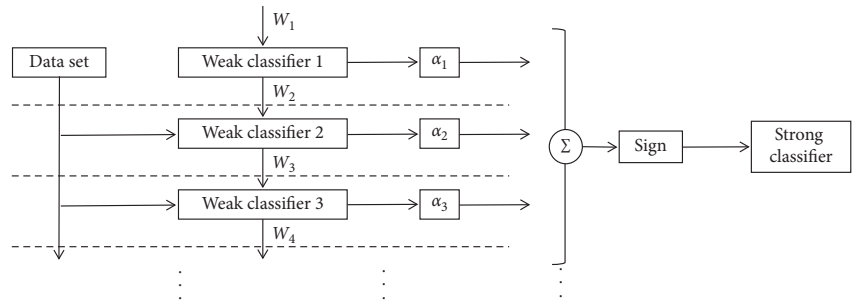}\\
\caption{The working principle diagram of AdaBoost \cite{chen2021emotion}.}
\label{fig:adaboost}
\end{figure*}

The Ensemble Learning-based recognition approaches are also an effective strategy for getting ideal performance in the EEG-based emotion recognition tasks. Ensemble Learning follows the idea of `two heads are better than one' by taking advantage of multiple models' decision boundaries. For example, \citet{mehmood2017optimal} used four Ensemble Learning strategies (Bagging, Boosting, Stacking, and Voting) to integrate the abilities of multiple machine learning models and then obtained the best recognition effect based on the Voting approach. Stacking is to use the training data to build several base learners and use the probability output of these learners as a new training set to learn a meta learner. The meta learner learns to organize the input and assign weights to the base learners. For example, \citet{yin2017recognition} proposed one Deep Learning-based stacking algorithm. The constructed network combines multi-layer stacked AutoEncoders. Each corresponds to a feature subset of multiple time-frequency-domain features. Hence, multiple types of higher-level features can be extracted from those subnetworks that promote the generalization capability and the robustness against data imbalance. \citet{chen2021emotion} proposed to apply the Adaboost algorithm to elevate the recognition performance adaptively. As shown in Figure \ref{fig:adaboost}, it works based on the iteration mechanism. In each iteration, a weak classifier is included to be trained on the weighted samples, the importance (weight) of this weak classifier is determined in each iteration, and the sample weight is adjusted according to the classification results. Specifically, the misclassified samples will be assigned a higher weight in the next iteration to get more attention in model training, while the correctly classified samples' weights will decrease. At last, those weak classifiers are combined into one strong classifier by their weights to obtain an improved recognition performance while improving generalization ability and avoiding over-fitting.

\subsection{Route: \textbf{0\texorpdfstring{$\rightarrow$}.1\texorpdfstring{$\rightarrow$}.2\texorpdfstring{$\rightarrow$}.6\texorpdfstring{$\rightarrow$}.11}}
The acquired EEG signal and the corresponding labels might be noisy, imprecise, and uncertain, leading to precise modeling problems. Fuzzy Logic provides a foundation for approximate reasoning based on fuzzy set theory. Hence, in addition to those aforementioned technical routes, some studies adopted the Fuzzy Logic-based technical route, assigning the samples to multiple categories with a certain degree of membership. The Fuzzy C-Means and Fuzzy k-Means clustering methods are two representative Fuzzy Logic methods that have been implemented very early in EEG-based emotion recognition \cite{murugappan2007eeg}. Besides, \citet{matiko2014fuzzy} designed fuzzy classification rules based on the asymmetry theory of emotional activities in the left and right brain hemispheres. Then, the classifier outputs the type of emotion and the confidence levels according to various rules. Based on Dempster-Shafer's theory, \citet{soroush2018novel} improved the accuracy of recognition by fusing the feature subsets and multiple MLP classifiers. Additionally, fuzzy cognitive maps (FCMs), which combine aspects of Fuzzy Logic, neural networks, and nonlinear dynamical systems, also has been verified its effectiveness in EEG-based emotion recognition \cite{guo2019hybrid}. As a whole, Fuzzy Logic is rarely studied in this research field.

\subsection{Route: \textbf{3\texorpdfstring{$\rightarrow$}.9\texorpdfstring{$\rightarrow$}.11}}
From the previous related works, we can see that adopting Deep Learning is the trend in EEG-based emotion recognition. Nevertheless, deep neural network models contain more parameters and rely on sufficient labeled training data to optimize the parameters compared with shallow models. Consequently, we must face one central challenge in EEG-based emotion recognition: acquiring adequate and high-quality training data. Hence, a promising research route is studying the \textbf{`data-constrained learning'} to address the data limitation problems. Here we introduce two potential ways.
\subsubsection{Data Augmentation Techniques}
Some researchers focus on studying data augmentation (DA) techniques. We can generate new samples from the existing dataset to increase the number of training samples. Exposing the model to more variable representations of training samples makes it robust to data transformations that are likely to encounter in real applications. Furthermore, increasing the size of the training set facilitates training more complex models with many parameters and reduces over-fitting. Data augmentation is typically conducted in computer vision by applying geometric transformations, e.g., rotation, cropping, etc. Nevertheless, the EEG is non-stationary time series. The geometric transformation methods are not suitable for EEG. One naive way is adding random noise (e.g., Gaussian, Poisson, Salt, Pepper noise, etc.) to the raw EEG signals \cite{li2016emotion,wang2018data}. Sliding window-based approaches are also adopted for data augmentation. However, these approaches may introduce modeling and performance evaluation risks that we discuss carefully in Section 5.3. Deep Generative Learning-based data augmentation methods are recently drawing widespread attention, including the generative adversarial network (GAN) based approaches and the variational autoencoder (VAE) based approaches \cite{luo2020data}. For example, during the training process of the adversarial network, the generator tries to generate data that are similar to the real data until the discriminator can not distinguish the fake data. More related works about data augmentation for EEG can be found in the review paper published by \citet{lashgari2020data}.
\subsubsection{Few-shot Learning Techniques}
\begin{figure*}[!htbp]
\centering
\includegraphics[width=0.6\textwidth]{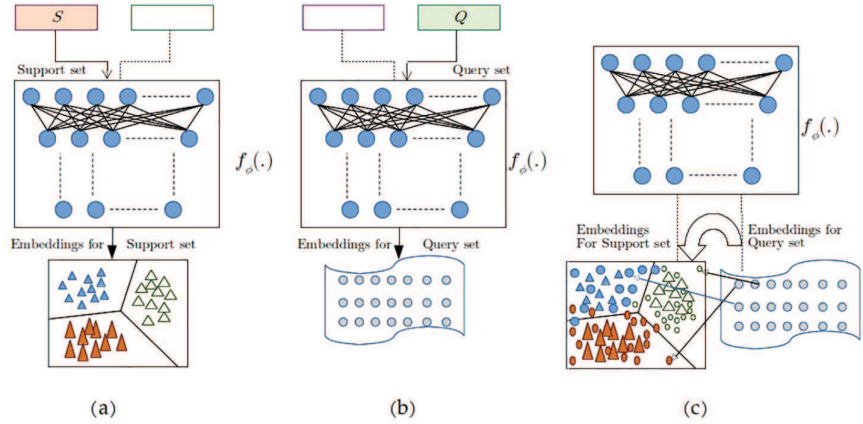}\\
\caption{General few-shot learning paradigm: (a) generating embeddings for the support set S, (b) generating embeddings for the query set Q, (c) Mapping query set embeddings using support sample\protect\footnotemark[15] \cite{bhosale2022calibration}.}
\label{fig:few-shot}
\end{figure*}

Few-shot learning also is potential for dealing with the data limitation problems and has been studied in recent related works \cite{bhosale2022calibration}. Few-shot learning is a class of machine learning techniques that build effective models that generalize well on classes unseen during the training process. It works well with limited samples and does not rely on re-training on the data belonging to the new classes. The few-shot learning can also be called a $N$-shot-$K$-way learning problem. Most few-shot learning techniques rely on metric learning. As shown in Figure \ref{fig:few-shot}, we need to construct the Support set and Query set, respectively. For the Support set, we sample $N$ samples for each of the $K$ classes. Further, we again sample the $K$ classes to construct the Query set. An embedding function is needed to project these samples to a latent space, in which the model is optimized to reduce the distance between the embeddings of query and support samples belonging to the same class while increasing the distance of the samples belonging to different categories. The Support and Query set construction process and optimization process iterate several times. After iteration, for testing, we only need a few samples belonging to the unseen class to form a support set, whereas the samples of the query set are the target to be inferred.

\footnotetext[15]{Reprinted from [Bhosale S, Chakraborty R, Kopparapu S K. Calibration free meta learning based approach for subject independent EEG emotion recognition. Biomedical Signal Processing and Control, 2022, 72: 103289] with the permission of Elsevier Publishing.}

\subsection{Route: \textbf{3\texorpdfstring{$\rightarrow$}.10\texorpdfstring{$\rightarrow$}.11} and \textbf{5\texorpdfstring{$\rightarrow$}.10\texorpdfstring{$\rightarrow$}.11}}
The processes involved in EEG-based emotion recognition studies are somewhat tedious. The domain knowledge hidden in this task is far beyond the machine learning specialists' knowledge range. Is there a way to automatically build robust recognition models on raw EEG data? In this regard, Automated Machine Learning (AutoML) is drawing attention in this domain. For AI-based approaches, an effective model is primarily decided by the model hyperparameters and the data representations. AutoML refers to end-to-end methodologies and tools for automatic optimization of data preprocessing, feature engineering, model selection, model building, and hyperparameter optimization \cite{he2021automl}. It aims to generate the models that provide the best classification performance and minimize the generalization error for a specific problem. Currently, a few researchers have started to introduce the AutoML techniques into EEG-based emotion recognition. For example, \citet {he2020strengthen} proposed one firefly integrated optimization algorithm (FIOA) to simultaneously realize the automatic parameter optimization, feature selection, and classifier selection. For Deep Learning-based technical routes, \citet{aquino2021optimization} proposed a fully-configurable optimization framework based on multi-objective optimization for Deep Learning architectures. It is not only capable of optimizing the model hyperparameters, but it can also adapt the model architecture, e.g., inserting or removing layers. At present, the main problem it encounters is that the computation burden is high. For example, the neural architecture search algorithm NASNet that was proposed by Google takes 28 days of training on 800 GPUs. Such high computational costs make search algorithms impractical for most researchers. It is encouraged that researchers are devoted to reducing the cost of AutoML training. We believe introducing AutoML techniques into various EEG modeling tasks will be very promising in the future.

\section{Performance Evaluation}
\subsection{Benchmark dataset}
The proposed recognition algorithms and models should be verified on EEG data with emotional ratings or labels. However, it is impossible for some researchers, especially those in computer science, to build a professional experimental environment and design a scientific user experimental paradigm that needs specialized knowledge of psychology. Most researchers interested in studying recognition models choose to verify their ideas and compare with related works on the recognized benchmark dataset. Hence, developing open-source EEG dataset that can help evaluate recognition models' performance is something the field urgently needs and well worth studying.

\begin{table}[!htbp]
\tiny
\caption{List of Recognized Benchmark Emotional EEG Dataset}
\label{table_dataset}
\begin{tabular}{p{0.5cm}p{1.5cm}p{2cm}p{1cm}p{1.5cm}p{5cm}p{2.5cm}}
\toprule
\textbf{Year} & \textbf{Data Set} & \textbf{Participants} & \textbf{EEG Device} & \textbf{Stimulus} & \textbf{Acquired Data Modalities} & \textbf{Quantification of Emotion} \\
\midrule
    2011 & DEAP \cite{koelstra2012deap} & 32 (16 male, 16 female) & Biosemi ActiveTwo & 40 video clips & 32-channel EEG, Peripheral physiological signals (Galvanic Skin Response, Blood Volume Pulse, respiration, Skin temperature, Electromyography, Electro-Oculogram), Face video & continuous type (Arousal, Valence, Dominance, Liking)\\
    2012 & MAHNOB-HCI \cite{soleymani2011multimodal} & 27 (11 male, 16 female) & Biosemi ActiveTwo & 20 video clips & 32-channel EEG, Peripheral physiological signals (Galvanic Skin Response, Respiration, Skin temperature, Electrocardiograph), Face and body video, Eye-tracking data, Audio & discrete type (9 types), continuous type (Arousal, Valence, Dominance)\\
    2017 & SEED \cite{zheng2017identifying} & 15 (7 male, 8 female) & ESI NeuroScan & 15 video clips & 62-channel EEG, Peripheral physiological signals (Electromyography, Electro-Oculogram), Face video & discrete type (Arousal-negative, Arousal-neutral, Arousal-positive)\\
    2018 & DREAMER \cite{2017DREAMER} & 25 (14 male, 11 female) & Emotiv EPOC & 18 video clips & 14-channel EEG, Peripheral physiological signals (Electrocardiogram) & continuous type (Arousal, Valence, Dominance)\\
    2019 & SEED-IV \cite{zheng2018emotionmeter} & 15 (7 male, 8 female) & ESI NeuroScan & 72 video clips & 62-channel EEG,  Eye-tracking data & discrete type (happy, sad, fear, and neutral) \\
    2019 & MPED \cite{song2019mped} & 23 (10 male, 13 female) & ESI NeuroScan & 28 video clips & 62-channel EEG, Peripheral physiological signals (Electrocardiogram, Respiration, Galvanic Skin Response & discrete type (joy, funny, anger, fear, disgust, sad and neurality) \\
\bottomrule
\end{tabular}
\end{table}

\small
Among them, the \textbf{D}ataset for \textbf{E}motion Analysis using \textbf{E}EG, \textbf{P}hysiological and video signals (\textbf{DEAP}) is mostly used and cited, which was collected and opened by researchers from Queen Mary University of London, the University of Geneva in Switzerland, etc. \cite{koelstra2012deap}. Thirty-two participants were recruited for the emotional EEG induction experiment. The EEG and several kinds of peripheral physiological signals were acquired while watching forty 60s long music movie clips. Then the subjective emotional experience in induction experiments was self-evaluated and rated on assessment scales that cover multiple emotional dimensions, including Arousal, Valence, Like, Dominance and Familiarity. The ratings are taken as the emotional ratings and labels of the EEG samples for model optimization. Another well-recognized benchmark dataset is the \textbf{MAHNOB-HCI} multi-modal dataset. It not only records the physiological and eye-
tracking activities of participants during the emotion induction experiments, but also the videos (face and body) and the audios are also synchronously recorded. This dataset is developed for emotion detection and implicit tagging studies \cite{soleymani2011multimodal}. \textbf{S}JTU \textbf{E}motion \textbf{E}EG \textbf{D}ataset (\textbf{SEED}) \cite{zheng2017identifying} also has a great community influence that was released by the BCMI Research Center in Shanghai Jiaotong University (SJTU). In the experiment, 15 movie clips with three types of emotional states were adopted to induce the specific emotions. Each genre had five clips and each for about 4 minutes. Regarding the influence of cultural background of language on emotional stimulation effect, only Chinese movies are selected in SEED for Chinese subjects. 15 Chinese participants participated in the EEG acquisition experiments for three different periods. This experimental design allows evaluating the algorithm's robustness when the data are acquired in different periods. It is believed that the EEG patterns and the subjective emotional experience may be unstable across different periods, which is a kind of 'domain shift'\protect\footnotemark[16] phenomenon in data science that brings great challenges in intelligent modeling \cite{ma2019reducing}. Recently, \textbf{DREAMER} is also adopted as a benchmark dataset for evaluating performance in some researches. It contains the EEG and electrocardiogram (ECG) signals that are simultaneously collected during the audio-visual emotion induction experiments. Twenty-three participants were recruited to participate in the experiment. After each trial, self-assessment in terms of Valence, Arousal, and Dominance is required. It is worth mentioning that all the signals were captured using portable, wearable, wireless and low-cost equipment that has the potential to verify recognition methods in everyday applications \cite{2017DREAMER}. \textbf{SEED-IV} is an extended version of the SEED data set, which is also released by SJTU \cite{zheng2018emotionmeter}, and also is well recognized in recent related works. It follows the experimental paradigm adopted in the SEED, and it records the 62-channel EEG from selected 15 subjects across three testing sessions. They choose 72 film clips with four different emotional labels (neutral, sad, fear, and happy). Each subject watched six film clips for each emotion class in each session, resulting in 24 trials. \textbf{M}ulti-modal \textbf{P}hysiological \textbf{E}motion \textbf{D}atabase (\textbf{MPED}) is one less well known and studied dataset that worth of mentioning here [219]. It is a multi-modal dataset that records the 62-channel EEG, the ECG, the respiration, and the galvanic skin response from 23 Chinese student volunteers. 28 Chinese video clips with seven types of discrete emotion (joy, funny, anger, fear, disgust, sadness, and neutrality) are selected from 1500 video clips, including film clips, TV News, and TV shows. The experiment is divided into two sessions with an interval of at least 24-hours. Each volunteer watched 14 video clips in each session, resulting in 14 trials.

\footnotetext[16]{We discuss the domain shift problem in EEG based emotion recognition in Section 4.5}

\subsection{Evaluation Metric}
\begin{figure*}[!htbp]
\centering
\includegraphics[width=0.3\textwidth]{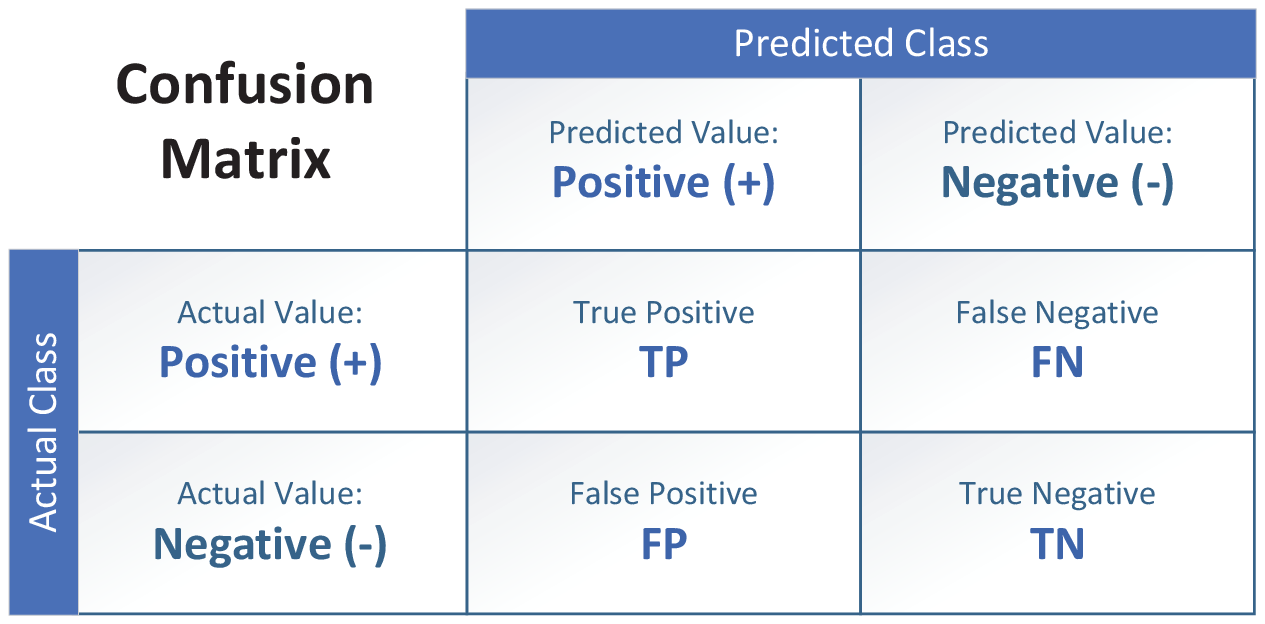}\\
\caption{Binary classification confusion matrix.}
\label{fig:metric}
\end{figure*}

All the proposed recognition methods should be evaluated on benchmark datasets by comparing the predicted emotion ratings/labels with the ground truth. For a classification modeling task, the built classifiers are typically assessed based on a confusion matrix-based approach, as shown in Figure \ref{fig:metric}, based on which four classification metrics can be derived for performance comparison, namely Precision, Sensitivity, Specificity, and F-score. Regarding the binary classification problem for the sake of simplicity, the four metrics are calculated as the following Formula \ref{eq8}. The TP, FP, TN, and FN are the abbreviations for True Positive, False Positive, True Negative, and False Negative, respectively. We usually adopt these metrics to measure the performance of a classification algorithm. True Positive is used to measure the number of actual positives (e.g., the emotion of happiness) which are correctly identified. Similarly, a True Negative is used to measure the number of actual negatives (e.g., the emotion of sadness) which are correctly identified. False Positive is the number of true negatives misclassified as positives. False Negative is the number of true positives incorrectly identified as the negatives.

\begin{scriptsize}
\begin{equation}
\label{eq8}
\left\{
\begin{aligned}
& Precision = \frac{TP}{TP+FP}, \\
& Sensitivity = \frac{TP}{TP+FN}, \\
& Specificity = \frac{TN}{TN+FP}, \\
& F_{score} = \frac{2*Precision*Sensitivity}{Precision+Sensitivity} \\
\end{aligned}
\right.
\end{equation}
\end{scriptsize}

For a regression modeling task, the built model is typically evaluated by computing the deviation between the predicted emotional rating and the subjects' reported rating. Researchers usually report the mean squared error (MSE) and the mean absolute error (MAE), as well as the coefficient of determination ($R^{2}$) \cite{koelstra2013fusion}, as in the Formula \ref{eq9}, where $y_i$ is the ground truth rating for a sample $i$, $\bar{y}$ is the mean ground truth rating of all samples and $\hat{y}_i$ is the rating as estimated by the regressor for sample i. $R^2$-score range from zero to one, a higher value indicates a higher consistency between the model prediction and the ground truth.

\begin{scriptsize}
\begin{equation}
\label{eq9}
\left\{
\begin{aligned}
& MSE = \sqrt{\frac{\sum_{i=1}^{n}(y_i-\hat{y}_i)^2}{n}}, \\
& MAE = \frac{1}{n}\sum_{i=1}^{n}(y_i-\hat{y}_i)^2, \\
& R^2 = 1-\frac{\sum_{i=1}^{n}(y_i-\bar{y}_i)^2}{\sum_{i=1}^{n}(y_i-\hat{y}_i)^2}\\
\end{aligned}
\right.
\end{equation}
\end{scriptsize}

\subsection{Data Split and Validation Strategy}
The strategy of constructing the training and test samples has not been thoroughly studied and discussed in related works. Hence, it may be difficult to reproduce some research results, which may be considered weird tricks and mislead some researchers. From our point of view, the data split and validation strategy should be designed based on the research objectives. The strategy guides to validate whether or not the research objectives can be achieved.

Hence, we will discuss this issue from the perspective of the research objectives. Generally, the research objectives in EEG-based emotion recognition can be divided into two main categories, namely subject-dependent modeling and subject-independent modeling. Subject-dependent modeling assumes that it is impossible to build one universal model to recognize each subject well. In this setting, the model training is conducted for each subject, and testing is performed on the same subject's data, which measures how well the recognition model performs on data with intra-subject variabilities. The researchers split the dataset in each subject scope when adopting this strategy. Nevertheless, there would be a large deviation in the distributions of EEGs collected from different subjects. Hence, subject-independent modeling and evaluation aim to build a universal and robust recognition model for various data domains. The training is performed on a group of subjects, and the evaluation is conducted on data from one or more unseen subjects. This evaluation strategy measures how well the built model handles the problem of inter-subject variabilities. 

Regardless of research objectives, they all can adopt the k-fold cross-validation scheme for overall evaluation that splits the whole dataset into k independent parts without overlap. Each time, one part of the entire data is selected out as the test set for evaluating the classifier, and the remaining parts of data are used as the training set for model building. This process is repeated k times. The final overall evaluation of the recognition method can be derived by averaging the results obtained on each fold.

\begin{figure*}[!htbp]
\centering
\includegraphics[width=0.6\textwidth]{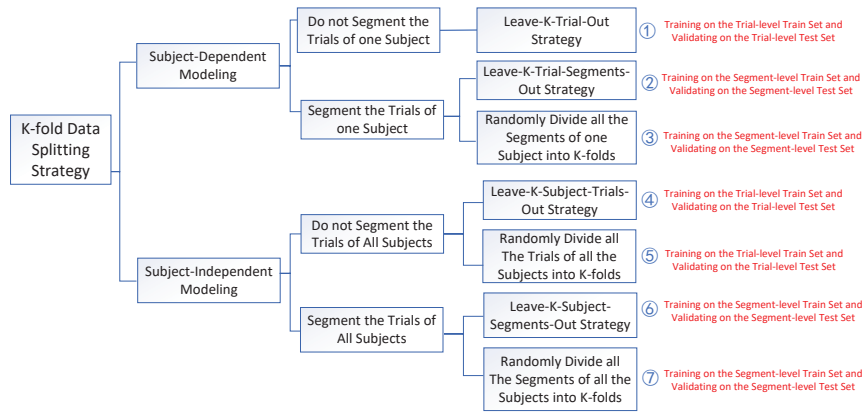}\\
\caption{The possible k-fold data split strategy adopted in related works}
\label{fig:k-fold}
\end{figure*}

We illustrate the possible splitting strategies for the two kinds of research objectives in Figure \ref{fig:k-fold}. Seven possible K-fold splitting strategies may be adopted in related works. Usually, we do not segment the original trial data both for the subject-dependent and subject-independent modeling and adopt the `leave-K-trial-out' and `leave-K-subject-trials-out' strategies, respectively. Hence, the 1st, 4th, and 5th splitting strategies are all reasonable and recommended, and they reflect the recognition performance in real-world settings. However, we find some related works proposed to segment the trials. For example, for a 60-second long trial, we can divide the entire trial into 15 segments with a 4-second long non-overlap sliding window. They build and validate the model on the segments, as shown in the 2nd, 3rd, 6th, and 7th strategies. The main reason behind it is the number of EEG trials of one subject is usually no more than 100 considering the factor of user fatigue, which is not enough for optimizing a complex recognition model, especially in the case of training deep neural networks. In these approaches, the segments that belong to the same trial are all assigned with the same trial label. From our point of view, these splitting approaches may be somewhat unsuitable, which has been illustrated in Figure 16, where the subjects' emotional state may be continuously evolving, and the label of the trial is the overall emotional experience of the subject during a trial. Hence, assigning the segments with the same label may introduce lots of noisy labels, greatly influencing the model training. These segment-based approaches only hold when the emotional stimuli are strong enough to consistently evoke the subject's target emotion. Nevertheless, we have to point out that the 3rd and 7th splitting strategies are wrong approaches that the researchers of this field should not consider. To be specific, if we randomly divide the segments into K folds, the segments from the same trial may be assigned to the training and test sets simultaneously, introducing information leakage from the training set to the test set. It will lead to the high recognition performance on the test set, as the segments of the same trial are somewhat correlated in the data distribution that the trained model has remembered in the training phase.

\begin{landscape}
\tiny
\begin{longtable}{p{1cm}p{2.5cm}p{2.5cm}p{8cm}p{6cm}}
\caption{List of relevant representative research works in recent 5 years (from 2016 to 2021 year)} \\
\toprule
    Year & Author & Experimental Data Set & Methodology & Experimental Results\\
\midrule
    2016 & \citet{atkinson2016improving} & DEAP & kernel classifiers combines feature selection method based on minimum-Redundancy-Maximum-Relevance (MRMR) & 2 classes, subject-independent modeling (Valence: 0.7314, Arousal: 0.7306)   \\
    2016 & \citet{thammasan2016application} & EEGs induced by MIDI audio materials & Sliding window based signal segmentation, modeling based on DBN and handcraft features of fractal dimension, power spectral and discrete wavelet transform & 2 classes, subject-dependent modeling (Valence: 0.8824, Arousal: 0.8242) \\
    2016 & \citet{li2016emotion} & DEAP & one CNN+RNN hybrid Deep Learning framework combines wavelet scalogram representation for multi-channel EEGs & 2 classes (Valence: 0.7206, Arousal: 0.7412) \\
    2016 & \citet{lan2016real} & EEGs induced by IADS audio materials & SVM classifier combines statistical features, fractal dimension, power spectral and higher-order crossing feature & 4 classes (pleasant/happy/frightened/angry) (cross-session: 24.97\%$\sim$ 49.63\%, intra-session: 46.56\%$\sim$81.41\%) \\
    2016 & \citet{ackermann2016eeg} & DEAP & handcraft features (statistics, STFT, HOC, hilbert-huang spectrum) combines with MRMR dimension reduction method and Random Forest/SVM classifiers & 3 classes (Anger/Surprise/Other): $\simeq$52\% \\
    2017 & \citet{tripathi2017using} & DEAP & 9 kinds of time-domain features combine with DNN and CNN models & 2 classes, subject-dependent modeling (CNN-Valence: 81.41\%, CNN-Arousal: 73.36\%) \\
    2017 & \citet{mohammadi2017wavelet} & DEAP & feature extraction by discrete wavelet transform and classify by KNN/SVM & 2 classes, subject-dependent modeling (Valence: 86.75\%, Arousal: 84.05\%) \\
    2017 & \citet{yin2017recognition} & DEAP & ensemble classifiers based on multiple stacked auto-encoders combine with features of statistics and PSD & 2 classes, subject-dependent modeling (Valence: 0.7243, Arousal: 0.6901) \\
    2017 & \citet{chen2017subject} & DEAP & a three-stage decision framework based on partioning the subjects into different groups & 2 classes (Valence: 0.4357, Arousal: 0.7757) \\
    2018 & \citet{liu2018real} & EEGs induced by movie clips & power spectral feature extraction by STFT combines with LDA based feature selection and SVM based classifier & positive/negative emotion: 86.63\%, intra positive emotion(amusement/joy/tenderness): 86.43\%, intra negative emotion(anger/disgust/fear/sadness): 65.09\% \\
    2018 & \citet{li2018hierarchical} & SEED & 2D map of differential entropy feature combines with hierarchical convolutional neural network & 2 classes (subject dependent: 0.882, subject independent: 0.802) \\
    2018 & \citet{katsigiannis2018dreamer} & DREAMER & artifacts removal by ASR and CAR method, PSD features extraction combines with SVM classifier & 2 classes (Valence: 62.49\%, Arousal: 62.17\%) \\
    2018 & \citet{Li2018Exploring} & DEAP, SEED & 18 kinds of handcrafted EEG features combine with multiple feature selection methods and SVM classifier & 2 classes, subject-independent modeling (DEAP-Valence: 0.7167, DEAP-Arousal: 0.7154, SEED: 0.8234) \\
    2018 & \citet{mert2018emotion} & DEAP & multivariate empirical mode decomposition based multiple features extraction combines with ICA based feature dimension reduction and ANN classifier & 2 classes (ANN-Valence: 0.7287, ANN-Arousal: 0.7500) \\
    2018 & \citet{wang2018emotionet} & DEAP & an emotional EEG-specific 3D-CNN (EmotioNet) using a simultaneous temporal-spatial feature detection, the input EEGs are arranged according to the topological structure & 2 classes, subject-independent modeling (Valence: 72.1\%, Arousal: 73.1\%) \\
    2018 & \citet{salama2018eeg} & DEAP & a 3D-CNN is employed for extracting the spatiotemporal features, the input EEGs are randomly arranged and one data augmentation scheme is adopted & 2 classes, subject-dependent modeling (Valence: 87.44\%, Arousal: 88.49\%) \\
    2018 & \citet{zhang2018cascade,yang2018emotion} & DEAP & cascade and parallel hybrid models integrate CNN and RNN, in which the input is the raw EEG signal arranged according to electrode topology & 2 classes, subject-dependent modeling(Valence: 90.80\%, Arousal: 91.03\%) \\
    2018 & \citet{lan2018domain} & DEAP, SEED & transfer learning methods (MIDA, TCA) combines with differential entropy feature and logistic regression classifier & 3 classes, subject-independent modeling (DEAP-MIDA: 0.4893, DEAP-TCA: 0.4722, SEED-MIDA: 0.7247, SEED-TCA: 0.7167) \\
    2018 & \citet{li2018novel} & SEED & a bi-hemispheres domain adversarial neural network (BiDANN) that takes into the cerebral hemisphere asymmetry information as well as the domain adaptation & 3 classes (subject-dependent modeling: 92.38\%, subject-independent modeling: 83.28\%) \\
    2018 & \citet{luo2018wgan} & DEAP, SEED & Wasserstein generative adversarial network domain adaptation (WGANDA) framework based transfer learning combines with differential entropy feature & 2 classes, subject-independent modeling (DEAP-Valence: 0.6799, DEAP-Arousal: 0.6685, SEED: 0.8707) \\
    2018 & \citet{li2018cross} & SEED & a deep adaptation network (DAN) based on adversarial network is proposed to solve the cross-subject recognition problem & 2 classes, subject-independent modeling: 0.8381 \\
    2018 & \citet{yang2018recurrence}  & EEGs induced by movie clips & a channel-frequency convolutional neural network (CFCNN) combined with recurrence quantification analysis (RQA) & 3 classes (happiness/sadness/fear), subject-dependent modeling: 92.24\% \\
    2018 & \citet{song2018eeg} & SEED, DREAMER & a dynamical graph convolutional neural network dynamically learns graph adjacency weight matrix and classifies emotions based on EEG features of DE, PSD, DASM, RASM and DCAU & subject-dependent modeling (SEED: 90.4\%, DREAMER-Valence: 86.23\%, DREAMER-Arousal: 84.54\%), subject-independent modeling (SEED: 79.95\%) \\
    2019 & \citet{li2019eeg} & SEED & time-domain and frequency-domain EEG features combine with graph regularized sparse linear regression (GRSLR) model & 3 classes, subject-independent modeling: 0.8841 \\
    2019 & \citet{ma2019emotion} & DEAP & a multimodal residual LSTM classifier based on raw EEG and physiological signal & 2 classes, subject-dependent modeling (DEAP-Valence: 92.30\%, DEAP-Arousal: 92.87\%) \\
    2019 & \citet{zhang2019spatial} & SEED & adopting multi-directional SRNN layer and bidirectional TRNN layer to learn spatiotemporal dependencies in layers  & 3 classes, subject-independent modeling: 89.50\% \\
    2019 & \citet{li2019regional} & SEED & a hierarchical spatio-temporal neural network model based on LSTM and attention mechanism that acquires the intrinsic spatial relationship and time dependencies & 3 classes, subject-independent modeling: 93.38\% \\
    2019 & \citet{zhang2019spatial} & SEED & a Riemannian fusion network to learn the most discriminative and complementary spatial and temporal information & 3 classes (subject-dependent modeling: 93.72\%, subject-independent modeling: 84.16\%) \\
    2019 & \citet{guo2019hybrid} & DEAP & fuzzy cognitive maps (FCM) combines with SVM & 2 classes, subject-dependent modeling: 73.32\% \\
    2020 & \citet{li2020latent} & DEAP, SEED & a raw EEG decoding approach based on stacked AE (non-generative model), RBM (generative model), and VAE (generative model), then further utilized the LSTM for sequence learning  & 2 classes, subject-independent modeling (DEAP-Valence: 0.7167, DEAP-Arousal: 0.7243, SEED: 0.8429) \\
    2020 & \citet{cho2020spatio} & DEAP & 2 types of 3D-CNN models (C3D, R(2 + 1)D) are verified, the input raw EEGs are arranged into 2D frames according to 1electrode topology, the interpolated 2D EEG frames are further concatenated into 3D cubes  & 2 classes, subject-independent modeling (DEAP-Valence: 99.11\%, DEAP-Arousal: 99.74\%) \\
    2020 & \citet{chao2019emotion} & DEAP & a framework based on multi-band feature matrix (MFM) and capsule network (CapsNet) & 2 classes (Valence: 0.6673, Arousal: 0.6828, Dominance: 0.6725) \\
    2020 & \citet{zhong2020eeg} & SEED & a regularized GNN model (RGNN) that adopts two kinds of regularizers: NodeDAT and EmotionDL & 3 classes, subject-dependent modeling: 94.24\%, subject-independent modeling: 85.30\% \\
    2020 & \citet{zhang2020rfnet} & SEED & an end-to-end Riemannian fusion network (RFNet) that captures spatial information from the Riemannian manifold and temporal information from the Euclidean space & 3 classes, subject-dependent modeling: 0.9372 \\
    2020 & \citet{ding2020tsception} & EEGs induced by VR scenes & a temporal-spacial 1D convolution based model inspired by the Inception block of GoogleNet & Arousal-2 classes, subject-dependent-session-independent modeling: 86.03\% \\
    2020 & \citet{zhang2020variational} & SEED, MPED & a heuristic variational pathway reasoning (VPR) method for mining salient connections information embedded in the multi-channel EEGs that is conducive to recognition & 3 classes, subject-dependent modeling (SEED: 94.30\%, MPED: 75.06\%) \\
    2020 & \citet{luo2020eeg} & DEAP, SEED & a spiking neural network (SNN) based framework & subject-independent modeling (DEAP-Valence: 78\%, DEAP-Arousal: 74\%, SEED: 96.67\%) \\
    2020 & \citet{li2020novel} & SEED & a bi-hemispheric discrepancy model(BiHDM) learns the asymmetric characteristics between hemispheres & 3 classes (subject-dependent modeling: 93.12\%, subject-independent modeling: 85.40\%) \\
    2020 & \citet{cui2020eeg} & DEAP, DREAMER & an end-to-end regional-asymmetric CNN (RACNN) consists of temporal (1D-CNN), regional (2D-CNN) and asymmetric feature extractors (ADL) & 2 classes, subject-dependent modeling (DEAP-Valence: 96.65\%, DEAP-Arousal: 97.11\%, DREAMER-Valence: 95.55\%, DREAMER-Arousal: 97.01\%) \\
    2020 & \citet{cimtay2020investigating} & SEED, DEAP, LUMED & a pretrained state-of-the-art CNN model InceptionResnetV2 & subject-independent modeling (SEED-2 classes: 86.56\%, SEED-3 classes: 78.34\%, DEAP-2 classes: 72.81, LUMED-2 classes: 81.8\%) \\
    2020 & \citet{duan2020ultra} & DEAP & a transfer learning method named Meta Update Strategy (MUPS-EEG) that involves a meta representation learning phase followed by meta adaptation to target subject & Arousal-2 classes, subject-independent modelling: 66.5\% \\
    2020 & \citet{duan2020meta} & SEED, DEAP & a generalized optimization-based meta learning approach under constrained transfer learning setting that trains on large standard datasets of known subjects and then quickly adapt to a new subject with minimal target data. & subject-independent modelling (DEAP-Arousal-2 classes: 67.5\%, SEED-3 classes: 78.6\%) \\
    2020 & \citet{wei2020eeg} & SEED & an emotion recognition system based on Simple Recurrent Units (SRU) network, Dual-tree Complex Wavelet Transform (DT-CWT), differential entropy (DE), and ensemble learning & subject-dependent modelling (SEED-3 classes: 83.13\%) \\
    2020 & \citet{wang2020emotion} & SEED, DEAP &  an electrode-frequency distribution maps (EFDMs) with short-time Fourier transform (STFT) , a Residual block-based deep convolutional neural network, and deep model transfer & subject-independent modelling (SEED-3 classes: 90.59\%, DEAP-Valence-3 classes: 82.84\%) \\
    2020 & \citet{tao2020eeg} & DEAP, DREAMER & an attention-based convolutional recurrent neural network (ACRNN) with channel-wise attention and an extended self-attention to extract discriminative spatiotemporal information & 3 classes, subject-dependent modelling (DEAP-Valence: 93.72\%, DEAP-Arousal:93.38\%, DREAMER-Valence: 97.93\%, DREAMER-Arousal: 97.98\%, DREAMER-dominance: 98.23\%) \\
    2020 & \citet{jia2020sst} & SEED, SEED-IV & a novel spatial-spectral-temporal based attention 3D dense network (SST-EmotionNet) works with 3D differential entropy representations & SEED-3 classes: 96.02\%, SEED-IV-4 classes: 84.92\% \\
    2020  & \citet{luo2020data} & SEED, DEAP & a method based on two deep generative models, variational autoencoder(VAE) and generative adversarial network (GAN), and two data augmentation strategies & subject-dependent modelling (SEED-3 classes: 93.5\%, DEAP-4 classes: 50.8\%) \\
    2020 & \citet{cheng2020emotion} & DEAP, DREAMER &  a multi-grained cascade Forest model (gcForest) with input of 2D frame representation & subject-dependent modelling (DEAP-Valence-2 classes: 97.69\%, DEAP-Arousal-2 classes: 97.53\%, DREAMER-Valence-3 classes: 89.03\%, DREAMER-Arousal-3 classes: 90.41\%, DREAMER-Dominance-3 classes: 89.89\%) \\
    2020 & \citet{liu2020multi} & DEAP, DREAMER & an effective multi-level features guided capsule network (MLF-CapsNet) & 2 classes, subject-dependent
    modelling (DEAP-Valence: 97.97\%, DEAP-Arousal: 98.31\%, DEAP-Dominance: 98.32\%, DREAMER-Valence: 94.59\%, DREAMER-Arousal: 95.26\%, DREAMER-Dominance: 95.13\%) \\
    2020 & \citet{he2020strengthen} & LabEdata, DEAP & a novel firefly integrated optimization algorithm that automatically optimize feature selection, parameter setting and classifier selection & 2 classes, subject-dependent modelling (LabEdata: 95\%, DEAP: 92\%) \\
    2021 & \citet{wang2021deep} & SEED & the multi-source selection is employed to obtain the source domain mostly correlated with new subject, the deep multi-source adaptation transfer network (DMATN) model is used to map correlated source domain and the target domain (new subject) into reproducing kernel Hilbert space & 3 classes, subject-independent modelling: 84.23\% \\
    2021 & \citet{huang2021differences} & DEAP & a bi-hemisphere discrepancy convolutional neural network model (BiDCNN) built on 3 kinds of feature matrices & 2 classes, subject-dependent modeling (Valence: 94.38\%, Arousal: 94.72\%), subject-independent modeling (Valence: 68.14\%, Arousal: 63.94\%)\\
    2021 & \citet{yin2021eeg} & DEAP & a GCNN-LSTM hybrid model, named ECGGCNN & 2 classes, subject-dependent modeling (Valence: 90.45\%, Arousal: 90.60\%), subject-independent modeling (Valence: 84.81\%, Arousal: 85.27\%) \\
    2021 & \citet{chen2021emotion} & DEAP & Adaboost based ensemble learning method & 2 classes, subject-independent modeling (Valence: 85.57\%, Arousal: 88.36\%) \\
    2021 & \citet{zhang2021distilling} & SEED & a LSTM-Capsule based knowledge distillation framework & 3 classes, subject-independent modeling: 91.07\% \\
    2021 & \citet{zhao2021plug} & SEED & a novel plug-and-play domain adaptation (PPDA) method that subject-invariant representations and private components of source subjects are separately captured by a shared encoder and private encoders & subject-independent modelling (SEED-3 classes: 86.7\%) \\
    2021 & \citet{fernandez2021cross} & SEED & a CNN-based network with differential Entropy input and three hidden layers each followed by stratified normalization method & subject-independent modelling (SEED-2 classes: 91.6\%, SEED-3 classes: 79.6\%) \\
    2021 & \citet{cai2021cross} & SEED & a domain adaptation method, named the maximum classifier difference for domain adversarial neural networks (MCD\_DA) &
    3 classes, subject-independent modelling: 88.33\% \\
    2021 & \citet{islam2021eeg} & DEAP & a deep Convolutional Neural Network (CNN) with Pearson’s Correlation Coefficient (PCC) featured images of channel correlation of EEG sub-bands & subject-dependent modelling (valence-2 classes: 78.22\%, arousal-2 classes: 74.92\%, valence-3 classes: 70.23\%, arousal-3 classes: 70.25\%) \\
    2021 & \citet{yin2020locally} & DEAP, MAHNOB-HCI &  a new locally-robust feature selection (LRFS) method cooperate with SVM and ensemble learning & 2 classes, subject-independent modeling (DEAP-Arousal: 65\%, DEAP-Valence: 68\%, MAHNOB-HCI-Arousal: 67\%, MAHNOB-HCI-Valence: 70\%) \\
    2021 & \citet{ding2021eeg} & SEED & a Task-specific Domain Adversarial Neural Network (T-DANN) that transfers knowledge from either one subject to predict on another subject or knowledge from one phase to predict on another phase within the same subject & 3 classes, subject-independent modelling (cross-subject: 74.19\%, cross-phase: 85.13\%) \\
    2021 & \citet{wang2021cross} & SEED, DEAP & a novel method called `few-label adversarial domain adaption' (FLADA) that works on a small target data & subject-independent modelling (DEAP-2 classes:68.0\%, SEED-3 classes: 89.32\%) \\
    2021 & \citet{liu2021domain} & DEAP & an extended domain adaptation method by introducing subject clustering (DASC) & 2 classes, subject-independent modelling (DEAP-Valence: 73.9\%, DEAP-Arousal: 68.8\%) \\
    2022 & \citet{bhosale2022calibration} &  DEAP & a few-shot adaptation method based on meta learning without requiring any fine-tuning of the pre-trained model & 2 classes, subject-independent modelling (Valence: 76.46\%, Arousal: 75.81\%, Dominance: 70.63\%) \\
    
\bottomrule
\end{longtable}
\end{landscape}

\section{Conclusion and Discussion}
We choose to outline the review from the perspective of researchers who try to take the first step on this topic. Hence, we review not only the overall current situation in the EEG based emotion recognition research but also provide a tutorial to guide the researchers to start from a very beginning, as well as illustrate the theoretical basis and the research motivation, which will help the readers to understand why those techniques are employed. For this prospect, we introduce the preliminaries and basic knowledge of this field. Firstly, we present the definition and quantification methods of emotion. It is the prerequisite for affective computing, and it determines the objective of the modeling tasks (regression, clustering, classification). Then we illustrate the specificities and neural correlates of EEG in the emotional process, and we demonstrate the feasibility of EEG in studying the emotion recognition technologies. Before reviewing the technical routes, we also exhibit the classical research methodologies for EEG-based emotion detection studies, which helps the readers understand the goal of this field and the mainstream methodologies in the past quickly. The section of `Preliminaries and Basic Knowledge' guides the readers to understand the following sections' contents better.

Then, we devote much effort to guiding the newcomers of the EEG preprocessing and the feature engineering methods, which is the basis for most classical methodologies. The remaining parts of this paper mainly focus on the pattern recognition technical routes applied in the field, in which we summarize the mainstream and latest technical routes involved in this field and review plenty of representative works under each route. Finally, we discuss the evaluation methods adopted in this field. In addition to the benchmark datasets, we discuss the potential influence of different data split strategies in modeling and validation, which is a topic that many researchers care about. Considering the rapid development of Deep Learning and its successful application in this field, we select to review as many Deep Learning-based approaches as possible, and the selected works are within the scope of the recent three years.  We tend to summarize representative works in this field and conduct empirical comparisons for closely related approaches from a descriptive perspective. We try to list these works in a structured table (Table 4), which presents the methodologies, the validation strategies, and the achieved performance in a direct way.


Though there have been many achievements in this field, there still exists several problems and challenges need to be further studied and resolved, as follows:

\begin{itemize}[leftmargin=*]
\item There is still room for research to explore effective EEG representation (transformation) approaches, which rely on EEG preprocessing and feature engineering. EEG preprocessing makes the emotion-related information (components)  effectively filtered out from the multi-channel EEGs that contain redundant noisy components. Feature engineering helps to determine the critical variables related to emotions. Current widely utilized features cover various aspects, such as time-frequency characteristics, nonlinear dynamical system characteristics, etc. The feature extraction process may incur a high overhead and depends on subtle parameter settings, especially in nonlinear dynamical system feature computation. Nevertheless, the extracted features at a high cost contain many redundant and irrelevant variables that contribute slightly to the performance improvement. For example, \citet{Li2018Exploring} explored a variety of EEG features in cross-subject emotion recognition, and experimental results indicate that only one or two key features lead to comparative performance to that obtained on the whole feature set. Hence, research on the critical EEG features and variables is still worth conducting. The decided scope of EEG features and variables helps reduce the computation cost in EEG representation, meanwhile, improve the recognition effect. Besides, the decided critical EEG channels help mitigate the difficulties in user experiment, e.g., only selecting to attach fewer electrodes on the cortex help to improve the user engagement and reduce the pre-experiment preparation. The critical EEG features and variables also could provide a new perspective to analyze the mechanism of the emotion cognition process.

Though several studies mentioned above have taken prior knowledge into the model design. Take the BiDHM, BiDANN, and BiDCNN shown in Figure \ref{fig:BiHDM} and Figure \ref{fig:BiDCNN} for example, the asymmetry correlation information between the bi-hemispheres is utilized, there still exists gaps between the true emotion process manifested by the EEG and the information processing modeled by the existing models. It will be better to integrate prior knowledge about the users' gender, age, physical condition, mental condition, and prior knowledge about the emotional stimuli, the semantic context, the environment, and the knowledge graph about emotion into the model design. Prior knowledge-guided recognition models must obtain an enhanced and robust performance on cross-domain learning and few-shot learning settings. Meanwhile, prior knowledge introduces interpretability to the constructed models and the obtained results.

\item Although abundant cutting-edge artificial intelligence models have been studied, developing computational methods for emotion recognition needs a deeper understanding of emotion processes and their neural basis. Psychophysiology-inspired, biology-inspired, and brain-inspired cognitive models based on the principle of how the human brain works in the emotion cognition process should also be taken into consideration for us. The popular Deep Learning models only are a less precise mathematical abstraction of the brain functions. They have limitations in online learning, small-sample learning, modeling the information interaction between different brain regions. The biologically inspired methods are built on the architecture of the neocortex and try to model the process of how the human brain handles complex information about vision, audio, behavior, and emotion. These biologically inspired methods (e.g., the hierarchical temporal memory model based on neocortex theory \cite{cui2017htm} and the spiking neural network) are intuitively suitable for modeling brain imaging data and other behavioral data controlled by the neocortex. For example, \citet{luo2020eeg} proposed one spiking neural network (SNN) based model, which makes full use of the spatiotemporal features of the EEG signal. As one kind of brain-inspired computing model, the SNN is able to encode the neural data through the synapses, neurons, and spiking activity. In addition, the Deep Learning model is perceived as one black-box that is hard to understand why they get specific decisions \cite{pouyanfar2018survey}. How to resolve the problem of the weak statistical interpretability should be taken into consideration for future works, e.g., through the `inceptions' techniques adopted in Google Brain \cite{mordvintsev2015inceptionism}, and the model-agnostic explanation approach (LIME) \cite{ribeiro2016model}.

\item Several works mentioned above face the problem of lengthy signal modeling. The EEG signal acquired in one trial with a high sampling rate will be extremely long. As a result, the model computation burden will be largely increased, especially for RNN based models. Hence, researchers should not only focus on the metric of recognition accuracy but also the computational efficiency should also be reported. At present, the researches mainly focus on the offline data processing scene. Therefore, the algorithm suitable for real-time emotion recognition and monitoring should be extensively explored.

\item The `domain shift' problem and `transfer learning' will still be the hot research topics in the next few years. Here we summarize several potential research approaches in these topics. Currently, researchers mainly focus on studying these topics within one single dataset setting, which we can call the intra-dataset-inter-subject modeling problem, in which the user difference in cultures, ages, gender, and physiology will degrade the performance. Nevertheless, developing domain adaptation techniques in inter-dataset-inter-subject settings is a more challenging task that deserves careful exploration. As we know, only a few open-source datasets about emotion recognition are available nowadays. If more EEG datasets uncorrelated with emotion recognition can be simultaneously employed, a robust model with more complex structure and more parameters could be trained. Considering the recent advances in the large-scale pre-trained model in natural language processing (e.g., GPT, BERT) \cite{qiu2020pre}, we believe it is possible to develop the EEG-oriented pre-trained model on large-scale open-source EEG datasets, which are not limited to emotion recognition. Nevertheless, the EEG data of multi-sources is acquired with different devices, different experiment designs, different types of stimuli, etc. These factors could further increase the discrepancies caused by inter-subject/session variability. Hence, research on 'modeling on multiple source domains' is another direction worthy of further exploration. In addition, current research on the domain shift problem mainly focuses on domain adaptation of the extracted EEG features. There are lots of room for performance enhancement on the cross-subject, cross-dataset, or cross-session emotion recognition tasks if the raw EEG signals could be calibrated and aligned to common data space in advance before adopting the traditional transfer learning approaches. Researchers who are interested in this research topic can refer to the raw EEG alignment methods based on Riemannian geometry that have been adopted in the motor imagery BCI \cite{zanini2017transfer}.

\item How effective those models perform in an open environment is still unknown. As in real-world scenarios, the people are continuously in a dynamic condition that they may seldomly be calm, which is quite different from the controlled experimental environment. Realtime EEG signals are inevitably influenced by continually evolving activities, including physical factors such as body movements and environmental noise and psychological factors such as mental workloads and attention. The robustness of the emotion recognition system will be affected when people are executing psychological or physical activities. It raises a great challenge to develop recognition models that can capture robust and distinctive emotion-related features from real-time and dynamic EEGs that generalize well under various people states. Researchers should devote themselves to developing open-environment EEG datasets, the corresponding algorithms, and the evaluation criteria.

\item The reviewed works in this paper were designed on a single EEG modality. In recent years, increasing published articles manifest a shift of research interest from unimodal to multi-modal information-based emotion recognition tasks, in which the multi-modal approaches fuse two or more modalities for emotion recognition. This shift is based on some problems that unimodal systems mainly faced. Firstly, the unimodal data may be missing or inconsecutive for some reason, e.g., the monitored signal may be blocked or affected by external obstacles, noise, or device instability. In such a circumstance, the data from other modalities complement the single modality properly. Secondly, the exterior behavioral information sometimes may not be consistent with the actual affective state. An individual may conceal their real feelings under social masks. For example, the same facial expressions may represent different psychological activities, so that the single data modality may be insufficient for an accurate recognition task. Last but not least, the recognition performance may be promoted when multi-modal information is fused and utilized, and this is also the ultimate goal of multi modalities-based approaches. For example, \citet{ma2019emotion} developed one LSTM based multi-modal recognition framework that successfully learned the joint information from the original EEG and physiological data, and thus significantly improving the recognition effect of the DEAP dataset. Researchers should take as many data modalities as possible into emotion recognition studies, including the EEG, the facial expression, the gesture, the gait, the peripheral physiological signal, the eye movement, etc., to build a comprehensive recognition model. We will devote ourselves to reviewing relevant multi-modal fusion studies in future work.
\end{itemize}

\begin{acks}
This work was supported in part by the Major Science and Technology Innovation Projects of Key R\&D Programs of Shandong Province (grant No. 2019JZZY010108 and grant No.2019JZZY010113), the Natural Science Foundation of China (grant No. U1636203 and grant No. 62006212), the Natural Science Foundation of Shandong Province (`Research on Cross-domain Emotion Recognition Based on Large-scale Pre-trained EEG Model'), the Natural Science Foundation of China (`Research on pre-trained EEG model for EEG-based cross-domain emotion recognition task'), the fund of State Key Lab. for Novel Software Technology in Nanjing University (grant No. KFKT2021B41), and the Industrial Science and Technology Research Project of Henan Province (grant No. 222102210031). This work was also supported by the Academy of Finland (grants 336033, 315896), Business Finland (grant 884/31/2018), and EU H2020 (grant 101016775).
\end{acks}

\bibliographystyle{unsrtnat}

\begin{spacing}{0.2} 
    \bibliography{references}

\begin{thebibliography}{231}
\providecommand{\natexlab}[1]{#1}
\providecommand{\url}[1]{\texttt{#1}}
\expandafter\ifx\csname urlstyle\endcsname\relax
  \providecommand{\doi}[1]{doi: #1}\else
  \providecommand{\doi}{doi: \begingroup \urlstyle{rm}\Url}\fi

\bibitem[Picard(2000)]{picard2000affective}
Rosalind~W Picard.
\newblock \emph{Affective computing}.
\newblock MIT press, 2000.

\bibitem[Picard(2001)]{picard2001building}
Rosalind~W Picard.
\newblock Building hal: Computers that sense, recognize, and respond to human
  emotion.
\newblock In \emph{Human Vision and Electronic Imaging VI}, volume 4299, pages
  518--523. International Society for Optics and Photonics, 2001.

\bibitem[Dunne et~al.(2021)Dunne, Morris, and Harper]{dunne2021survey}
Rob Dunne, Tim Morris, and Simon Harper.
\newblock A survey of ambient intelligence.
\newblock \emph{ACM Computing Surveys (CSUR)}, 54\penalty0 (4):\penalty0 1--27,
  2021.

\bibitem[Adolphs and Anderson(2018)]{adolphs2018neuroscience}
Ralph Adolphs and David~J Anderson.
\newblock \emph{The neuroscience of emotion: A new synthesis}.
\newblock Princeton University Press, 2018.

\bibitem[Cannon(1927)]{cannon1927james}
Walter~B Cannon.
\newblock The james-lange theory of emotions: A critical examination and an
  alternative theory.
\newblock \emph{The American journal of psychology}, 39\penalty0
  (1/4):\penalty0 106--124, 1927.

\bibitem[Halim and Rehan(2020)]{halim2020identification}
Zahid Halim and Mahma Rehan.
\newblock On identification of driving-induced stress using
  electroencephalogram signals: A framework based on wearable safety-critical
  scheme and machine learning.
\newblock \emph{Information Fusion}, 53:\penalty0 66--79, 2020.

\bibitem[Rozgic et~al.(2014)Rozgic, Vazquez-Reina, Crystal, Srivastava, Tan,
  and Berka]{rozgic2014multi}
Viktor Rozgic, Amelio Vazquez-Reina, Michael Crystal, Amit Srivastava, Veasna
  Tan, and Chris Berka.
\newblock Multi-modal prediction of ptsd and stress indicators.
\newblock In \emph{2014 IEEE International Conference on Acoustics, Speech and
  Signal Processing (ICASSP)}, pages 3636--3640. IEEE, 2014.

\bibitem[Valstar et~al.(2013)Valstar, Schuller, Smith, Eyben, Jiang, Bilakhia,
  Schnieder, Cowie, and Pantic]{valstar2013avec}
Michel Valstar, Bj{\"o}rn Schuller, Kirsty Smith, Florian Eyben, Bihan Jiang,
  Sanjay Bilakhia, Sebastian Schnieder, Roddy Cowie, and Maja Pantic.
\newblock Avec 2013: the continuous audio/visual emotion and depression
  recognition challenge.
\newblock In \emph{Proceedings of the 3rd ACM international workshop on
  Audio/visual emotion challenge}, pages 3--10, 2013.

\bibitem[Cai et~al.(2020)Cai, Qu, Li, Zhang, Hu, and Hu]{cai2020feature}
Hanshu Cai, Zhidiao Qu, Zhe Li, Yi~Zhang, Xiping Hu, and Bin Hu.
\newblock Feature-level fusion approaches based on multimodal eeg data for
  depression recognition.
\newblock \emph{Information Fusion}, 59:\penalty0 127--138, 2020.

\bibitem[Beck(1979)]{beck1979cognitive}
Aaron~T Beck.
\newblock \emph{Cognitive therapy and the emotional disorders}.
\newblock Penguin, 1979.

\bibitem[G{\"u}m{\"u}sl{\"u} et~al.(2020)G{\"u}m{\"u}sl{\"u}, Erol~Barkana, and
  K{\"o}se]{gumuslu2020emotion}
Elif G{\"u}m{\"u}sl{\"u}, Duygun Erol~Barkana, and Hatice K{\"o}se.
\newblock Emotion recognition using eeg and physiological data for
  robot-assisted rehabilitation systems.
\newblock In \emph{Companion Publication of the 2020 International Conference
  on Multimodal Interaction}, pages 379--387, 2020.

\bibitem[Chang et~al.(2017)Chang, Huang, and Wu]{chang2017personalized}
Hong-Yi Chang, Shih-Chang Huang, and Jia-Hao Wu.
\newblock A personalized music recommendation system based on
  electroencephalography feedback.
\newblock \emph{Multimedia Tools and Applications}, 76\penalty0 (19):\penalty0
  19523--19542, 2017.

\bibitem[Ramirez et~al.(2018)Ramirez, Planas, Escude, Mercade, and
  Farriols]{ramirez2018eeg}
Rafael Ramirez, Josep Planas, Nuria Escude, Jordi Mercade, and Cristina
  Farriols.
\newblock Eeg-based analysis of the emotional effect of music therapy on
  palliative care cancer patients.
\newblock \emph{Frontiers in psychology}, 9:\penalty0 254, 2018.

\bibitem[Soleymani et~al.(2011)Soleymani, Lichtenauer, Pun, and
  Pantic]{soleymani2011multimodal}
Mohammad Soleymani, Jeroen Lichtenauer, Thierry Pun, and Maja Pantic.
\newblock A multimodal database for affect recognition and implicit tagging.
\newblock \emph{IEEE transactions on affective computing}, 3\penalty0
  (1):\penalty0 42--55, 2011.

\bibitem[Moshfeghi and Jose(2013)]{moshfeghi2013effective}
Yashar Moshfeghi and Joemon~M Jose.
\newblock An effective implicit relevance feedback technique using affective,
  physiological and behavioural features.
\newblock In \emph{Proceedings of the 36th international ACM SIGIR conference
  on Research and development in information retrieval}, pages 133--142, 2013.

\bibitem[Koelstra and Patras(2013)]{koelstra2013fusion}
Sander Koelstra and Ioannis Patras.
\newblock Fusion of facial expressions and eeg for implicit affective tagging.
\newblock \emph{Image and Vision Computing}, 31\penalty0 (2):\penalty0
  164--174, 2013.

\bibitem[Arapakis et~al.(2009)Arapakis, Moshfeghi, Joho, Ren, Hannah, and
  Jose]{arapakis2009enriching}
Ioannis Arapakis, Yashar Moshfeghi, Hideo Joho, Reede Ren, David Hannah, and
  Joemon~M Jose.
\newblock Enriching user profiling with affective features for the improvement
  of a multimodal recommender system.
\newblock In \emph{Proceedings of the ACM international conference on image and
  video retrieval}, pages 1--8, 2009.

\bibitem[Moshfeghi(2012)]{moshfeghi2012role}
Yashar Moshfeghi.
\newblock \emph{Role of emotion in information retrieval}.
\newblock PhD thesis, University of Glasgow, 2012.

\bibitem[Yadollahi et~al.(2017)Yadollahi, Shahraki, and
  Zaiane]{yadollahi2017current}
Ali Yadollahi, Ameneh~Gholipour Shahraki, and Osmar~R Zaiane.
\newblock Current state of text sentiment analysis from opinion to emotion
  mining.
\newblock \emph{ACM Computing Surveys (CSUR)}, 50\penalty0 (2):\penalty0 1--33,
  2017.

\bibitem[Alakus et~al.(2020)Alakus, Gonen, and Turkoglu]{alakus2020database}
Talha~Burak Alakus, Murat Gonen, and Ibrahim Turkoglu.
\newblock Database for an emotion recognition system based on eeg signals and
  various computer games--gameemo.
\newblock \emph{Biomedical Signal Processing and Control}, 60:\penalty0 101951,
  2020.

\bibitem[Stavroulia et~al.(2019)Stavroulia, Christofi, Baka, Michael-Grigoriou,
  Magnenat-Thalmann, and Lanitis]{stavroulia2019assessing}
Kalliopi~Evangelia Stavroulia, Maria Christofi, Evangelia Baka, Despina
  Michael-Grigoriou, Nadia Magnenat-Thalmann, and Andreas Lanitis.
\newblock Assessing the emotional impact of virtual reality-based teacher
  training.
\newblock \emph{The International Journal of Information and Learning
  Technology}, 2019.

\bibitem[Vesisenaho et~al.(2019)Vesisenaho, Juntunen, Johanna, Fagerlund,
  Miakush, Parviainen, et~al.]{vesisenaho2019virtual}
Mikko Vesisenaho, Merja Juntunen, P~Johanna, Janne Fagerlund, Iryna Miakush,
  Tiina Parviainen, et~al.
\newblock Virtual reality in education: Focus on the role of emotions and
  physiological reactivity.
\newblock \emph{Journal For Virtual Worlds Research}, 12\penalty0 (1), 2019.

\bibitem[Lotte et~al.(2018)Lotte, Bougrain, Cichocki, Clerc, Congedo,
  Rakotomamonjy, and Yger]{lotte2018review}
Fabien Lotte, Laurent Bougrain, Andrzej Cichocki, Maureen Clerc, Marco Congedo,
  Alain Rakotomamonjy, and Florian Yger.
\newblock A review of classification algorithms for eeg-based brain--computer
  interfaces: a 10 year update.
\newblock \emph{Journal of neural engineering}, 15\penalty0 (3):\penalty0
  031005, 2018.

\bibitem[Alarcao and Fonseca(2017)]{alarcao2017emotions}
Soraia~M Alarcao and Manuel~J Fonseca.
\newblock Emotions recognition using eeg signals: A survey.
\newblock \emph{IEEE Transactions on Affective Computing}, 10\penalty0
  (3):\penalty0 374--393, 2017.

\bibitem[Dadebayev et~al.(2021)Dadebayev, Goh, and Tan]{dadebayev2021eeg}
Didar Dadebayev, Wei~Wei Goh, and Ee~Xion Tan.
\newblock Eeg-based emotion recognition: Review of commercial eeg devices and
  machine learning techniques.
\newblock \emph{Journal of King Saud University-Computer and Information
  Sciences}, 2021.

\bibitem[Suhaimi et~al.(2020)Suhaimi, Mountstephens, and Teo]{suhaimi2020eeg}
Nazmi~Sofian Suhaimi, James Mountstephens, and Jason Teo.
\newblock Eeg-based emotion recognition: A state-of-the-art review of current
  trends and opportunities.
\newblock \emph{Computational intelligence and neuroscience}, 2020, 2020.

\bibitem[Arya et~al.(2021)Arya, Kumar, and Bhushan]{arya2021affect}
Resham Arya, Ashok Kumar, and Megha Bhushan.
\newblock Affect recognition using brain signals: A survey.
\newblock In \emph{Computational Methods and Data Engineering}, pages 529--552.
  Springer, 2021.

\bibitem[Rahman et~al.(2021)Rahman, Sarkar, Hossain, Hossain, Islam, Hossain,
  Quinn, and Moni]{rahman2021recognition}
Md~Mustafizur Rahman, Ajay~Krishno Sarkar, Md~Amzad Hossain, Md~Selim Hossain,
  Md~Rabiul Islam, Md~Biplob Hossain, Julian~MW Quinn, and Mohammad~Ali Moni.
\newblock Recognition of human emotions using eeg signals: A review.
\newblock \emph{Computers in Biology and Medicine}, 136:\penalty0 104696, 2021.

\bibitem[Craik et~al.(2019)Craik, He, and Contreras-Vidal]{craik2019deep}
Alexander Craik, Yongtian He, and Jose~L Contreras-Vidal.
\newblock Deep learning for electroencephalogram (eeg) classification tasks: a
  review.
\newblock \emph{Journal of neural engineering}, 16\penalty0 (3):\penalty0
  031001, 2019.

\bibitem[Ekman(2009)]{ekman2009darwin}
Paul Ekman.
\newblock Darwin's contributions to our understanding of emotional expressions.
\newblock \emph{Philosophical Transactions of the Royal Society B: Biological
  Sciences}, 364\penalty0 (1535):\penalty0 3449--3451, 2009.

\bibitem[Ekman(1999)]{ekman1999basic}
Paul Ekman.
\newblock Basic emotions.
\newblock In Tim Dalgleish and Mick Power, editors, \emph{Handbook of Cognition
  and Emotion}, pages 45--60. Hoboken, United States: John Wiley \& Sons Inc.,
  1999.

\bibitem[Plutchik(2003)]{plutchik2003emotions}
Robert Plutchik.
\newblock \emph{Emotions and Life: Perspectives from Psychology, Biology, and
  Evolution}.
\newblock Washington, United States: American Psychological Association, 2003.

\bibitem[Plutchik and Kellerman(2013)]{plutchik2013theories}
Robert Plutchik and Henry Kellerman.
\newblock \emph{Theories of emotion}, volume~1.
\newblock Academic Press, 2013.

\bibitem[Parrott and Gerrod(2001)]{Parrott2001Emotions}
Parrott and W.~Gerrod.
\newblock \emph{Emotions in Social Psychology: Essential Readings}.
\newblock Oxfordshire, United Kingdom: Taylor \& Francis Group, 2001.

\bibitem[Russell(1979)]{Russell1979Affective}
James~A Russell.
\newblock Affective space is bipolar.
\newblock \emph{Journal of Personality and Social Psychology}, 37\penalty0
  (3):\penalty0 345--356, 1979.

\bibitem[Langeslag(2018)]{langeslag2018effects}
Sandra~JE Langeslag.
\newblock Effects of organization and disorganization on pleasantness,
  calmness, and the frontal negativity in the event-related potential.
\newblock \emph{PloS one}, 13\penalty0 (8):\penalty0 e0202726, 2018.

\bibitem[Schmidt and Trainor(2001)]{schmidt2001frontal}
Louis~A Schmidt and Laurel~J Trainor.
\newblock Frontal brain electrical activity (eeg) distinguishes valence and
  intensity of musical emotions.
\newblock \emph{Cognition \& Emotion}, 15\penalty0 (4):\penalty0 487--500,
  2001.

\bibitem[Adolphs et~al.(1994)Adolphs, Tranel, Damasio, and
  Damasio]{adolphs1994impaired}
Ralph Adolphs, Daniel Tranel, Hanna Damasio, and Antonio Damasio.
\newblock Impaired recognition of emotion in facial expressions following
  bilateral damage to the human amygdala.
\newblock \emph{Nature}, 372\penalty0 (6507):\penalty0 669--672, 1994.

\bibitem[Lench et~al.(2011)Lench, Flores, and Bench]{lench2011discrete}
Heather~C Lench, Sarah~A Flores, and Shane~W Bench.
\newblock Discrete emotions predict changes in cognition, judgment, experience,
  behavior, and physiology: a meta-analysis of experimental emotion
  elicitations.
\newblock \emph{Psychological bulletin}, 137\penalty0 (5):\penalty0 834, 2011.

\bibitem[Lindquist et~al.(2012)Lindquist, Wager, Kober, Bliss-Moreau, and
  Barrett]{lindquist2012brain}
Kristen~A Lindquist, Tor~D Wager, Hedy Kober, Eliza Bliss-Moreau, and
  Lisa~Feldman Barrett.
\newblock The brain basis of emotion: a meta-analytic review.
\newblock \emph{The Behavioral and brain sciences}, 35\penalty0 (3):\penalty0
  121, 2012.

\bibitem[LeDoux(2012)]{ledoux2012rethinking}
Joseph LeDoux.
\newblock Rethinking the emotional brain.
\newblock \emph{Neuron}, 73\penalty0 (4):\penalty0 653--676, 2012.

\bibitem[Britton et~al.(2006)Britton, Phan, Taylor, Welsh, Berridge, and
  Liberzon]{britton2006neural}
Jennifer~C Britton, K~Luan Phan, Stephan~F Taylor, Robert~C Welsh, Kent~C
  Berridge, and Israel Liberzon.
\newblock Neural correlates of social and nonsocial emotions: An fmri study.
\newblock \emph{Neuroimage}, 31\penalty0 (1):\penalty0 397--409, 2006.

\bibitem[Gordon et~al.(1996)Gordon, Hart~Jr, Lesser, and
  Arroyo]{gordon1996mapping}
Barry Gordon, John Hart~Jr, Ronald~P Lesser, and Santiago Arroyo.
\newblock Mapping cerebral sites for emotion and emotional expression with
  direct cortical electrical stimulation and seizure discharges.
\newblock \emph{Progress in brain research}, 107:\penalty0 617--622, 1996.

\bibitem[P{\'e}ron et~al.(2010)P{\'e}ron, Biseul, Leray, Vicente, Le~Jeune,
  Drapier, Drapier, Sauleau, Haegelen, and V{\'e}rin]{peron2010subthalamic}
J~P{\'e}ron, I~Biseul, E~Leray, S~Vicente, F~Le~Jeune, S~Drapier, D~Drapier,
  P~Sauleau, C~Haegelen, and M~V{\'e}rin.
\newblock Subthalamic nucleus stimulation affects fear and sadness recognition
  in parkinson's disease.
\newblock \emph{Neuropsychology}, 24\penalty0 (1):\penalty0 1--8, 2010.

\bibitem[Satow et~al.(2003)Satow, Usui, Matsuhashi, Yamamoto, Begum, Shibasaki,
  Ikeda, Mikuni, Miyamoto, and Hashimoto]{satow2003mirth}
T~Satow, K~Usui, M~Matsuhashi, J~Yamamoto, T~Begum, Hiroshi Shibasaki, A~Ikeda,
  N~Mikuni, S~Miyamoto, and N~Hashimoto.
\newblock Mirth and laughter arising from human temporal cortex.
\newblock \emph{Journal of Neurology, Neurosurgery \& Psychiatry}, 74\penalty0
  (7):\penalty0 1004--1005, 2003.

\bibitem[Vaca et~al.(2011)Vaca, L{\"u}ders, Basha, and Miller]{vaca2011mirth}
Guadalupe Fern{\'a}ndez-Baca Vaca, Hans~O L{\"u}ders, Maysaa~Merhi Basha, and
  Jonathan~P Miller.
\newblock Mirth and laughter elicited during brain stimulation.
\newblock \emph{Epileptic Disorders}, 13\penalty0 (4):\penalty0 435--440, 2011.

\bibitem[Davidson(1992)]{Davidson1992Anterior}
R.~J. Davidson.
\newblock Anterior cerebral asymmetry and the nature of emotion.
\newblock \emph{Brain and Cognition}, 20\penalty0 (1):\penalty0 125--151, 1992.

\bibitem[Davidson(2004)]{Davidson2004What}
R.~J. Davidson.
\newblock What does the prefrontal cortex "do" in affect: perspectives on
  frontal eeg asymmetry research.
\newblock \emph{Biological Psychology}, 67\penalty0 (1):\penalty0 219--234,
  2004.

\bibitem[Koelsch(2014)]{koelsch2014brain}
Stefan Koelsch.
\newblock Brain correlates of music-evoked emotions.
\newblock \emph{Nature Reviews Neuroscience}, 15\penalty0 (3):\penalty0
  170--180, 2014.

\bibitem[Palmiero and Piccardi(2017)]{palmiero2017frontal}
Massimiliano Palmiero and Laura Piccardi.
\newblock Frontal eeg asymmetry of mood: A mini-review.
\newblock \emph{Frontiers in Behavioral Neuroscience}, 11:\penalty0 224, 2017.

\bibitem[Sammler et~al.(2007)Sammler, Grigutsch, Fritz, and
  Koelsch]{sammler2007music}
Daniela Sammler, Maren Grigutsch, Thomas Fritz, and Stefan Koelsch.
\newblock Music and emotion: electrophysiological correlates of the processing
  of pleasant and unpleasant music.
\newblock \emph{Psychophysiology}, 44\penalty0 (2):\penalty0 293--304, 2007.

\bibitem[Daly et~al.(2014)Daly, Malik, Hwang, Roesch, Weaver, Kirke, Williams,
  Miranda, and Nasuto]{daly2014neural}
Ian Daly, Asad Malik, Faustina Hwang, Etienne Roesch, James Weaver, Alexis
  Kirke, Duncan Williams, Eduardo Miranda, and Slawomir~J Nasuto.
\newblock Neural correlates of emotional responses to music: an eeg study.
\newblock \emph{Neuroscience letters}, 573:\penalty0 52--57, 2014.

\bibitem[Aftanas et~al.(2006)Aftanas, Reva, Savotina, and
  Makhnev]{aftanas2006neurophysiological}
LI~Aftanas, NV~Reva, LN~Savotina, and VP~Makhnev.
\newblock Neurophysiological correlates of induced discrete emotions in humans:
  an individually oriented analysis.
\newblock \emph{Neuroscience and Behavioral Physiology}, 36\penalty0
  (2):\penalty0 119--130, 2006.

\bibitem[Koelstra et~al.(2012)Koelstra, Muhl, Soleymani, Lee, Yazdani,
  Ebrahimi, Pun, Nijholt, and Patras]{koelstra2012deap}
Sander Koelstra, Christian Muhl, Mohammad Soleymani, Jong-Seok Lee, Ashkan
  Yazdani, Touradj Ebrahimi, Thierry Pun, Anton Nijholt, and Ioannis Patras.
\newblock Deap: A database for emotion analysis; using physiological signals.
\newblock \emph{IEEE Transactions on Affective Computing}, 3\penalty0
  (1):\penalty0 18--31, 2012.

\bibitem[Balconi and Lucchiari(2008)]{balconi2008consciousness}
Michela Balconi and Claudio Lucchiari.
\newblock Consciousness and arousal effects on emotional face processing as
  revealed by brain oscillations. a gamma band analysis.
\newblock \emph{International Journal of Psychophysiology}, 67\penalty0
  (1):\penalty0 41--46, 2008.

\bibitem[Yang et~al.(2020)Yang, Tong, Shu, Zhuang, Yan, and Zeng]{yang2020high}
Kai Yang, Li~Tong, Jun Shu, Ning Zhuang, Bin Yan, and Ying Zeng.
\newblock High gamma band eeg closely related to emotion: evidence from
  functional network.
\newblock \emph{Frontiers in human neuroscience}, 14:\penalty0 89, 2020.

\bibitem[Hagemann et~al.(1999)Hagemann, Naumann, L{\"u}rken, Becker, Maier, and
  Bartussek]{hagemann1999eeg}
Dirk Hagemann, Ewald Naumann, Alexander L{\"u}rken, Gabriele Becker, Stefanie
  Maier, and Dieter Bartussek.
\newblock Eeg asymmetry, dispositional mood and personality.
\newblock \emph{Personality and Individual Differences}, 27\penalty0
  (3):\penalty0 541--568, 1999.

\bibitem[Coan and Allen(2004)]{coan2004frontal}
James~A Coan and John~JB Allen.
\newblock Frontal eeg asymmetry as a moderator and mediator of emotion.
\newblock \emph{Biological Psychology}, 67\penalty0 (1-2):\penalty0 7--50,
  2004.

\bibitem[Li and Lu(2009)]{li2009emotion}
Mu~Li and Bao-Liang Lu.
\newblock Emotion classification based on gamma-band eeg.
\newblock In \emph{2009 Annual International Conference of the IEEE Engineering
  in medicine and biology society}, pages 1223--1226. IEEE, 2009.

\bibitem[Zheng and Lu(2015)]{zheng2015investigating}
Weilong Zheng and Baoliang Lu.
\newblock Investigating critical frequency bands and channels for eeg-based
  emotion recognition with deep neural networks.
\newblock \emph{IEEE Transactions on Autonomous Mental Development}, 7\penalty0
  (3):\penalty0 162--175, 2015.

\bibitem[Li et~al.(2018{\natexlab{a}})Li, Zhang, and He]{li2018hierarchical}
Jinpeng Li, Zhaoxiang Zhang, and Huiguang He.
\newblock Hierarchical convolutional neural networks for eeg-based emotion
  recognition.
\newblock \emph{Cognitive Computation}, pages 1--13, 2018{\natexlab{a}}.

\bibitem[Li et~al.(2018{\natexlab{b}})Li, Song, Zhang, Zhang, Hou, and
  Hu]{Li2018Exploring}
Xiang. Li, Dawei. Song, Peng. Zhang, Yazhou. Zhang, Yuexian. Hou, and Bin. Hu.
\newblock Exploring eeg features in cross-subject emotion recognition.
\newblock \emph{Frontiers in Neuroscience}, 12:\penalty0 162,
  2018{\natexlab{b}}.

\bibitem[Naser and Saha(2021)]{naser2021influence}
Daimi~Syed Naser and Goutam Saha.
\newblock Influence of music liking on eeg based emotion recognition.
\newblock \emph{Biomedical Signal Processing and Control}, 64:\penalty0 102251,
  2021.

\bibitem[Zhuang et~al.(2018)Zhuang, Zeng, Yang, Zhang, Tong, and
  Yan]{zhuang2018investigating}
Ning Zhuang, Ying Zeng, Kai Yang, Chi Zhang, Li~Tong, and Bin Yan.
\newblock Investigating patterns for self-induced emotion recognition from eeg
  signals.
\newblock \emph{Sensors}, 18\penalty0 (3):\penalty0 841, 2018.

\bibitem[Constantinescu et~al.(2017)Constantinescu, Wolters, Moore, and
  MacPherson]{constantinescu2017cluster}
Alexandra~C Constantinescu, Maria Wolters, Adam Moore, and Sarah~E MacPherson.
\newblock A cluster-based approach to selecting representative stimuli from the
  international affective picture system (iaps) database.
\newblock \emph{Behavior research methods}, 49\penalty0 (3):\penalty0 896--912,
  2017.

\bibitem[Greco et~al.(2016)Greco, Valenza, Citi, and
  Scilingo]{greco2016arousal}
Alberto Greco, Gaetano Valenza, Luca Citi, and Enzo~Pasquale Scilingo.
\newblock Arousal and valence recognition of affective sounds based on
  electrodermal activity.
\newblock \emph{IEEE Sensors Journal}, 17\penalty0 (3):\penalty0 716--725,
  2016.

\bibitem[Zhang et~al.(2017)Zhang, Shu, Xu, and Liao]{zhang2017affective}
Wenzhuo Zhang, Lin Shu, Xiangmin Xu, and Dan Liao.
\newblock Affective virtual reality system (avrs): design and ratings of
  affective vr scenes.
\newblock In \emph{2017 International Conference on Virtual Reality and
  Visualization (ICVRV)}, pages 311--314. IEEE, 2017.

\bibitem[Katsigiannis and Ramzan(2018)]{katsigiannis2018dreamer}
Stamos Katsigiannis and Naeem Ramzan.
\newblock Dreamer: A database for emotion recognition through eeg and ecg
  signals from wireless low-cost off-the-shelf devices.
\newblock \emph{IEEE Journal of Biomedical and Health Informatics}, 22\penalty0
  (1):\penalty0 98--107, 2018.

\bibitem[Petrantonakis and Hadjileontiadis(2010)]{petrantonakis2010emotion}
Panagiotis~C Petrantonakis and Leontios~J Hadjileontiadis.
\newblock Emotion recognition from brain signals using hybrid adaptive
  filtering and higher order crossings analysis.
\newblock \emph{IEEE Transactions on Affective Computing}, 1\penalty0
  (2):\penalty0 81--97, 2010.

\bibitem[Lan et~al.(2016)Lan, Sourina, Wang, and Liu]{lan2016real}
Zirui Lan, Olga Sourina, Lipo Wang, and Yisi Liu.
\newblock Real-time eeg-based emotion monitoring using stable features.
\newblock \emph{The Visual Computer}, 32\penalty0 (3):\penalty0 347--358, 2016.

\bibitem[Olofsson et~al.(2008)Olofsson, Nordin, Sequeira, and
  Polich]{olofsson2008affective}
Jonas~K Olofsson, Steven Nordin, Henrique Sequeira, and John Polich.
\newblock Affective picture processing: an integrative review of erp findings.
\newblock \emph{Biological psychology}, 77\penalty0 (3):\penalty0 247--265,
  2008.

\bibitem[Bernat et~al.(2001)Bernat, Bunce, and Shevrin]{bernat2001event}
Edward Bernat, Scott Bunce, and Howard Shevrin.
\newblock Event-related brain potentials differentiate positive and negative
  mood adjectives during both supraliminal and subliminal visual processing.
\newblock \emph{International Journal of Psychophysiology}, 42\penalty0
  (1):\penalty0 11--34, 2001.

\bibitem[Frantzidis et~al.(2010)Frantzidis, Bratsas, Papadelis, Konstantinidis,
  Pappas, and Bamidis]{frantzidis2010toward}
Christos~A Frantzidis, Charalampos Bratsas, Christos~L Papadelis, Evdokimos
  Konstantinidis, Costas Pappas, and Panagiotis~D Bamidis.
\newblock Toward emotion aware computing: an integrated approach using
  multichannel neurophysiological recordings and affective visual stimuli.
\newblock \emph{IEEE Transactions on Information Technology in Biomedicine},
  14\penalty0 (3):\penalty0 589--597, 2010.

\bibitem[Wang et~al.(2014)Wang, Nie, and Lu]{wang2014emotional}
Xiaowei Wang, Dan Nie, and Baoliang Lu.
\newblock Emotional state classification from eeg data using machine learning
  approach.
\newblock \emph{Neurocomputing}, 129:\penalty0 94--106, 2014.

\bibitem[Lin et~al.(2009)Lin, Wang, Wu, Jeng, and Chen]{Lin2009EEG}
Yuan-Pin Lin, Chi-Hong Wang, Tien-Lin Wu, Shyh-Kang Jeng, and Jyh-Horng Chen.
\newblock Eeg-based emotion recognition in music listening: A comparison of
  schemes for multiclass support vector machine.
\newblock In \emph{Proceedings of the 2009 IEEE international conference on
  acoustics, speech and signal processing}, pages 489--492. IEEE, 2009.

\bibitem[Liu et~al.(2018)Liu, Yu, Zhao, Song, Ge, and Shi]{liu2018real}
Yong-Jin Liu, Minjing Yu, Guozhen Zhao, Jinjing Song, Yan Ge, and Yuanchun Shi.
\newblock Real-time movie-induced discrete emotion recognition from eeg
  signals.
\newblock \emph{IEEE Transactions on Affective Computing}, 9\penalty0
  (4):\penalty0 550--562, 2018.

\bibitem[Sorkhabi(2014)]{sorkhabi2014emotion}
Majid~Memarian Sorkhabi.
\newblock Emotion detection from eeg signals with continuous wavelet analyzing.
\newblock \emph{American Journal of Computing Research Repository}, 2\penalty0
  (4):\penalty0 66--70, 2014.

\bibitem[Mohammadi et~al.(2017)Mohammadi, Frounchi, and
  Amiri]{mohammadi2017wavelet}
Zeynab Mohammadi, Javad Frounchi, and Mahmood Amiri.
\newblock Wavelet-based emotion recognition system using eeg signal.
\newblock \emph{Neural Computing and Applications}, 28\penalty0 (8):\penalty0
  1985--1990, 2017.

\bibitem[Subasi(2005)]{subasi2005automatic}
A.~Subasi.
\newblock Automatic recognition of alertness level from eeg by using neural
  network and wavelet coefficients.
\newblock \emph{Expert systems with applications}, 28\penalty0 (4):\penalty0
  701--711, 2005.

\bibitem[Gandhi et~al.(2011)Gandhi, Panigrahi, and
  Anand]{gandhi2011comparative}
T.~Gandhi, B.~K. Panigrahi, and S.~Anand.
\newblock A comparative study of wavelet families for eeg signal
  classification.
\newblock \emph{Neurocomputing}, 74\penalty0 (17):\penalty0 3051--3057, 2011.

\bibitem[Mert and Akan(2018)]{mert2018emotion}
Ahmet Mert and Aydin Akan.
\newblock Emotion recognition from eeg signals by using multivariate empirical
  mode decomposition.
\newblock \emph{Pattern Analysis and Applications}, 21\penalty0 (1):\penalty0
  81--89, 2018.

\bibitem[Stam(2005)]{Stam2005Nonlinear}
C~J Stam.
\newblock Nonlinear dynamical analysis of eeg and meg.
\newblock \emph{Clinical Neurophysiology}, 116\penalty0 (10):\penalty0
  2266--2301, 2005.

\bibitem[Kulish et~al.(2006)Kulish, Sourin, and Sourina]{Kulish2006Human}
Vladimir Kulish, Alexei Sourin, and Olga Sourina.
\newblock Human electroencephalograms seen as fractal time series: Mathematical
  analysis and visualization.
\newblock \emph{Computers in Biology and Medicine}, 36\penalty0 (3):\penalty0
  291--302, 2006.

\bibitem[Wang et~al.(2011)Wang, Sourina, and Nguyen]{wang2011fractal}
Qiang Wang, Olga Sourina, and Minh~Khoa Nguyen.
\newblock Fractal dimension based neurofeedback in serious games.
\newblock \emph{The Visual Computer}, 27\penalty0 (4):\penalty0 299--309, 2011.

\bibitem[Liu et~al.(2010)Liu, Sourina, and Nguyen]{Liu2010Real}
Yisi Liu, Olga Sourina, and Minh~Khoa Nguyen.
\newblock Real-time eeg-based human emotion recognition and visualization.
\newblock In \emph{Proceedings of the 2010 International Conference on
  Cyberworlds}, pages 262--269. IEEE, 2010.

\bibitem[Hosseinifard et~al.(2013)Hosseinifard, Moradi, and
  Rostami]{hosseinifard2013classifying}
Behshad Hosseinifard, Mohammad~Hassan Moradi, and Reza Rostami.
\newblock Classifying depression patients and normal subjects using machine
  learning techniques and nonlinear features from eeg signal.
\newblock \emph{Computer Methods and Programs in Biomedicine}, 109\penalty0
  (3):\penalty0 339--345, 2013.

\bibitem[Liu and Sourina(2014)]{Liu2014EEG}
Yisi Liu and Olga Sourina.
\newblock Eeg-based subject-dependent emotion recognition algorithm using
  fractal dimension.
\newblock In \emph{Proceedings of the 2014 IEEE International Conference on
  Systems, Man, and Cybernetics}, pages 3166--3171. IEEE, 2014.

\bibitem[Eckmann et~al.(1995)Eckmann, Kamphorst, Ruelle,
  et~al.]{eckmann1995recurrence}
Jean-Pierre Eckmann, S~Oliffson Kamphorst, David Ruelle, et~al.
\newblock Recurrence plots of dynamical systems.
\newblock \emph{World Scientific Series on Nonlinear Science Series A},
  16:\penalty0 441--446, 1995.

\bibitem[Yu et~al.(2016)Yu, Li, Song, Zhao, Zhang, Hou, and Hu]{yu2016encoding}
Guangliang Yu, Xiang Li, Dawei Song, Xiaozhao Zhao, Peng Zhang, Yuexian Hou,
  and Bin Hu.
\newblock Encoding physiological signals as images for affective state
  recognition using convolutional neural networks.
\newblock In \emph{2016 38th Annual International Conference of the IEEE
  Engineering in Medicine and Biology Society (EMBC)}, pages 812--815. IEEE,
  2016.

\bibitem[Yang et~al.(2018{\natexlab{a}})Yang, Gao, Wang, Li, Han, Marwan, and
  Kurths]{yang2018recurrence}
Yu-Xuan Yang, Zhong-Ke Gao, Xin-Min Wang, Yan-Li Li, Jing-Wei Han, Norbert
  Marwan, and J{\"u}rgen Kurths.
\newblock A recurrence quantification analysis-based channel-frequency
  convolutional neural network for emotion recognition from eeg.
\newblock \emph{Chaos: An Interdisciplinary Journal of Nonlinear Science},
  28\penalty0 (8):\penalty0 085724, 2018{\natexlab{a}}.

\bibitem[Shi et~al.(2013)Shi, Jiao, and Lu]{shi2013differential}
Li-Chen Shi, Ying-Ying Jiao, and Bao-Liang Lu.
\newblock Differential entropy feature for eeg-based vigilance estimation.
\newblock In \emph{2013 35th Annual International Conference of the IEEE
  Engineering in Medicine and Biology Society (EMBC)}, pages 6627--6630. IEEE,
  2013.

\bibitem[Garc{\'\i}a-Mart{\'\i}nez et~al.(2019)Garc{\'\i}a-Mart{\'\i}nez,
  Martinez-Rodrigo, Alcaraz, and Fern{\'a}ndez-Caballero]{garcia2019review}
Beatriz Garc{\'\i}a-Mart{\'\i}nez, Arturo Martinez-Rodrigo, Raul Alcaraz, and
  Antonio Fern{\'a}ndez-Caballero.
\newblock A review on nonlinear methods using electroencephalographic
  recordings for emotion recognition.
\newblock \emph{IEEE Transactions on Affective Computing}, 2019.

\bibitem[Jones and Fox(1992)]{Jones1992Electroencephalogram}
Nancy~Aaron Jones and Nathan~A Fox.
\newblock Electroencephalogram asymmetry during emotionally evocative films and
  its relation to positive and negative affectivity.
\newblock \emph{Brain and Cognition}, 20\penalty0 (2):\penalty0 280--299, 1992.

\bibitem[Mathersul et~al.(2008)Mathersul, Williams, Hopkinson, and
  Kemp]{Mathersul2008Investigating}
D~Mathersul, L.~M. Williams, P.~J. Hopkinson, and A.~H. Kemp.
\newblock Investigating models of affect: relationships among eeg alpha
  asymmetry, depression, and anxiety.
\newblock \emph{Emotion}, 8\penalty0 (4):\penalty0 560--572, 2008.

\bibitem[Greve et~al.(2013)Greve, Van~der Haegen, Cai, Stufflebeam, Sabuncu,
  Fischl, and Brysbaert]{greve2013surface}
Douglas~N Greve, Lise Van~der Haegen, Qing Cai, Steven Stufflebeam, Mert~R
  Sabuncu, Bruce Fischl, and Marc Brysbaert.
\newblock A surface-based analysis of language lateralization and cortical
  asymmetry.
\newblock \emph{Journal of cognitive neuroscience}, 25\penalty0 (9):\penalty0
  1477--1492, 2013.

\bibitem[Thammasan et~al.(2016{\natexlab{a}})Thammasan, Moriyama, Fukui, and
  Numao]{Thammasan2016Continuous}
Nattapong Thammasan, Koichi Moriyama, Ken~Ichi Fukui, and Masayuki Numao.
\newblock Continuous music-emotion recognition based on electroencephalogram.
\newblock \emph{IEICE Transactions on Information and Systems}, 99\penalty0
  (4):\penalty0 1234--1241, 2016{\natexlab{a}}.

\bibitem[Huang et~al.(2012)Huang, Guan, Ang, Zhang, and
  Pan]{Huang2012Asymmetric}
Dong Huang, Cuntai Guan, Kai~Keng Ang, Haihong Zhang, and Yaozhang Pan.
\newblock Asymmetric spatial pattern for eeg-based emotion detection.
\newblock In \emph{Proceedings of the 2012 International Joint Conference on
  Neural Networks (IJCNN)}, pages 1--7. IEEE, 2012.

\bibitem[Petrantonakis and Hadjileontiadis(2012)]{Petrantonakis2012Adaptive}
Panagiotis~C. Petrantonakis and Leontios~J. Hadjileontiadis.
\newblock Adaptive emotional information retrieval from eeg signals in the
  time-frequency domain.
\newblock \emph{IEEE Transactions on Signal Processing}, 60\penalty0
  (5):\penalty0 2604--2616, 2012.

\bibitem[Bullmore and Sporns(2012)]{bullmore2012economy}
Ed~Bullmore and Olaf Sporns.
\newblock The economy of brain network organization.
\newblock \emph{Nature Reviews Neuroscience}, 13\penalty0 (5):\penalty0
  336--349, 2012.

\bibitem[Wang et~al.(2019)Wang, El-Fiqi, Hu, and Abbass]{wang2019convolutional}
Min Wang, Heba El-Fiqi, Jiankun Hu, and Hussein~A Abbass.
\newblock Convolutional neural networks using dynamic functional connectivity
  for eeg-based person identification in diverse human states.
\newblock \emph{IEEE Transactions on Information Forensics and Security},
  14\penalty0 (12):\penalty0 3259--3272, 2019.

\bibitem[Lee and Hsieh(2014)]{lee2014classifying}
Youyun Lee and Shulan Hsieh.
\newblock Classifying different emotional states by means of eeg-based
  functional connectivity patterns.
\newblock \emph{PloS one}, 9\penalty0 (4):\penalty0 e95415, 2014.

\bibitem[Betzel and Bassett(2017)]{betzel2017multi}
Richard~F Betzel and Danielle~S Bassett.
\newblock Multi-scale brain networks.
\newblock \emph{Neuroimage}, 160:\penalty0 73--83, 2017.

\bibitem[Rotem-Kohavi et~al.(2017)Rotem-Kohavi, Oberlander, and
  Virji-Babul]{rotem2017infants}
N~Rotem-Kohavi, TF~Oberlander, and N~Virji-Babul.
\newblock Infants and adults have similar regional functional brain
  organization for the perception of emotions.
\newblock \emph{Neuroscience letters}, 650:\penalty0 118--125, 2017.

\bibitem[Chen et~al.(2018)Chen, Wang, and Hua]{chen2018assessment}
Jichi Chen, Hong Wang, and Chengcheng Hua.
\newblock Assessment of driver drowsiness using electroencephalogram signals
  based on multiple functional brain networks.
\newblock \emph{International Journal of Psychophysiology}, 133:\penalty0
  120--130, 2018.

\bibitem[Kinney-Lang et~al.(2019)Kinney-Lang, Yoong, Hunter, Tallur, Shetty,
  McLellan, Chin, and Escudero]{kinney2019analysis}
Eli Kinney-Lang, Michael Yoong, Matthew Hunter, Krishnaraya~Kamath Tallur, Jay
  Shetty, Ailsa McLellan, Richard~FM Chin, and Javier Escudero.
\newblock Analysis of eeg networks and their correlation with cognitive
  impairment in preschool children with epilepsy.
\newblock \emph{Epilepsy \& Behavior}, 90:\penalty0 45--56, 2019.

\bibitem[Morabito et~al.(2015)Morabito, Campolo, Labate, Morabito, Bonanno,
  Bramanti, De~Salvo, Marra, and Bramanti]{morabito2015longitudinal}
Francesco~Carlo Morabito, Maurizio Campolo, Domenico Labate, Giuseppe Morabito,
  Lilla Bonanno, Alessia Bramanti, Simona De~Salvo, Angela Marra, and Placido
  Bramanti.
\newblock A longitudinal eeg study of alzheimer's disease progression based on
  a complex network approach.
\newblock \emph{International journal of neural systems}, 25\penalty0
  (02):\penalty0 1550005, 2015.

\bibitem[De~Haan et~al.(2009)De~Haan, Pijnenburg, Strijers, van~der Made,
  van~der Flier, Scheltens, and Stam]{de2009functional}
Willem De~Haan, Yolande~AL Pijnenburg, Rob~LM Strijers, Yolande van~der Made,
  Wiesje~M van~der Flier, Philip Scheltens, and Cornelis~J Stam.
\newblock Functional neural network analysis in frontotemporal dementia and
  alzheimer's disease using eeg and graph theory.
\newblock \emph{BMC neuroscience}, 10\penalty0 (1):\penalty0 1--12, 2009.

\bibitem[Franciotti et~al.(2019)Franciotti, Falasca, Arnaldi, Fam{\`a},
  Babiloni, Onofrj, Nobili, and Bonanni]{franciotti2019cortical}
Raffaella Franciotti, Nicola~Walter Falasca, Dario Arnaldi, Francesco Fam{\`a},
  Claudio Babiloni, Marco Onofrj, Flavio~Mariano Nobili, and Laura Bonanni.
\newblock Cortical network topology in prodromal and mild dementia due to
  alzheimer’s disease: graph theory applied to resting state eeg.
\newblock \emph{Brain topography}, 32\penalty0 (1):\penalty0 127--141, 2019.

\bibitem[Varela et~al.(2001)Varela, Lachaux, Rodriguez, and
  Martinerie]{varela2001brainweb}
Francisco Varela, Jean-Philippe Lachaux, Eugenio Rodriguez, and Jacques
  Martinerie.
\newblock The brainweb: phase synchronization and large-scale integration.
\newblock \emph{Nature reviews neuroscience}, 2\penalty0 (4):\penalty0
  229--239, 2001.

\bibitem[Chen et~al.(2019)Chen, Song, and Li]{chen2019deep}
He~Chen, Yan Song, and Xiaoli Li.
\newblock A deep learning framework for identifying children with adhd using an
  eeg-based brain network.
\newblock \emph{Neurocomputing}, 356:\penalty0 83--96, 2019.

\bibitem[Guyon and Elisseeff(2003)]{guyon2003introduction}
Isabelle Guyon and Andr{\'e} Elisseeff.
\newblock An introduction to variable and feature selection.
\newblock \emph{Journal of Machine Learning Research}, 3\penalty0
  (Mar):\penalty0 1157--1182, 2003.

\bibitem[Huang et~al.(2006)Huang, Cai, and Xu]{huang2006filter}
Jinjie Huang, Yunze Cai, and Xiaoming Xu.
\newblock A filter approach to feature selection based on mutual information.
\newblock In \emph{Proceedings of 5th IEEE International Conference on
  Cognitive Informatics}, volume~1, pages 84--89. New York: IEEE Press, 2006.
\newblock \doi{10.1109/COGINF.2006.365681}.

\bibitem[Maldonado and Weber(2008)]{Maldonado2008A}
Sebastián Maldonado and Richard Weber.
\newblock A wrapper method for feature selection using support vector machines.
\newblock \emph{Information Sciences}, 179\penalty0 (13):\penalty0 2208--2217,
  2008.
\newblock \doi{10.1016/j.ins.2009.02.014}.

\bibitem[Ding and Peng(2005)]{ding2005minimum}
Chris Ding and Hanchuan Peng.
\newblock Minimum redundancy feature selection from microarray gene expression
  data.
\newblock \emph{Journal of bioinformatics and computational biology},
  3\penalty0 (02):\penalty0 185--205, 2005.

\bibitem[Atkinson and Campos(2016)]{atkinson2016improving}
John Atkinson and Daniel Campos.
\newblock Improving bci-based emotion recognition by combining eeg feature
  selection and kernel classifiers.
\newblock \emph{Expert Systems with Applications}, 47:\penalty0 35--41, 2016.

\bibitem[Guyon et~al.(2002)Guyon, Weston, Barnhill, and Vapnik]{Guyon2002Gene}
Isabelle Guyon, Jason Weston, Stephen Barnhill, and Vladimir Vapnik.
\newblock Gene selection for cancer classification using support vector
  machines.
\newblock \emph{Machine Learning}, 46\penalty0 (1):\penalty0 389--422, 2002.
\newblock \doi{10.1023/A:1012487302797}.

\bibitem[Ng(2004)]{Ng2004Feature}
Andrew~Y. Ng.
\newblock Feature selection, {L1} vs. {L2} regularization, and rotational
  invariance.
\newblock In \emph{Proceedings of the 21st International Conference on Machine
  Learning}, pages 78--85. New York: ACM Press, 2004.
\newblock \doi{10.1145/1015330.1015435}.

\bibitem[Haufe et~al.(2014)Haufe, Meinecke, G{\"o}rgen, D{\"a}hne, Haynes,
  Blankertz, and Bie{\ss}mann]{Haufe2014On}
Stefan Haufe, Frank Meinecke, Kai G{\"o}rgen, Sven D{\"a}hne, J.~D. Haynes,
  Benjamin Blankertz, and Felix Bie{\ss}mann.
\newblock On the interpretation of weight vectors of linear models in
  multivariate neuroimaging.
\newblock \emph{Neuroimage}, 87\penalty0 (2):\penalty0 96--110, 2014.
\newblock \doi{10.1016/j.neuroimage.2013.10.067}.

\bibitem[Pham et~al.(2015)Pham, Tran, Ma, and Tran]{pham2015enhancing}
Trung~Duy Pham, Dat Tran, Wanli Ma, and Nga~Thuy Tran.
\newblock Enhancing performance of eeg-based emotion recognition systems using
  feature smoothing.
\newblock In \emph{International Conference on Neural Information Processing},
  pages 95--102. Springer, 2015.

\bibitem[Tang et~al.(2017)Tang, Wang, Tan, and Miao]{tang2017eeg}
Cheng Tang, Di~Wang, Ah-Hwee Tan, and Chunyan Miao.
\newblock Eeg-based emotion recognition via fast and robust feature smoothing.
\newblock In \emph{International Conference on Brain Informatics}, pages
  83--92. Springer, 2017.

\bibitem[Zhang et~al.(2020{\natexlab{a}})Zhang, Chen, Tan, Chen, Chen, Li,
  Yang, Su, Huang, and Che]{zhang2020investigation}
Yaqing Zhang, Jinling Chen, Jen~Hong Tan, Yuxuan Chen, Yunyi Chen, Dihan Li,
  Lei Yang, Jian Su, Xin Huang, and Wenliang Che.
\newblock An investigation of deep learning models for eeg-based emotion
  recognition.
\newblock \emph{Frontiers in Neuroscience}, 14, 2020{\natexlab{a}}.

\bibitem[Chung and Yoon(2012)]{Chung2012Affective}
Seong~Youb Chung and Hyun~Joong Yoon.
\newblock Affective classification using bayesian classifier and supervised
  learning.
\newblock In \emph{Proceedings of the 2012 12th International Conference on
  Control, Automation and Systems}, pages 1768--1771. IEEE, 2012.

\bibitem[Ackermann et~al.(2016)Ackermann, Kohlschein, Bitsch, Wehrle, and
  Jeschke]{ackermann2016eeg}
Pascal Ackermann, Christian Kohlschein, J{\'o}~Agila Bitsch, Klaus Wehrle, and
  Sabina Jeschke.
\newblock Eeg-based automatic emotion recognition: Feature extraction,
  selection and classification methods.
\newblock In \emph{Proceedings of the 2016 18th International Conference on
  e-Health Networking, Applications and Services (Healthcom)}, pages 1--6.
  IEEE, 2016.

\bibitem[Bhatti et~al.(2016)Bhatti, Majid, Anwar, and Khan]{bhatti2016human}
Adnan~Mehmood Bhatti, Muhammad Majid, Syed~Muhammad Anwar, and Bilal Khan.
\newblock Human emotion recognition and analysis in response to audio music
  using brain signals.
\newblock \emph{Computers in Human Behavior}, 65:\penalty0 267--275, 2016.

\bibitem[Murugappan et~al.(2008)Murugappan, Rizon, Nagarajan, Yaacob, Zunaidi,
  and Hazry]{Murugappan2008Lifting}
M~Murugappan, M~Rizon, R~Nagarajan, S~Yaacob, I~Zunaidi, and D~Hazry.
\newblock Lifting scheme for human emotion recognition using eeg.
\newblock In \emph{Proceedings of the 2008 International symposium on
  information technology}, volume~2, pages 1--7. IEEE, 2008.

\bibitem[Jenke et~al.(2014)Jenke, Peer, and Buss]{jenke2014feature}
Robert Jenke, Angelika Peer, and Martin Buss.
\newblock Feature extraction and selection for emotion recognition from eeg.
\newblock \emph{IEEE Transactions on Affective Computing}, 5\penalty0
  (3):\penalty0 327--339, 2014.

\bibitem[Li et~al.(2019{\natexlab{a}})Li, Zheng, Cui, Zong, and Ge]{li2019eeg}
Yang Li, Wenming Zheng, Zhen Cui, Yuan Zong, and Sheng Ge.
\newblock Eeg emotion recognition based on graph regularized sparse linear
  regression.
\newblock \emph{Neural Processing Letters}, 49\penalty0 (2):\penalty0 555--571,
  2019{\natexlab{a}}.

\bibitem[Cheng et~al.(2020)Cheng, Chen, Li, Liu, Song, Liu, and
  Chen]{cheng2020emotion}
Juan Cheng, Meiyao Chen, Chang Li, Yu~Liu, Rencheng Song, Aiping Liu, and Xun
  Chen.
\newblock Emotion recognition from multi-channel eeg via deep forest.
\newblock \emph{IEEE Journal of Biomedical and Health Informatics}, 25\penalty0
  (2):\penalty0 453--464, 2020.

\bibitem[Scarselli and Tsoi(1998)]{scarselli1998universal}
Franco Scarselli and Ah~Chung Tsoi.
\newblock Universal approximation using feedforward neural networks: A survey
  of some existing methods, and some new results.
\newblock \emph{Neural networks}, 11\penalty0 (1):\penalty0 15--37, 1998.

\bibitem[Thammasan et~al.(2016{\natexlab{b}})Thammasan, Fukui, and
  Numao]{thammasan2016application}
Nattapong Thammasan, Ken-ichi Fukui, and Masayuki Numao.
\newblock Application of deep belief networks in eeg-based dynamic
  music-emotion recognition.
\newblock In \emph{Proceedings of the 2016 International Joint Conference on
  Neural Networks (IJCNN)}, pages 881--888. IEEE, 2016{\natexlab{b}}.

\bibitem[Hinton et~al.(2006)Hinton, Osindero, and Teh]{hinton2006fast}
Geoffrey~E Hinton, Simon Osindero, and Yee-Whye Teh.
\newblock A fast learning algorithm for deep belief nets.
\newblock \emph{Neural computation}, 18\penalty0 (7):\penalty0 1527--1554,
  2006.

\bibitem[Tripathi et~al.(2017)Tripathi, Acharya, Sharma, Mittal, and
  Bhattacharya]{tripathi2017using}
Samarth Tripathi, Shrinivas Acharya, Ranti~Dev Sharma, Sudhanshi Mittal, and
  Samit Bhattacharya.
\newblock Using deep and convolutional neural networks for accurate emotion
  classification on deap dataset.
\newblock In \emph{Proceedings of the 21st AAAI Conference on Artificial
  Intelligence}, pages 4746--4752. AAAI Press, 2017.

\bibitem[Chao et~al.(2019)Chao, Dong, Liu, and Lu]{chao2019emotion}
Hao Chao, Liang Dong, Yongli Liu, and Baoyun Lu.
\newblock Emotion recognition from multiband eeg signals using capsnet.
\newblock \emph{Sensors}, 19\penalty0 (9):\penalty0 2212, 2019.

\bibitem[Sabour et~al.(2017)Sabour, Frosst, and Hinton]{sabour2017dynamic}
Sara Sabour, Nicholas Frosst, and Geoffrey~E Hinton.
\newblock Dynamic routing between capsules.
\newblock In \emph{Proceedings of the 31st International Conference on Neural
  Information Processing Systems}, pages 3859--3869, 2017.

\bibitem[Liu et~al.(2020)Liu, Ding, Li, Cheng, Song, Wan, and
  Chen]{liu2020multi}
Yu~Liu, Yufeng Ding, Chang Li, Juan Cheng, Rencheng Song, Feng Wan, and Xun
  Chen.
\newblock Multi-channel eeg-based emotion recognition via a multi-level
  features guided capsule network.
\newblock \emph{Computers in Biology and Medicine}, 123:\penalty0 103927, 2020.

\bibitem[Such et~al.(2017)Such, Sah, Dominguez, Pillai, Zhang, Michael, Cahill,
  and Ptucha]{such2017robust}
Felipe~Petroski Such, Shagan Sah, Miguel~Alexander Dominguez, Suhas Pillai,
  Chao Zhang, Andrew Michael, Nathan~D Cahill, and Raymond Ptucha.
\newblock Robust spatial filtering with graph convolutional neural networks.
\newblock \emph{IEEE Journal of Selected Topics in Signal Processing},
  11\penalty0 (6):\penalty0 884--896, 2017.

\bibitem[Song et~al.(2018)Song, Zheng, Song, and Cui]{song2018eeg}
Tengfei Song, Wenming Zheng, Peng Song, and Zhen Cui.
\newblock Eeg emotion recognition using dynamical graph convolutional neural
  networks.
\newblock \emph{IEEE Transactions on Affective Computing}, 11\penalty0
  (3):\penalty0 532--541, 2018.

\bibitem[Zhang et~al.(2020{\natexlab{b}})Zhang, Cui, Xu, Zheng, and
  Yang]{zhang2020variational}
Tong Zhang, Zhen Cui, Chunyan Xu, Wenming Zheng, and Jian Yang.
\newblock Variational pathway reasoning for eeg emotion recognition.
\newblock In \emph{Proceedings of the AAAI Conference on Artificial
  Intelligence}, pages 2709--2716, 2020{\natexlab{b}}.

\bibitem[Adolphs(2002)]{adolphs2002neural}
Ralph Adolphs.
\newblock Neural systems for recognizing emotion.
\newblock \emph{Current opinion in neurobiology}, 12\penalty0 (2):\penalty0
  169--177, 2002.

\bibitem[Bullmore and Sporns(2009)]{bullmore2009complex}
Ed~Bullmore and Olaf Sporns.
\newblock Complex brain networks: graph theoretical analysis of structural and
  functional systems.
\newblock \emph{Nature reviews neuroscience}, 10\penalty0 (3):\penalty0
  186--198, 2009.

\bibitem[Li et~al.(2020{\natexlab{a}})Li, Wang, Zheng, Zong, Qi, Cui, Zhang,
  and Song]{li2020novel}
Yang Li, Lei Wang, Wenming Zheng, Yuan Zong, Lei Qi, Zhen Cui, Tong Zhang, and
  Tengfei Song.
\newblock A novel bi-hemispheric discrepancy model for eeg emotion recognition.
\newblock \emph{IEEE Transactions on Cognitive and Developmental Systems},
  13\penalty0 (2):\penalty0 354--367, 2020{\natexlab{a}}.

\bibitem[Li et~al.(2018{\natexlab{c}})Li, Zheng, Cui, Zhang, and
  Zong]{li2018novel}
Yang Li, Wenming Zheng, Zhen Cui, Tong Zhang, and Yuan Zong.
\newblock A novel neural network model based on cerebral hemispheric asymmetry
  for eeg emotion recognition.
\newblock In \emph{IJCAI}, pages 1561--1567, 2018{\natexlab{c}}.

\bibitem[Huang et~al.(2021)Huang, Chen, Liu, Zheng, Tian, and
  Jiang]{huang2021differences}
Dongmin Huang, Sentao Chen, Cheng Liu, Lin Zheng, Zhihang Tian, and Dazhi
  Jiang.
\newblock Differences first in asymmetric brain: A bi-hemisphere discrepancy
  convolutional neural network for eeg emotion recognition.
\newblock \emph{Neurocomputing}, 448:\penalty0 140--151, 2021.

\bibitem[Cui et~al.(2020)Cui, Liu, Zhang, Chen, Wang, and Chen]{cui2020eeg}
Heng Cui, Aiping Liu, Xu~Zhang, Xiang Chen, Kongqiao Wang, and Xun Chen.
\newblock Eeg-based emotion recognition using an end-to-end regional-asymmetric
  convolutional neural network.
\newblock \emph{Knowledge-Based Systems}, 205:\penalty0 106243, 2020.

\bibitem[Lawhern et~al.(2018)Lawhern, Solon, Waytowich, Gordon, Hung, and
  Lance]{lawhern2018eegnet}
Vernon~J Lawhern, Amelia~J Solon, Nicholas~R Waytowich, Stephen~M Gordon,
  Chou~P Hung, and Brent~J Lance.
\newblock Eegnet: a compact convolutional neural network for eeg-based
  brain--computer interfaces.
\newblock \emph{Journal of neural engineering}, 15\penalty0 (5):\penalty0
  056013, 2018.

\bibitem[Wang et~al.(2018{\natexlab{a}})Wang, Huang, McCane, and
  Neo]{wang2018emotionet}
Yi~Wang, Zhiyi Huang, Brendan McCane, and Phoebe Neo.
\newblock Emotionet: A 3-d convolutional neural network for eeg-based emotion
  recognition.
\newblock In \emph{2018 International Joint Conference on Neural Networks
  (IJCNN)}, pages 1--7. IEEE, 2018{\natexlab{a}}.

\bibitem[Islam et~al.(2021)Islam, Islam, Rahman, Mondal, Singha, Ahmad, Awal,
  Islam, and Moni]{islam2021eeg}
Md~Rabiul Islam, Md~Milon Islam, Md~Mustafizur Rahman, Chayan Mondal,
  Suvojit~Kumar Singha, Mohiuddin Ahmad, Abdul Awal, Md~Saiful Islam, and
  Mohammad~Ali Moni.
\newblock Eeg channel correlation based model for emotion recognition.
\newblock \emph{Computers in Biology and Medicine}, 136:\penalty0 104757, 2021.

\bibitem[Salama et~al.(2018)Salama, El-Khoribi, Shoman, and
  Shalaby]{salama2018eeg}
Elham~S Salama, Reda~A El-Khoribi, Mahmoud~E Shoman, and Mohamed A~Wahby
  Shalaby.
\newblock Eeg-based emotion recognition using 3d convolutional neural networks.
\newblock \emph{Int. J. Adv. Comput. Sci. Appl}, 9\penalty0 (8):\penalty0
  329--337, 2018.

\bibitem[Cho and Hwang(2020)]{cho2020spatio}
Jungchan Cho and Hyoseok Hwang.
\newblock Spatio-temporal representation of an electoencephalogram for emotion
  recognition using a three-dimensional convolutional neural network.
\newblock \emph{Sensors}, 20\penalty0 (12):\penalty0 3491, 2020.

\bibitem[Ding et~al.(2020)Ding, Robinson, Zeng, Chen, Wai, Lee, and
  Guan]{ding2020tsception}
Yi~Ding, Neethu Robinson, Qiuhao Zeng, Duo Chen, Aung Aung~Phyo Wai, Tih-Shih
  Lee, and Cuntai Guan.
\newblock Tsception: a deep learning framework for emotion detection using eeg.
\newblock In \emph{2020 International Joint Conference on Neural Networks
  (IJCNN)}, pages 1--7. IEEE, 2020.

\bibitem[Wei et~al.(2020)Wei, Chen, Song, Lou, and Li]{wei2020eeg}
Chen Wei, Lan-lan Chen, Zhen-zhen Song, Xiao-guang Lou, and Dong-dong Li.
\newblock Eeg-based emotion recognition using simple recurrent units network
  and ensemble learning.
\newblock \emph{Biomedical Signal Processing and Control}, 58:\penalty0 101756,
  2020.

\bibitem[Ji et~al.(2012)Ji, Xu, Yang, and Yu]{ji20123d}
Shuiwang Ji, Wei Xu, Ming Yang, and Kai Yu.
\newblock 3d convolutional neural networks for human action recognition.
\newblock \emph{IEEE transactions on pattern analysis and machine
  intelligence}, 35\penalty0 (1):\penalty0 221--231, 2012.

\bibitem[Jia et~al.(2020)Jia, Lin, Cai, Chen, Gou, and Wang]{jia2020sst}
Ziyu Jia, Youfang Lin, Xiyang Cai, Haobin Chen, Haijun Gou, and Jing Wang.
\newblock Sst-emotionnet: Spatial-spectral-temporal based attention 3d dense
  network for eeg emotion recognition.
\newblock In \emph{Proceedings of the 28th ACM International Conference on
  Multimedia}, pages 2909--2917, 2020.

\bibitem[Li et~al.(2016)Li, Song, Zhang, Yu, Hou, and Hu]{li2016emotion}
Xiang Li, Dawei Song, Peng Zhang, Guangliang Yu, Yuexian Hou, and Bin Hu.
\newblock Emotion recognition from multi-channel eeg data through convolutional
  recurrent neural network.
\newblock In \emph{2016 IEEE international conference on bioinformatics and
  biomedicine (BIBM)}, pages 352--359. IEEE, 2016.

\bibitem[Zhang et~al.(2018)Zhang, Yao, Zhang, Wang, Chen, Boots, and
  Benatallah]{zhang2018cascade}
Dalin Zhang, Lina Yao, Xiang Zhang, Sen Wang, Weitong Chen, Robert Boots, and
  Boualem Benatallah.
\newblock Cascade and parallel convolutional recurrent neural networks on
  eeg-based intention recognition for brain computer interface.
\newblock In \emph{Proceedings of the AAAI Conference on Artificial
  Intelligence}, 2018.

\bibitem[Yang et~al.(2018{\natexlab{b}})Yang, Wu, Qiu, Wang, and
  Chen]{yang2018emotion}
Yilong Yang, Qingfeng Wu, Ming Qiu, Yingdong Wang, and Xiaowei Chen.
\newblock Emotion recognition from multi-channel eeg through parallel
  convolutional recurrent neural network.
\newblock In \emph{2018 International Joint Conference on Neural Networks
  (IJCNN)}, pages 1--7. IEEE, 2018{\natexlab{b}}.

\bibitem[Tao et~al.(2020)Tao, Li, Song, Cheng, Liu, Wan, and Chen]{tao2020eeg}
Wei Tao, Chang Li, Rencheng Song, Juan Cheng, Yu~Liu, Feng Wan, and Xun Chen.
\newblock Eeg-based emotion recognition via channel-wise attention and self
  attention.
\newblock \emph{IEEE Transactions on Affective Computing}, 2020.

\bibitem[Li et~al.(2019{\natexlab{b}})Li, Zheng, Wang, Zong, and
  Cui]{li2019regional}
Yang Li, Wenming Zheng, Lei Wang, Yuan Zong, and Zhen Cui.
\newblock From regional to global brain: A novel hierarchical spatial-temporal
  neural network model for eeg emotion recognition.
\newblock \emph{IEEE Transactions on Affective Computing}, 2019{\natexlab{b}}.

\bibitem[Zhang et~al.(2019)Zhang, Zheng, Cui, Zong, and Li]{zhang2019spatial}
Tong Zhang, Wenming Zheng, Zhen Cui, Yuan Zong, and Yang Li.
\newblock Spatial-temporal recurrent neural network for emotion recognition.
\newblock \emph{IEEE transactions on cybernetics}, 49\penalty0 (3):\penalty0
  839--847, 2019.

\bibitem[Lew et~al.(2020)Lew, Wang, Shylouskaya, Zhang, Lim, Ang, and
  Tan]{lew2020eeg}
Wai-Cheong~Lincoln Lew, Di~Wang, Katsiaryna Shylouskaya, Zhuo Zhang, Joo-Hwee
  Lim, Kai~Keng Ang, and Ah-Hwee Tan.
\newblock Eeg-based emotion recognition using spatial-temporal representation
  via bi-gru.
\newblock In \emph{2020 42nd Annual International Conference of the IEEE
  Engineering in Medicine \& Biology Society (EMBC)}, pages 116--119. IEEE,
  2020.

\bibitem[Li et~al.(2020{\natexlab{b}})Li, Zhao, Song, Zhang, Pan, Wu, Huo, Niu,
  and Wang]{li2020latent}
Xiang Li, Zhigang Zhao, Dawei Song, Yazhou Zhang, Jingshan Pan, Lu~Wu, Jidong
  Huo, Chunyang Niu, and Di~Wang.
\newblock Latent factor decoding of multi-channel eeg for emotion recognition
  through autoencoder-like neural networks.
\newblock \emph{Frontiers in neuroscience}, 14:\penalty0 87,
  2020{\natexlab{b}}.

\bibitem[Xing et~al.(2019)Xing, Li, Xu, Shu, Hu, and Xu]{xing2019sae}
Xiaofen Xing, Zhenqi Li, Tianyuan Xu, Lin Shu, Bin Hu, and Xiangmin Xu.
\newblock Sae+ lstm: A new framework for emotion recognition from multi-channel
  eeg.
\newblock \emph{Frontiers in neurorobotics}, 13:\penalty0 37, 2019.

\bibitem[Yin et~al.(2021)Yin, Zheng, Hu, Zhang, and Cui]{yin2021eeg}
Yongqiang Yin, Xiangwei Zheng, Bin Hu, Yuang Zhang, and Xinchun Cui.
\newblock Eeg emotion recognition using fusion model of graph convolutional
  neural networks and lstm.
\newblock \emph{Applied Soft Computing}, 100:\penalty0 106954, 2021.

\bibitem[Zhang and Etemad(2020)]{zhang2020rfnet}
Guangyi Zhang and Ali Etemad.
\newblock Rfnet: Riemannian fusion network for eeg-based brain-computer
  interfaces.
\newblock \emph{arXiv preprint arXiv:2008.08633}, 2020.

\bibitem[Kalunga et~al.(2015)Kalunga, Chevallier, Barth{\'e}lemy, Djouani,
  Hamam, and Monacelli]{kalunga2015euclidean}
Emmanuel~K Kalunga, Sylvain Chevallier, Quentin Barth{\'e}lemy, Karim Djouani,
  Yskandar Hamam, and Eric Monacelli.
\newblock From euclidean to riemannian means: Information geometry for ssvep
  classification.
\newblock In \emph{International Conference on Geometric Science of
  Information}, pages 595--604. Springer, 2015.

\bibitem[Kim and Andr{\'e}(2008)]{kim2008emotion}
Jonghwa Kim and Elisabeth Andr{\'e}.
\newblock Emotion recognition based on physiological changes in music
  listening.
\newblock \emph{IEEE Transactions on Pattern Analysis and Machine
  Intelligence}, 30\penalty0 (12):\penalty0 2067--2083, 2008.

\bibitem[AlZoubi et~al.(2012)AlZoubi, D'Mello, and Calvo]{alzoubi2012detecting}
Omar AlZoubi, Sidney~K D'Mello, and Rafael~A Calvo.
\newblock Detecting naturalistic expressions of nonbasic affect using
  physiological signals.
\newblock \emph{IEEE Transactions on Affective Computing}, 3\penalty0
  (3):\penalty0 298--310, 2012.

\bibitem[Hamann and Canli(2004)]{Hamann2004Individual}
S~Hamann and T~Canli.
\newblock Individual differences in emotion processing.
\newblock \emph{Current Opinion in Neurobiology}, 14\penalty0 (2):\penalty0
  233--238, 2004.

\bibitem[Flores-Guti{\'e}rrez et~al.(2009)Flores-Guti{\'e}rrez, D{\'\i}az,
  Barrios, Guevara, del R{\'\i}o-Portilla, Corsi-Cabrera, and del
  Flores-Guti{\'e}rrez]{flores2009differential}
Enrique~O Flores-Guti{\'e}rrez, Jos{\'e}-Luis D{\'\i}az, Fernando~A Barrios,
  Miguel~{\'A}ngel Guevara, Yolanda del R{\'\i}o-Portilla, Mar{\'\i}a
  Corsi-Cabrera, and Enrique~O del Flores-Guti{\'e}rrez.
\newblock Differential alpha coherence hemispheric patterns in men and women
  during pleasant and unpleasant musical emotions.
\newblock \emph{International Journal of Psychophysiology}, 71\penalty0
  (1):\penalty0 43--49, 2009.

\bibitem[Bilalpur et~al.(2017)Bilalpur, Kia, Chua, and
  Subramanian]{bilalpur2017discovering}
Maneesh Bilalpur, Seyed~Mostafa Kia, Tat-Seng Chua, and Ramanathan Subramanian.
\newblock Discovering gender differences in facial emotion recognition via
  implicit behavioral cues.
\newblock In \emph{2017 Seventh International Conference on Affective Computing
  and Intelligent Interaction (ACII)}, pages 119--124. IEEE, 2017.

\bibitem[Goshvarpour and Goshvarpour(2019)]{goshvarpour2019eeg}
Atefeh Goshvarpour and Ateke Goshvarpour.
\newblock Eeg spectral powers and source localization in depressing, sad, and
  fun music videos focusing on gender differences.
\newblock \emph{Cognitive neurodynamics}, 13\penalty0 (2):\penalty0 161--173,
  2019.

\bibitem[Bradley et~al.(2001)Bradley, Codispoti, Sabatinelli, and
  Lang]{bradley2001emotion}
Margaret~M Bradley, Maurizio Codispoti, Dean Sabatinelli, and Peter~J Lang.
\newblock Emotion and motivation ii: sex differences in picture processing.
\newblock \emph{Emotion}, 1\penalty0 (3):\penalty0 300, 2001.

\bibitem[Lee et~al.(2005)Lee, Liu, Chan, Fang, and Gao]{lee2005neural}
TMC Lee, HL~Liu, CCH Chan, SY~Fang, and JH~Gao.
\newblock Neural activities associated with emotion recognition observed in men
  and women.
\newblock \emph{Molecular psychiatry}, 10\penalty0 (5):\penalty0 450--455,
  2005.

\bibitem[Raab et~al.(2016)Raab, Kirsch, and Mier]{raab2016understanding}
Kyeon Raab, Peter Kirsch, and Daniela Mier.
\newblock Understanding the impact of 5-httlpr, antidepressants, and acute
  tryptophan depletion on brain activation during facial emotion processing: A
  review of the imaging literature.
\newblock \emph{Neuroscience \& Biobehavioral Reviews}, 71:\penalty0 176--197,
  2016.

\bibitem[Fischer et~al.(2004)Fischer, Rodriguez~Mosquera, Van~Vianen, and
  Manstead]{fischer2004gender}
Agneta~H Fischer, Patricia~M Rodriguez~Mosquera, Annelies~EM Van~Vianen, and
  Antony~SR Manstead.
\newblock Gender and culture differences in emotion.
\newblock \emph{Emotion}, 4\penalty0 (1):\penalty0 87, 2004.

\bibitem[Zhu et~al.(2015)Zhu, Zheng, and Lu]{zhu2015cross}
Jia-Yi Zhu, Wei-Long Zheng, and Bao-Liang Lu.
\newblock Cross-subject and cross-gender emotion classification from eeg.
\newblock In \emph{World Congress on Medical Physics and Biomedical
  Engineering, June 7-12, 2015, Toronto, Canada}, pages 1188--1191. Springer,
  2015.

\bibitem[Pava et~al.(2018)Pava, {\'A}lvarez, Herrera,
  Castellanos-Dom{\'\i}nguez, and Orozco]{pava2018gender}
I~Pava, A~{\'A}lvarez, Paula Herrera, Germ{\'a}n Castellanos-Dom{\'\i}nguez,
  and A~Orozco.
\newblock Gender effects on an eeg-based emotion level classification system.
\newblock In \emph{Iberoamerican Congress on Pattern Recognition}, pages
  810--819. Springer, Cham, 2018.

\bibitem[Huang(2004)]{huang2004native}
Y.~Huang.
\newblock Native assessment of international affective picture system.
\newblock \emph{Chinese Mental Health Journal}, 2004.

\bibitem[Kurbalija et~al.(2018)Kurbalija, Ivanovi{\'c}, Radovanovi{\'c}, Geler,
  Dai, and Zhao]{kurbalija2018emotion}
Vladimir Kurbalija, Mirjana Ivanovi{\'c}, Milo{\v{s}} Radovanovi{\'c}, Zoltan
  Geler, Weihui Dai, and Weidong Zhao.
\newblock Emotion perception and recognition: an exploration of cultural
  differences and similarities.
\newblock \emph{Cognitive Systems Research}, 52:\penalty0 103--116, 2018.

\bibitem[Gan et~al.(2019)Gan, Liu, Luo, Wu, and Lu]{gan2019cross}
Lu~Gan, Wei Liu, Yun Luo, Xun Wu, and Bao-Liang Lu.
\newblock A cross-culture study on multimodal emotion recognition using deep
  learning.
\newblock In \emph{International Conference on Neural Information Processing},
  pages 670--680. Springer, 2019.

\bibitem[Mohammad and Nishida(2010)]{mohammad2010using}
Yasser Mohammad and Toyoaki Nishida.
\newblock Using physiological signals to detect natural interactive behavior.
\newblock \emph{Applied Intelligence}, 33\penalty0 (1):\penalty0 79--92, 2010.

\bibitem[Zanini et~al.(2017)Zanini, Congedo, Jutten, Said, and
  Berthoumieu]{zanini2017transfer}
Paolo Zanini, Marco Congedo, Christian Jutten, Salem Said, and Yannick
  Berthoumieu.
\newblock Transfer learning: A riemannian geometry framework with applications
  to brain--computer interfaces.
\newblock \emph{IEEE Transactions on Biomedical Engineering}, 65\penalty0
  (5):\penalty0 1107--1116, 2017.

\bibitem[Fernandez et~al.(2021)Fernandez, Guttenberg, Witkowski, and
  Pasquali]{fernandez2021cross}
Javier Fernandez, Nicholas Guttenberg, Olaf Witkowski, and Antoine Pasquali.
\newblock Cross-subject eeg-based emotion recognition through neural networks
  with stratified normalization.
\newblock \emph{Frontiers in neuroscience}, 15:\penalty0 11, 2021.

\bibitem[Zhou et~al.(2011)Zhou, Qu, Helander, and Jiao]{zhou2011affect}
Feng Zhou, Xingda Qu, Martin~G Helander, and Jianxin~Roger Jiao.
\newblock Affect prediction from physiological measures via visual stimuli.
\newblock \emph{International Journal of Human-Computer Studies}, 69\penalty0
  (12):\penalty0 801--819, 2011.

\bibitem[Bailenson et~al.(2008)Bailenson, Pontikakis, Mauss, Gross, Jabon,
  Hutcherson, Nass, and John]{bailenson2008real}
Jeremy~N Bailenson, Emmanuel~D Pontikakis, Iris~B Mauss, James~J Gross, Maria~E
  Jabon, Cendri~AC Hutcherson, Clifford Nass, and Oliver John.
\newblock Real-time classification of evoked emotions using facial feature
  tracking and physiological responses.
\newblock \emph{International Journal of Human-Computer Studies}, 66\penalty0
  (5):\penalty0 303--317, 2008.

\bibitem[Chen et~al.(2017)Chen, Hu, Wang, Moore, Dai, Feng, and
  Ding]{chen2017subject}
Jing Chen, Bin Hu, Yue Wang, Philip Moore, Yongqiang Dai, Lei Feng, and Zhijie
  Ding.
\newblock Subject-independent emotion recognition based on physiological
  signals: a three-stage decision method.
\newblock \emph{BMC Medical Informatics and Decision Making}, 17\penalty0
  (3):\penalty0 167, 2017.

\bibitem[Liu et~al.(2021)Liu, Shen, Song, and Zhang]{liu2021domain}
Jin Liu, Xinke Shen, Sen Song, and Dan Zhang.
\newblock Domain adaptation for cross-subject emotion recognition by subject
  clustering.
\newblock In \emph{2021 10th International IEEE/EMBS Conference on Neural
  Engineering (NER)}, pages 904--908. IEEE, 2021.

\bibitem[Lan et~al.(2018)Lan, Sourina, Wang, Scherer, and
  M{\"u}ller-Putz]{lan2018domain}
Zirui Lan, Olga Sourina, Lipo Wang, Reinhold Scherer, and Gernot~R
  M{\"u}ller-Putz.
\newblock Domain adaptation techniques for eeg-based emotion recognition: a
  comparative study on two public datasets.
\newblock \emph{IEEE Transactions on Cognitive and Developmental Systems},
  11\penalty0 (1):\penalty0 85--94, 2018.

\bibitem[Luo et~al.(2018)Luo, Zhang, Zheng, and Lu]{luo2018wgan}
Yun Luo, Si-Yang Zhang, Wei-Long Zheng, and Bao-Liang Lu.
\newblock Wgan domain adaptation for eeg-based emotion recognition.
\newblock In \emph{International Conference on Neural Information Processing},
  pages 275--286. Springer, 2018.

\bibitem[Wang et~al.(2021{\natexlab{a}})Wang, Zhang, Xu, Ping, and
  Chu]{wang2021deep}
Fei Wang, Weiwei Zhang, Zongfeng Xu, Jingyu Ping, and Hao Chu.
\newblock A deep multi-source adaptation transfer network for cross-subject
  electroencephalogram emotion recognition.
\newblock \emph{Neural Computing and Applications}, pages 1--13,
  2021{\natexlab{a}}.

\bibitem[Cimtay and Ekmekcioglu(2020)]{cimtay2020investigating}
Yucel Cimtay and Erhan Ekmekcioglu.
\newblock Investigating the use of pretrained convolutional neural network on
  cross-subject and cross-dataset eeg emotion recognition.
\newblock \emph{Sensors}, 20\penalty0 (7):\penalty0 2034, 2020.

\bibitem[Wang et~al.(2020)Wang, Wu, Zhang, Xu, Zhang, Wu, and
  Coleman]{wang2020emotion}
Fei Wang, Shichao Wu, Weiwei Zhang, Zongfeng Xu, Yahui Zhang, Chengdong Wu, and
  Sonya Coleman.
\newblock Emotion recognition with convolutional neural network and eeg-based
  efdms.
\newblock \emph{Neuropsychologia}, 146:\penalty0 107506, 2020.

\bibitem[Li et~al.(2018{\natexlab{d}})Li, Jin, Zheng, and Lu]{li2018cross}
He~Li, Yi-Ming Jin, Wei-Long Zheng, and Bao-Liang Lu.
\newblock Cross-subject emotion recognition using deep adaptation networks.
\newblock In \emph{International conference on neural information processing},
  pages 403--413. Springer, 2018{\natexlab{d}}.

\bibitem[Cai et~al.(2021)Cai, Guo, Yang, Chen, and Xu]{cai2021cross}
Ziliang Cai, Miaomiao Guo, Xinsheng Yang, Xintong Chen, and Guizhi Xu.
\newblock Cross-subject electroencephalogram emotion recognition based on
  maximum classifier discrepancy.
\newblock \emph{Sheng wu yi xue gong cheng xue za zhi= Journal of biomedical
  engineering= Shengwu yixue gongchengxue zazhi}, 38\penalty0 (3):\penalty0
  455--462, 2021.

\bibitem[Zhao et~al.(2021)Zhao, Yan, and Lu]{zhao2021plug}
Li-Ming Zhao, Xu~Yan, and Bao-Liang Lu.
\newblock Plug-and-play domain adaptation for cross-subject eeg-based emotion
  recognition.
\newblock In \emph{Proceedings of the 35th AAAI Conference on Artificial
  Intelligence}. sn, 2021.

\bibitem[Ding et~al.(2021)Ding, Kimura, Fukui, and Numao]{ding2021eeg}
Ke-Ming Ding, Tsukasa Kimura, Ken-ichi Fukui, and Masayuki Numao.
\newblock Eeg emotion enhancement using task-specific domain adversarial neural
  network.
\newblock In \emph{2021 International Joint Conference on Neural Networks
  (IJCNN)}, pages 1--8. IEEE, 2021.

\bibitem[Wang et~al.(2021{\natexlab{b}})Wang, Liu, Ruan, Wang, and
  Wang]{wang2021cross}
Yingdong Wang, Jiatong Liu, Qunsheng Ruan, Shuocheng Wang, and Chen Wang.
\newblock Cross-subject eeg emotion classification based on few-label
  adversarial domain adaption.
\newblock \emph{Expert Systems with Applications}, 185:\penalty0 115581,
  2021{\natexlab{b}}.

\bibitem[Zhang and Etemad(2021)]{zhang2021distilling}
Guangyi Zhang and Ali Etemad.
\newblock Distilling eeg representations via capsules for affective computing.
\newblock \emph{arXiv preprint arXiv:2105.00104}, 2021.

\bibitem[Zhong et~al.(2020)Zhong, Wang, and Miao]{zhong2020eeg}
Peixiang Zhong, Di~Wang, and Chunyan Miao.
\newblock Eeg-based emotion recognition using regularized graph neural
  networks.
\newblock \emph{IEEE Transactions on Affective Computing}, 2020.

\bibitem[Duan et~al.(2020{\natexlab{a}})Duan, Chauhan, Shaikh, Chu, and
  Srihari]{duan2020ultra}
Tiehang Duan, Mihir Chauhan, Mohammad~Abuzar Shaikh, Jun Chu, and Sargur
  Srihari.
\newblock Ultra efficient transfer learning with meta update for cross subject
  eeg classification.
\newblock \emph{arXiv preprint arXiv:2003.06113}, 2020{\natexlab{a}}.

\bibitem[Duan et~al.(2020{\natexlab{b}})Duan, Shaikh, Chauhan, Chu, Srihari,
  Pathak, and Srihari]{duan2020meta}
Tiehang Duan, Mohammad~Abuzar Shaikh, Mihir Chauhan, Jun Chu, Rohini~K Srihari,
  Archita Pathak, and Sargur~N Srihari.
\newblock Meta learn on constrained transfer learning for low resource cross
  subject eeg classification.
\newblock \emph{IEEE Access}, 8:\penalty0 224791--224802, 2020{\natexlab{b}}.

\bibitem[Jim{\'e}nez-Guarneros and
  G{\'o}mez-Gil(2021)]{jimenez2021standardization}
Magdiel Jim{\'e}nez-Guarneros and Pilar G{\'o}mez-Gil.
\newblock Standardization-refinement domain adaptation method for cross-subject
  eeg-based classification in imagined speech recognition.
\newblock \emph{Pattern Recognition Letters}, 141:\penalty0 54--60, 2021.

\bibitem[Soleymani et~al.(2012)Soleymani, Pantic, and
  Pun]{soleymani2012multimodal}
Mohammad Soleymani, Maja Pantic, and Thierry Pun.
\newblock Multimodal emotion recognition in response to videos.
\newblock \emph{IEEE Transactions on Affective Computing}, 3\penalty0
  (2):\penalty0 211--223, 2012.

\bibitem[Yin et~al.(2020)Yin, Liu, Chen, Zhao, and Wang]{yin2020locally}
Zhong Yin, Lei Liu, Jianing Chen, Boxi Zhao, and Yongxiong Wang.
\newblock Locally robust eeg feature selection for individual-independent
  emotion recognition.
\newblock \emph{Expert Systems with Applications}, 162:\penalty0 113768, 2020.

\bibitem[Wan et~al.(2021)Wan, Yang, Huang, Zeng, and Liu]{wan2021review}
Zitong Wan, Rui Yang, Mengjie Huang, Nianyin Zeng, and Xiaohui Liu.
\newblock A review on transfer learning in eeg signal analysis.
\newblock \emph{Neurocomputing}, 421:\penalty0 1--14, 2021.

\bibitem[Mehmood et~al.(2017)Mehmood, Du, and Lee]{mehmood2017optimal}
Raja~Majid Mehmood, Ruoyu Du, and Hyo~Jong Lee.
\newblock Optimal feature selection and deep learning ensembles method for
  emotion recognition from human brain eeg sensors.
\newblock \emph{IEEE Access}, 5:\penalty0 14797--14806, 2017.

\bibitem[Yin et~al.(2017)Yin, Zhao, Wang, Yang, and Zhang]{yin2017recognition}
Zhong Yin, Mengyuan Zhao, Yongxiong Wang, Jingdong Yang, and Jianhua Zhang.
\newblock Recognition of emotions using multimodal physiological signals and an
  ensemble deep learning model.
\newblock \emph{Computer Methods and Programs in Biomedicine}, 140:\penalty0
  93--110, 2017.

\bibitem[Chen et~al.(2021)Chen, Chang, and Guo]{chen2021emotion}
Yu~Chen, Rui Chang, and Jifeng Guo.
\newblock Emotion recognition of eeg signals based on the ensemble learning
  method: Adaboost.
\newblock \emph{Mathematical Problems in Engineering}, 2021, 2021.

\bibitem[Murugappan et~al.(2007)Murugappan, Rizon, Nagarajan, Yaacob, Zunaidi,
  and Hazry]{murugappan2007eeg}
Murugappn Murugappan, M~Rizon, R~Nagarajan, S~Yaacob, I~Zunaidi, and D~Hazry.
\newblock Eeg feature extraction for classifying emotions using fcm and fkm.
\newblock \emph{International journal of Computers and Communications},
  1\penalty0 (2):\penalty0 21--25, 2007.

\bibitem[Matiko et~al.(2014)Matiko, Beeby, and Tudor]{matiko2014fuzzy}
Joseph~W Matiko, Stephen~P Beeby, and John Tudor.
\newblock Fuzzy logic based emotion classification.
\newblock In \emph{Proceedings of the 2014 IEEE International Conference on
  Acoustics, Speech and Signal Processing (ICASSP)}, pages 4389--4393. IEEE,
  2014.

\bibitem[Soroush et~al.(2018)Soroush, Maghooli, Setarehdan, and
  Nasrabadi]{soroush2018novel}
Morteza~Zangeneh Soroush, Keivan Maghooli, Seyed~Kamaledin Setarehdan, and
  Ali~Motie Nasrabadi.
\newblock A novel method of eeg-based emotion recognition using nonlinear
  features variability and dempster--shafer theory.
\newblock \emph{Biomedical Engineering: Applications, Basis and
  Communications}, 30\penalty0 (04):\penalty0 1850026, 2018.

\bibitem[Guo et~al.(2019)Guo, Chai, Candra, Guo, Song, Nguyen, and
  Su]{guo2019hybrid}
Kairui Guo, Rifai Chai, Henry Candra, Ying Guo, Rong Song, Hung Nguyen, and
  Steven Su.
\newblock A hybrid fuzzy cognitive map/support vector machine approach for
  eeg-based emotion classification using compressed sensing.
\newblock \emph{International Journal of Fuzzy Systems}, 21\penalty0
  (1):\penalty0 263--273, 2019.

\bibitem[Wang et~al.(2018{\natexlab{b}})Wang, Zhong, Peng, Jiang, and
  Liu]{wang2018data}
Fang Wang, Sheng-hua Zhong, Jianfeng Peng, Jianmin Jiang, and Yan Liu.
\newblock Data augmentation for eeg-based emotion recognition with deep
  convolutional neural networks.
\newblock In \emph{International Conference on Multimedia Modeling}, pages
  82--93. Springer, 2018{\natexlab{b}}.

\bibitem[Luo et~al.(2020{\natexlab{a}})Luo, Zhu, Wan, and Lu]{luo2020data}
Yun Luo, Li-Zhen Zhu, Zi-Yu Wan, and Bao-Liang Lu.
\newblock Data augmentation for enhancing eeg-based emotion recognition with
  deep generative models.
\newblock \emph{Journal of Neural Engineering}, 17\penalty0 (5):\penalty0
  056021, 2020{\natexlab{a}}.

\bibitem[Lashgari et~al.(2020)Lashgari, Liang, and Maoz]{lashgari2020data}
Elnaz Lashgari, Dehua Liang, and Uri Maoz.
\newblock Data augmentation for deep-learning-based electroencephalography.
\newblock \emph{Journal of Neuroscience Methods}, page 108885, 2020.

\bibitem[Bhosale et~al.(2022)Bhosale, Chakraborty, and
  Kopparapu]{bhosale2022calibration}
Swapnil Bhosale, Rupayan Chakraborty, and Sunil~Kumar Kopparapu.
\newblock Calibration free meta learning based approach for subject independent
  eeg emotion recognition.
\newblock \emph{Biomedical Signal Processing and Control}, 72:\penalty0 103289,
  2022.

\bibitem[He et~al.(2021)He, Zhao, and Chu]{he2021automl}
Xin He, Kaiyong Zhao, and Xiaowen Chu.
\newblock Automl: A survey of the state-of-the-art.
\newblock \emph{Knowledge-Based Systems}, 212:\penalty0 106622, 2021.

\bibitem[He et~al.(2020)He, Tan, Ying, and Zhang]{he2020strengthen}
Hong He, Yonghong Tan, Jun Ying, and Wuxiong Zhang.
\newblock Strengthen eeg-based emotion recognition using firefly integrated
  optimization algorithm.
\newblock \emph{Applied Soft Computing}, 94:\penalty0 106426, 2020.

\bibitem[Aquino-Br{\'\i}tez et~al.(2021)Aquino-Br{\'\i}tez, Ortiz, Ortega,
  Le{\'o}n, Formoso, Gan, and Escobar]{aquino2021optimization}
Diego Aquino-Br{\'\i}tez, Andr{\'e}s Ortiz, Julio Ortega, Javier Le{\'o}n,
  Marco Formoso, John~Q Gan, and Juan~Jos{\'e} Escobar.
\newblock Optimization of deep architectures for eeg signal classification: An
  automl approach using evolutionary algorithms.
\newblock \emph{Sensors}, 21\penalty0 (6):\penalty0 2096, 2021.

\bibitem[Zheng et~al.(2017)Zheng, Zhu, and Lu]{zheng2017identifying}
Weilong Zheng, Jiayi Zhu, and Baoliang Lu.
\newblock Identifying stable patterns over time for emotion recognition from
  eeg.
\newblock \emph{arXiv preprint arXiv:1601.02197}, 2017.

\bibitem[Katsigiannis and Ramzan(2017)]{2017DREAMER}
S.~Katsigiannis and N.~Ramzan.
\newblock Dreamer: A database for emotion recognition through eeg and ecg
  signals from wireless low-cost off-the-shelf devices.
\newblock \emph{IEEE Journal of Biomedical \& Health Informatics}, pages 1--1,
  2017.

\bibitem[Zheng et~al.(2018)Zheng, Liu, Lu, Lu, and
  Cichocki]{zheng2018emotionmeter}
Wei-Long Zheng, Wei Liu, Yifei Lu, Bao-Liang Lu, and Andrzej Cichocki.
\newblock Emotionmeter: A multimodal framework for recognizing human emotions.
\newblock \emph{IEEE transactions on cybernetics}, 49\penalty0 (3):\penalty0
  1110--1122, 2018.

\bibitem[Song et~al.(2019)Song, Zheng, Lu, Zong, Zhang, and Cui]{song2019mped}
Tengfei Song, Wenming Zheng, Cheng Lu, Yuan Zong, Xilei Zhang, and Zhen Cui.
\newblock Mped: A multi-modal physiological emotion database for discrete
  emotion recognition.
\newblock \emph{IEEE Access}, 7:\penalty0 12177--12191, 2019.

\bibitem[Ma et~al.(2019{\natexlab{a}})Ma, Li, Zheng, and Lu]{ma2019reducing}
Bo-Qun Ma, He~Li, Wei-Long Zheng, and Bao-Liang Lu.
\newblock Reducing the subject variability of eeg signals with adversarial
  domain generalization.
\newblock In \emph{International Conference on Neural Information Processing},
  pages 30--42. Springer, 2019{\natexlab{a}}.

\bibitem[Ma et~al.(2019{\natexlab{b}})Ma, Tang, Zheng, and Lu]{ma2019emotion}
Jiaxin Ma, Hao Tang, Wei-Long Zheng, and Bao-Liang Lu.
\newblock Emotion recognition using multimodal residual lstm network.
\newblock In \emph{Proceedings of the 27th ACM International Conference on
  Multimedia}, pages 176--183, 2019{\natexlab{b}}.

\bibitem[Luo et~al.(2020{\natexlab{b}})Luo, Fu, Xie, Qin, Wu, Liu, Jiang, Cao,
  and Ding]{luo2020eeg}
Yuling Luo, Qiang Fu, Juntao Xie, Yunbai Qin, Guopei Wu, Junxiu Liu, Frank
  Jiang, Yi~Cao, and Xuemei Ding.
\newblock Eeg-based emotion classification using spiking neural networks.
\newblock \emph{IEEE Access}, 8:\penalty0 46007--46016, 2020{\natexlab{b}}.

\bibitem[Cui et~al.(2017)Cui, Ahmad, and Hawkins]{cui2017htm}
Yuwei Cui, Subutai Ahmad, and Jeff Hawkins.
\newblock The htm spatial pooler—a neocortical algorithm for online sparse
  distributed coding.
\newblock \emph{Frontiers in computational neuroscience}, 11:\penalty0 111,
  2017.

\bibitem[Pouyanfar et~al.(2018)Pouyanfar, Sadiq, Yan, Tian, Tao, Reyes, Shyu,
  Chen, and Iyengar]{pouyanfar2018survey}
Samira Pouyanfar, Saad Sadiq, Yilin Yan, Haiman Tian, Yudong Tao, Maria~Presa
  Reyes, Mei-Ling Shyu, Shu-Ching Chen, and SS~Iyengar.
\newblock A survey on deep learning: Algorithms, techniques, and applications.
\newblock \emph{ACM Computing Surveys (CSUR)}, 51\penalty0 (5):\penalty0 1--36,
  2018.

\bibitem[Mordvintsev et~al.(2015)Mordvintsev, Olah, and
  Tyka]{mordvintsev2015inceptionism}
Alexander Mordvintsev, Christopher Olah, and Mike Tyka.
\newblock Inceptionism: Going deeper into neural networks, 2015.
\newblock URL
  \url{https://research.googleblog.com/2015/06/inceptionism-going-deeper-into-neural.html}.

\bibitem[Ribeiro et~al.(2016)Ribeiro, Singh, and Guestrin]{ribeiro2016model}
Marco~Tulio Ribeiro, Sameer Singh, and Carlos Guestrin.
\newblock Model-agnostic interpretability of machine learning.
\newblock \emph{arXiv preprint arXiv:1606.05386}, 2016.

\bibitem[Qiu et~al.(2020)Qiu, Sun, Xu, Shao, Dai, and Huang]{qiu2020pre}
Xipeng Qiu, Tianxiang Sun, Yige Xu, Yunfan Shao, Ning Dai, and Xuanjing Huang.
\newblock Pre-trained models for natural language processing: A survey.
\newblock \emph{Science China Technological Sciences}, pages 1--26, 2020.

\end{thebibliography}
\end{spacing}


\end{document}